\documentclass[a4paper, 11pt]{article}
\usepackage[default]{lato}
\usepackage[T1]{fontenc}
\usepackage[left=1.5cm, right=1cm, top=0.5cm, bottom=1cm, includeheadfoot]{geometry}
\usepackage[english]{babel}
\usepackage[utf8]{inputenc}
\usepackage[authoryear,round]{natbib}
\usepackage{framed}
\usepackage{amsmath,amsfonts,amssymb,amsthm,mathtools,bbm}
\usepackage{booktabs, multirow}
\usepackage{adjustbox}
\usepackage{float}
\usepackage{listings}
\usepackage{color}
\usepackage{url}
\usepackage{hyperref}
\usepackage{xpatch}
\usepackage{xcolor}
\usepackage{listings}
\usepackage{realboxes}
\usepackage{marginnote}
\usepackage{changepage}
\usepackage{fancyhdr}
\usepackage{afterpage}
\usepackage{array}
\usepackage{algorithm}
\usepackage{algpseudocode}
\usepackage[many]{tcolorbox}
\usepackage[nodisplayskipstretch]{setspace}
\usepackage{longtable}
\usepackage{caption}
\newtheorem*{theorem}{Theorem}
\newtheorem*{lemma}{Lemma}

\date{}
\newcolumntype{P}[1]{>{\centering\arraybackslash}p{#1}}
\newcommand{\orcid}[1]{\href{https://orcid.org/#1}{\includegraphics[height=\fontcharht\font`A]{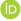}}}

\hypersetup{
	colorlinks=true,
	linkcolor=blue,
	citecolor=magenta,   
	urlcolor=cyan,
	pdfpagemode=FullScreen,
}

\DeclareMathOperator{\EX}{\mathbb{E}}
\DeclareMathAlphabet{\mathpzc}{OT1}{pzc}{m}{it}
\definecolor{applegreen}{rgb}{0.55, 0.71, 0.0}
\definecolor{darkolivegreen}{rgb}{0.33, 0.42, 0.18}
\definecolor{codecolors}{RGB}{ 219,225,226}
\definecolor{codecolorsinline}{RGB}{ 230.1000,  234.3000 , 235.0000}

\newcommand{\consolein}{\noindent\par\reversemarginpar\marginnote{\small\noindent\textcolor{applegreen}{In [~]}:}[0.4cm]}
\newcommand{\consoleout}{\noindent\par\reversemarginpar\marginnote{\small\noindent\textcolor{gray}{Out [~]}:}[0.4cm]}
\tcolorboxenvironment{lstlisting}{
	spartan,
	colframe=black,
	left=1mm,right=1mm,top=-2mm,bottom=-2mm,
	colback=codecolors
}

\lstdefinestyle{codeinput}{
	mathescape=true,
	keywordstyle=\color{blue}\bf,
	commentstyle=\color{darkolivegreen},
	stringstyle=\color{darkolivegreen},
	emphstyle=\color{magenta},	
	basicstyle=\footnotesize\ttfamily,
	breaklines=true
}

\lstdefinestyle{codeinputinline}{
	backgroundcolor = \color{codecolorsinline},
	mathescape=true,
	keywordstyle=\color{blue}\bf,
	commentstyle=\color{green},
	stringstyle=\color{darkolivegreen},
	emphstyle=\color{magenta},	
	basicstyle=\normalsize\ttfamily,
}

\lstdefinestyle{codeoutput}{
	basicstyle=\footnotesize\ttfamily,
	mathescape=true,
}

\makeatletter
\xpretocmd\lstinline{\Colorbox{codecolorsinline}\bgroup\appto\lst@DeInit{\egroup}}{}{}
\makeatother

\makeatletter
\renewcommand{\env@cases}[1][@{}l@{\quad}l@{}]{%
	\let\@ifnextchar\new@ifnextchar
	\left\lbrace
	\def\arraystretch{1.2}%
	\array{#1}%
}
\makeatother

\title{\textbf{MIXALIME}: MIXture models for ALlelic IMbalance Estimation in high-throughput sequencing data}
\author{
	\begin{tabular}{ccc}
		Georgy Meshcheryakov$^{1}$ \orcid{0000-0003-0751-8286} &
		Sergey Abramov$^{2}$ \orcid{0000-0002-3334-5334} & Aleksandr Boytsov$^{2,3,4}$  \\
		\href{mailto:iam@georgy.top}{\texttt{iam@georgy.top}} &
		\href{mailto:sabramov@altius.org}{\texttt{sabramov@altius.org}} & \href{mailto:boytsovs.av@phystech.edu}{\texttt{boytsovs.av@phystech.edu}}
	\end{tabular}\\~\\
	\begin{tabular}{ccc}
		Andrey I. Buyan$^{1}$  & Vsevolod J. Makeev$^{3,4}$ \orcid{0000-0001-9405-9748} &
		Ivan V. Kulakovskiy$^{1,4}$ \orcid{0000-0002-6554-8128} \\ \href{mailto:andreybuyanchik@gmail.com}{\texttt{andreybuyanchik@gmail.com}} & 
		\href{mailto:vsevolod.makeev@vigg.ru}{\texttt{vsevolod.makeev@vigg.ru}} & \href{mailto:ivan.kulakovskiy@gmail.com}{\texttt{ivan.kulakovskiy@gmail.com}}
	\end{tabular}
}

\begin{document}
	\maketitle
	\thispagestyle{empty}
	\begin{abstract}
		Modern high-throughput sequencing assays efficiently capture not only gene expression and different levels of gene regulation but also a multitude of genome variants. Focused analysis of alternative alleles of variable sites at homologous chromosomes of the human genome reveals allele-specific gene expression and allele-specific gene regulation by assessing allelic imbalance of read counts at individual sites. Here we formally describe an advanced statistical framework for detecting the allelic imbalance in allelic read counts at single-nucleotide variants detected in diverse omics studies (ChIP-Seq, ATAC-Seq, DNase-Seq, CAGE-Seq, and others). \textbf{MIXALIME} accounts for copy-number variants and aneuploidy, reference read mapping bias, and provides several scoring models to balance between sensitivity and specificity when scoring data with varying levels of experimental noise-caused overdispersion.
	\end{abstract}
	
	\footnotetext[1]{Institute of Protein Research, Russian Academy of Sciences, Puschino, Russia}
	\footnotetext[2]{Altius Institute for Biomedical Sciences, Seattle, WA, United States}
	\footnotetext[3]{Moscow Institute of Physics and Technology, Moscow, Russia}
	\footnotetext[4]{Vavilov Institute of General Genetics, Russian Academy of Sciences, Moscow, Russia}
	
	{\small\hfill{Code availability:  \href{https://github.com/autosome-ru/MixALime}{github.com/autosome-ru/MixALime}}}

	\section{Introduction}
	
	Let $\{c_i\}_{i=1}^{n}$ denote random variables (r.vs.) that model a single read emitted from a chromosome carrying a reference allele of single-nucleotide variant (SNV) in a high-throughput sequencing experiment with a total number of reads at the SNV being $n$. Straightforward reasoning implies that $c_i$ is a Bernoulli r.v. with some success probability $p$: $c_{i} \sim \mathpzc{Bernoulli}(p)$. Likewise, reads from an alternative allele are distributed as $\hat{c_i} \sim \mathpzc{Bernoulli}(1 - p)$;  $p$ is usually a known fixed value, e.g. in the case of a diploid genome without any copy-number variants $p = \frac{1}{2}$). The number of reads supporting a reference allele $x$ (or an alternative allele $y$) is distributed as a binomial random variable by a definition as a sum of i.i.d. Bernoulli r.vs.:
	\begin{equation}\label{eq:binomial_model}
		y \sim \mathpzc{Binom}(n, p),~~ x \sim \mathpzc{Binom}(n, 1 - p).
	\end{equation}
	The model assumption holds for the case when there is \underline{no} allele-specificity (AS), and we expect AS SNVs to deviate from this model, which can be tested with the simple two-tailed binomial test \citep{binomtest}. This reasoning sometimes suffices and is employed by existing methods such as \textbf{AlleleSeq} \citep{alleleseq}. A more robust approach is to assume that $p$ is not a fixed known value, but a $\mathpzc{Beta}(\alpha, \beta)$ r.v., where $\alpha, \beta$ are parameters to be estimated. Parameter $p$ can be marginalized out by integration and then $$y \sim \mathpzc{BetaBinom}(n, \alpha_x, \beta_x),~~x \sim \mathpzc{BetaBinom}(n, \alpha_y, \beta_y).$$
	For instance, this approach is followed by \textbf{StratAS} \citep{stratas} with parameters estimated locally in continuous genomic regions with copy numbers known from external annotation.
	
	One alternative approach is to assume that $y$ is distributed as a negative binomial random variable conditioned on the $x$ and vice-versa \citep{nb_pioneers}: $$y \sim \mathpzc{NB}(x, p), ~~ x \sim \mathpzc{NB}(y, 1 - p)$$
	There are numerous ways to motivate the choice of NB as a read counts distribution:
	\begin{enumerate}
		\item Non-rigorously, we know that there were at least $r = y$ failures among a total of $n=x+y$ read ''attempts''. This corresponds to the way NB is usually introduced intuitively. Of note, this interpretation is not formally correct, as the standard definition of the Negative Binomial assumes the trials are performed until a predefined number of failures is met. In turn, it would mean that the order of appear for reads supporting different alleles could affect the total variant coverage;
		\item Let's revisit the binomial distribution once more. The probability of observing $y$ given a total number of reads $n = x + y$ is $\binom{n}{y}p^y (1-p)^{n - y} = \binom{x + y}{y}p^y (1-p)^{x}$. After the substitution of $n = x +y$ this equation, however, is no longer a viable probability mass function (PMF) for a varying $x$: note that what was the $n$ parameter increases together with $x$, and, consequently, the value of the equation is non-zero for all $x \in \mathbbm{Z}_{+}$, whereas for the binomial random variable the support is bounded at the fixed value $n$. The valid PMF is easily obtained by estimating the normalizing constant: $c = \sum_{i=0}^\infty \binom{x + y}{y} p^y (1-p)^x = \frac{1}{1-p}$ and then:
		$$f(y|r=x, p) = \frac{\binom{x + y}{y} p^y (1-p)^x}{c} = \binom{x + y}{y} p^y (1-p)^{x+1} = \binom{x + y}{x} p^y (1-p)^{x+1}.$$ 
		That's very similar to the negative binomial PMF, albeit with a tiny difference -- the $r = x$ parameter is shifted by 1. This is actually reasonable as a read count on an allele is never below $1$.
		\item  A better formal substantiation for the Negative Binomial model in allelic read counting  is provided in Appendix~\ref{app:nb_origins}. Briefly, it arises from the assumption that the coverage $n$ is $\mathpzc{NB}$-distributed and the number of reads conditioned on the coverage is a binomial random variable.
	\end{enumerate}

	However, the naive binomial approach doesn't account for neither the problem of reference mapping bias nor the possible presence of CNVs. The less naive beta-binomial and negative-binomial approaches does so only indirectly by increasing the dispersion of the null distribution. Next, we describe a family of alternative approaches that tackle:
	\begin{itemize}
		\item reference mapping bias;
		\item CNVs and/or aneuploidy;
		\item other non-attributed e.g. experiment-specific sources of noise and variation in the underlying data.
	\end{itemize}
	
	All members of the proposed family are based on the negative binomial approach explained above. The necessity of choosing NB over binomial distribution is due to the assumption that the mean number of reads mapped to one allele (proportional to $r$) is linearly dependent on the read counts mapped to the other allele. We incorporate this assumption into the model by having $r$ linearly depend on the read count at the other allele:  
	\begin{equation}\label{eq:refbias}
		r(x, b, a) = b  x + a,
	\end{equation}
	where $b, a$ are some parameters that ought to be estimated from the data. The adequacy of this assumption is evaluated in Appendix~\ref{app:nb}.
	
\begin{figure}[t!]
		\centering
		\begin{tabular}{ll}
			\raisebox{7em}{\Huge A} &
			\begin{adjustbox}{max width=0.94\textwidth}
				\includegraphics{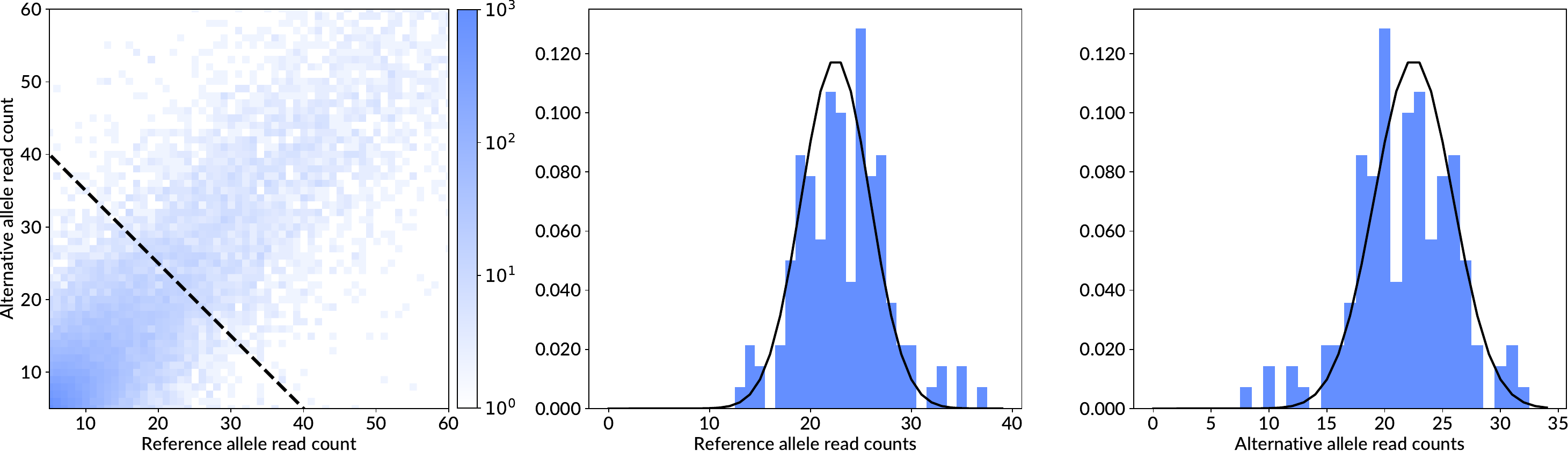}
			\end{adjustbox}\\
			\raisebox{7em}{\Huge B} &
			\begin{adjustbox}{max width=0.94\textwidth}
				\includegraphics{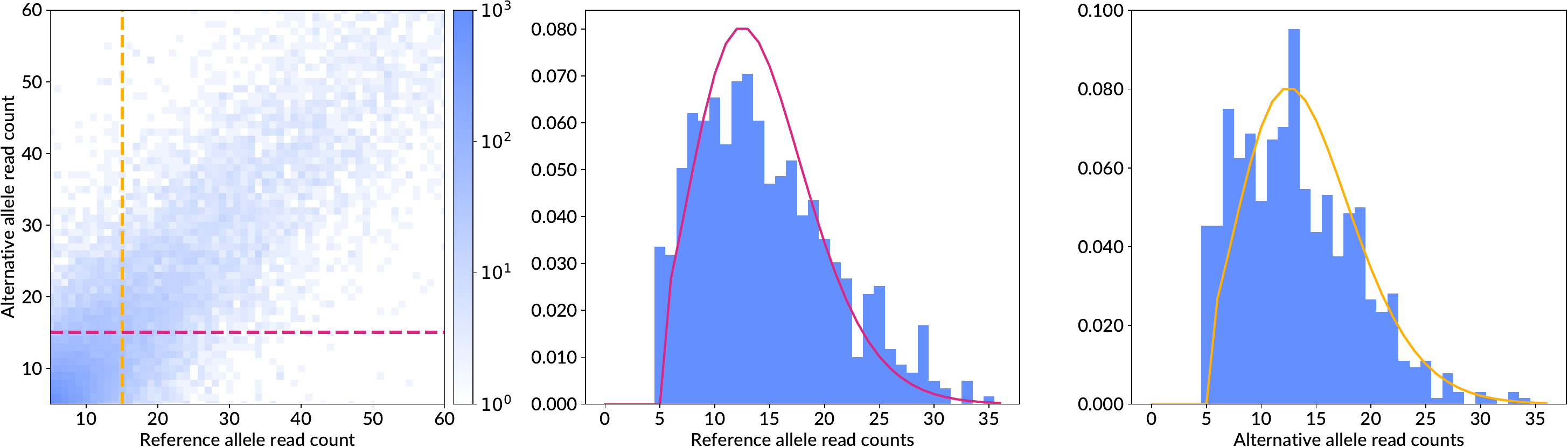}
			\end{adjustbox}
		\end{tabular}
		\caption{Heatmap and density plots for allelic read read counts. Heatmap is plotted for counts starting at $5$ (as counts below $5$ were filtered from the dataset as noisy observations). \textbf{A}. Binomial distributions $P(x| x + y = 45), P(y| x + y = 45)$ fitted to all counts that lie on the black dashed line. \textbf{B}. Negative binomial distributions $P(x|y = 15), P(y|x = 15)$ fitted to counts that line on the red and the orange dashed lines for reference and alternative allele read counts respectively.}
		\label{fig:slices}
	\end{figure}

	Note that naturally, we should've considered a hypothetical joint distribution of $x$ and $y$ with a probability mass function $f(x, y)$. This, though, is inconceivable for us and we limit ourselves to the conditional distributions $f(y|x=r_y), f(x|y=r_x)$, i.e. a distribution of reference allele read counts given alternative allele read counts and vice-versa. One can think of it as considering a distribution of read counts obtained by taking a horizontal slice at a given $r_x$ level (and a vertical slice for $x$ at a given $r_y$ level). This is illustrated by Figure~\ref{fig:slices}-B. In turn, we try to approximate the joint distribution  of $x, y$ by considering all possible horizontal and vertical slices with $r$ varying with each slice as in Equation~\ref{eq:refbias}, which effectively links ''slices'' together. The key feature of this approach is that it enables separate scoring of allelic imbalance favoring each of the two alleles. This way we model the reference mapping bias implicitly as a difference between $r$ parameters for reference and alternative distributions.
	
	\section{Negative binomial model}
	
	We assume linear reference bias as in Equation~\ref{eq:refbias} and a negative binomial distribution of $y$ for a given $x$, and, symmetrically, the same for $y$,
	
	\begin{equation}\label{eq:nbmodel}
		\begin{split}
			y \sim \mathpzc{LeftTruncatedNB}(r(x, b_x, a_x), p, l),\\
			x \sim \mathpzc{LeftTruncatedNB}(r(y, b_y, a_y), p, l),
		\end{split}
	\end{equation}
	where $\mathpzc{LeftTruncatedNB}$ is the left truncated at $l$ negative binomial distribution.
	
	As for the left truncation at $l$, we introduced it as the low-coverage SNVs are often noisy due to the SNP-calling errors and should be filtered out from the data, thus requiring to augment the  distribution function accordingly. If a PMF of some distribution $\mathcal{D}$ is $g(y)$, its cumulative distribution function (CDF) is $G(y)$, then PMF of the left truncated at $l$ version of $\mathcal{D}$ is $f(y)$ \citep{truncdist}:
	
	\begin{equation}\label{eq:trunc}
		f(y, l) = \frac{g(y) \mathbbm{1}_{y \ge l}}{1 - G(l)},
	\end{equation}
	where $\mathbbm{1}$ is an indicator function.

	\section{Beta negative binomial model}
	
	Similarly to a beta-binomial model, we can assume that $p \sim \mathpzc{Beta}(\alpha, \beta)$. We apply a convenient reparametrization of $\mathpzc{Beta}$ in terms of its mean $\mu$ and "concentration" $\kappa$ \citep{mixalime_bgrs}: 
	
	\begin{equation}\label{eq:beta}
		p \sim \mathpzc{Beta}(\mu, \kappa), ~\alpha = \mu \kappa,~ \beta = (1 - \mu) \kappa.
	\end{equation}
	Note that $\EX[p] = \mu$ and $var[p] = \frac{(1 - \mu)\kappa}{\kappa + 1}$. It follows that $\mu$ is indeed the mean of $p$ and $\kappa$ controls the variance of $p$. The higher the "concentration" $\kappa$, the lower the variance. Then, by combining the previous model from the Equation~\ref{eq:nbmodel} with Equation~\ref{eq:beta} and integrating out $p$ we obtain the following model:

	\begin{equation}\label{eq:bnb_model}
		\begin{split}
			y \sim \mathpzc{LeftTruncatedBetaNB}(r(x, b_x, a_x), \mu, \kappa, l),\\
			x \sim \mathpzc{LeftTruncatedBetaNB}(r(y, b_y, a_y), \mu, \kappa, l),
		\end{split}
	\end{equation}
	where $\mathpzc{LeftTruncatedBetaNB}$ is the left truncated at $l$ beta negative binomial distribution \citep{betanb}. Likewise, the computation of the truncated distribution is performed with the general Equation~\ref{eq:trunc}.

	Note that as $\kappa \rightarrow \infty$, the beta negative binomial model converges to the negative binomial model. In practice, due to the limitations of numerical algebra and finiteness of underlying data, the exact convergence does not happen. This might lead to a loss of sensitivity as the beta negative binomial distribution is heavy-tailed. Therefore, the beta negative binomial model should not be used as a general substitute for the regular negative binomial model, but can be used as a very conservative test.
	
	\section{Marginalized compound negative binomial model}
	When formulating NB (Equation~\ref{eq:nbmodel}) and BetaNB (Equation~\ref{eq:bnb_model}) models, we assumed that the read count at the preselected (fixed) allele is precisely known. In practice this is not true, as read counts are prone to errors, i.e. the read counts at a preselected allele should be a random variable themselves. We consider an assumption that measurements of an alternative allele read count $y$ are distributed as a zero-truncated binomial random variable (next $\mathpzc{ZTBin}$) to be a reasonable one. The zero-truncation is necessary to accommodate for the two facts: allele-specificity does not make sense at homozygous SNVs, and technically $y > 0 $ in a $NB$ distribution. 
	
	Let's consider the following model:
	
	\begin{equation}\label{eq:pre_mcnb}
		y \sim \mathpzc{NB}(\hat{x}, p), ~\hat{x} \sim \mathpzc{ZTBin}(r, 1 - p).
	\end{equation}
	It turns out that $\hat{x}$ can be marginalized out and a marginal distribution of $y$ can be obtained:
	\begin{equation}\label{eq:mcnb}
		f_{\mathpzc{MCNB}}(y|r, p) = \frac{r (p-1)^2 p^{r+y-1} \, _2F_1\left(1-r,y+1;2;-\frac{(1-p)^2}{p}\right)}{1 - p^r},
	\end{equation}
	where $ \, _2F_1$ is Gauss hypergeometric function defined as $\,_2F_1(a, b c;z) = \sum_{i=0}^\infty \beta_i, ~\beta_0 = 1, \frac{\beta_{i+1}}{\beta_i} = \frac{(i + a) (i + b)}{(i + c) (i + 1)} z^i$. We avoid computing it using the definition and instead use the reccurrent formulae for $f_{\mathpzc{MCNB}}$ that we've inferred (see Appendix~\ref{app:mcnb_loglik}).
	We shall call this law ''Marginalized compound negative binomial'' distribution or MCNB for short \citep{sbb_school}. For proof on why the Equation~\ref{eq:mcnb} holds, see Appendix~\ref{app:mcnb_proof}. So, the model proposed in this section is
	\begin{equation}
		\begin{split}
			y \sim \mathpzc{LeftTruncatedMCNB}(r(x, b_x, a_x), p, l),\\
			x \sim \mathpzc{LeftTruncatedMCNB}(r(y, b_y, a_y), p, l).
		\end{split}
	\end{equation}
	
	Note: see in ~\ref{table:moments}, that interpretation of the $r$ parameter for MCNB varies significantly from $NB$ and $BetaNB$. That's because for the latter two $r$ can be thought of as a number of counts of an alternative allele, whereas here $r$ was introduced as a total number $x + y$ of read counts. It becomes more evident as $\lim_{r\rightarrow \infty} \EX[x] = r p$, which is exactly 2 times greater than a mean of NB for $p = \frac{1}{2}$. We shall deal with this nuisance later in Section~\ref{sec:r}.

	\section{Mixture model}
	\begin{figure}[t!]
		\centering
		\begin{adjustbox}{max width=\textwidth}
			\includegraphics{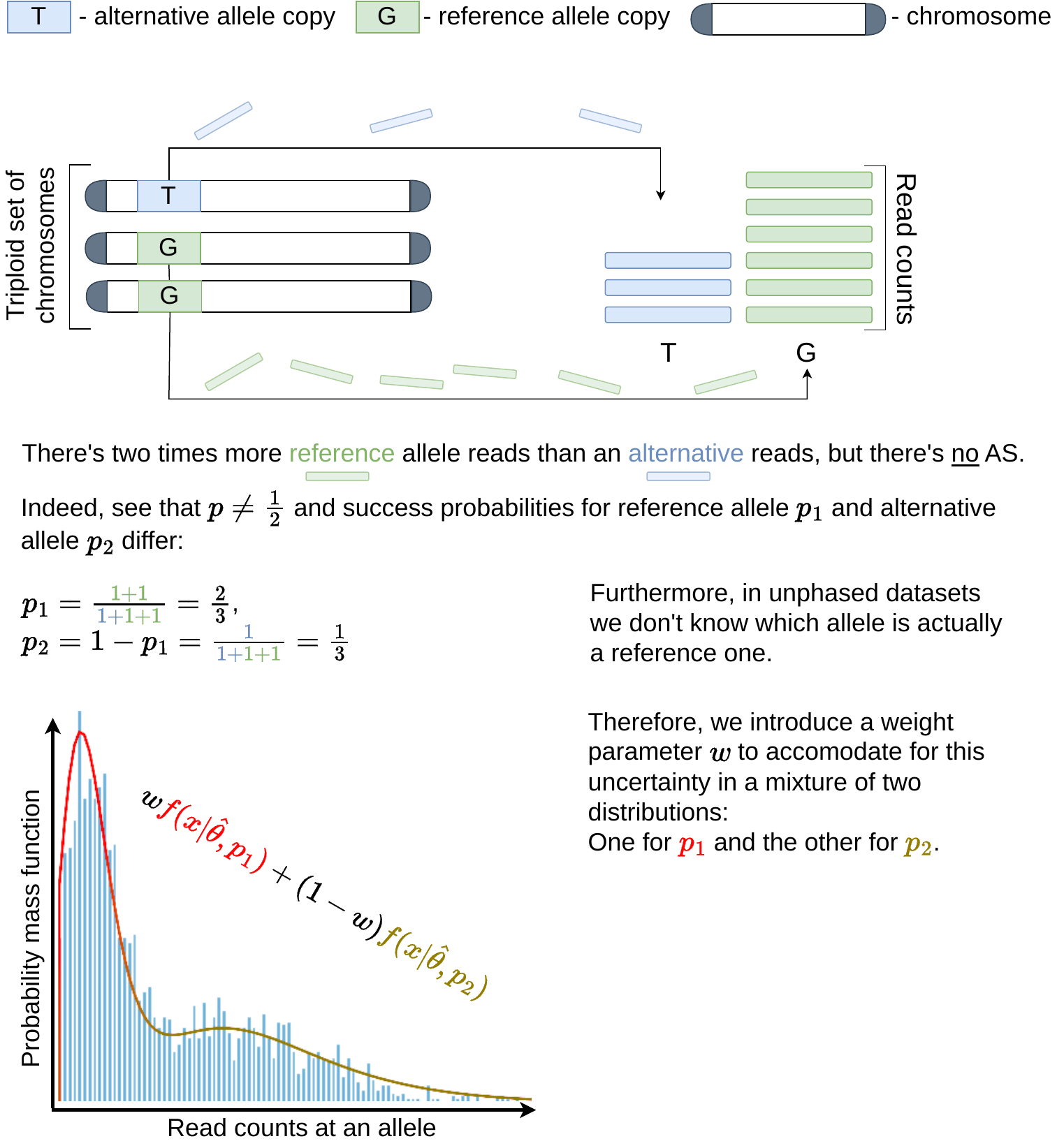}
		\end{adjustbox}
		\caption{Graphical representation of the mixture distribution idea for case when $BAD = 2$. Two components of the $f_{\mathpzc{Mix}}$ mixture are colored differently. The same logic applies to CNVs as well.}
		\label{fig:mixture}
	\end{figure}
	Suppose that $p \neq \frac{1}{2}$ (or $\mu \neq \frac{1}{2}$ in the case of $BetaNB$) which happens for SNVs located in CNVs or duplicated chromosomes. For instance, there might have 3 copies of a maternal allele and one copy of a paternal allele in a tetraploid organism, which results in $p_{m} = \frac{3}{4}$ and $p_{p} = \frac{1}{4} = 1 - p_{m}$. Most of the time, the completely phased personal genome and even partial haplotypes are not available, i.e. the exact number of copies of the reference and the alternative allele for any particular SNP remain unknown. However, it is often possible to estimate the ratio of the major to the minor allele copy numbers , that is the relative background allelic dosage (BAD), directly from SNP calls with an unsupervised approach \citep{babachi} or from an experimentally obtained CNV map \citep{adastra}. 
	
	We tackle this problem by assuming that each read is coming from the one (e.g. 'maternal') chromosome with probability $w$ and from the other chromosome (e.g. paternal) with probability $1-w$, where the balance between $w$ and $1-w$ reflects BAD. This is done naturally with the mixture distribution:
	
	$$f_{\mathpzc{Mix}}(x|p, \hat{\theta}) = w f(x|p, \hat{\theta}) + (1 - w) f(x|1 - p, \hat{\theta}),$$
	where $f$ is a distribution function of either NB, BetaNB or MCNB models, $\hat{\theta}$ is a parameter vector with $p$ excluded, $p = \frac{BAD}{BAD + 1}$ and $w$ is a weight in the mixture model (see Figure~\ref{fig:mixture}), an active parameter to be estimated from the data.

	\section{\texorpdfstring{Regularization by reparametrization of $r$}{Regularization by reparametrization of r}}\label{sec:r}
	\begin{table}[t!]
		\centering
		\begin{tabular}{c|c|c|c}
			& NB & MCNB & BetaNB\\
			\hline
			$\EX[x]$ & $\frac{r p}{1 - p}$ & $\frac{r p} {1 - p^r}$  & $\frac{r \mu \kappa}{(1-\mu)\kappa - 1}$\\
			$var[x]$ & $\frac{rp}{(1 - p)^2}$ & $\frac{p r \left(p^2+\left(p^2 (r-1)-p r-1\right) p^r+1\right)}{(1-p) \left(1 - p^r\right)^2}$ & $\frac{(\kappa-1) \kappa \mu r (\kappa (\mu-1)-r+1)}{(\kappa (\mu-1)+1)^2 (\kappa (\mu-1)+2)}$\\
			$\frac{var[x]}{\EX[x]}$ & $\frac{1}{1 - p}$ & $\frac{p r}{p^r-1}+\frac{p^2+1}{1-p}+p r$ & $\frac{(1-\kappa) (\kappa (\mu-1)-r+1)}{(\kappa (\mu-1)+1) (\kappa (\mu-1)+2)}$
		\end{tabular}
		\caption{Mean and variances for NB, MCNB and BetaNB distributions. For the derivation of MCNB moments, see Appendix~\ref{app:mcnb_props}.}
		\label{table:moments}
	\end{table}
	
	\begin{figure}[t!]
		\centering
		\begin{adjustbox}{max width=\textwidth}
			\includegraphics{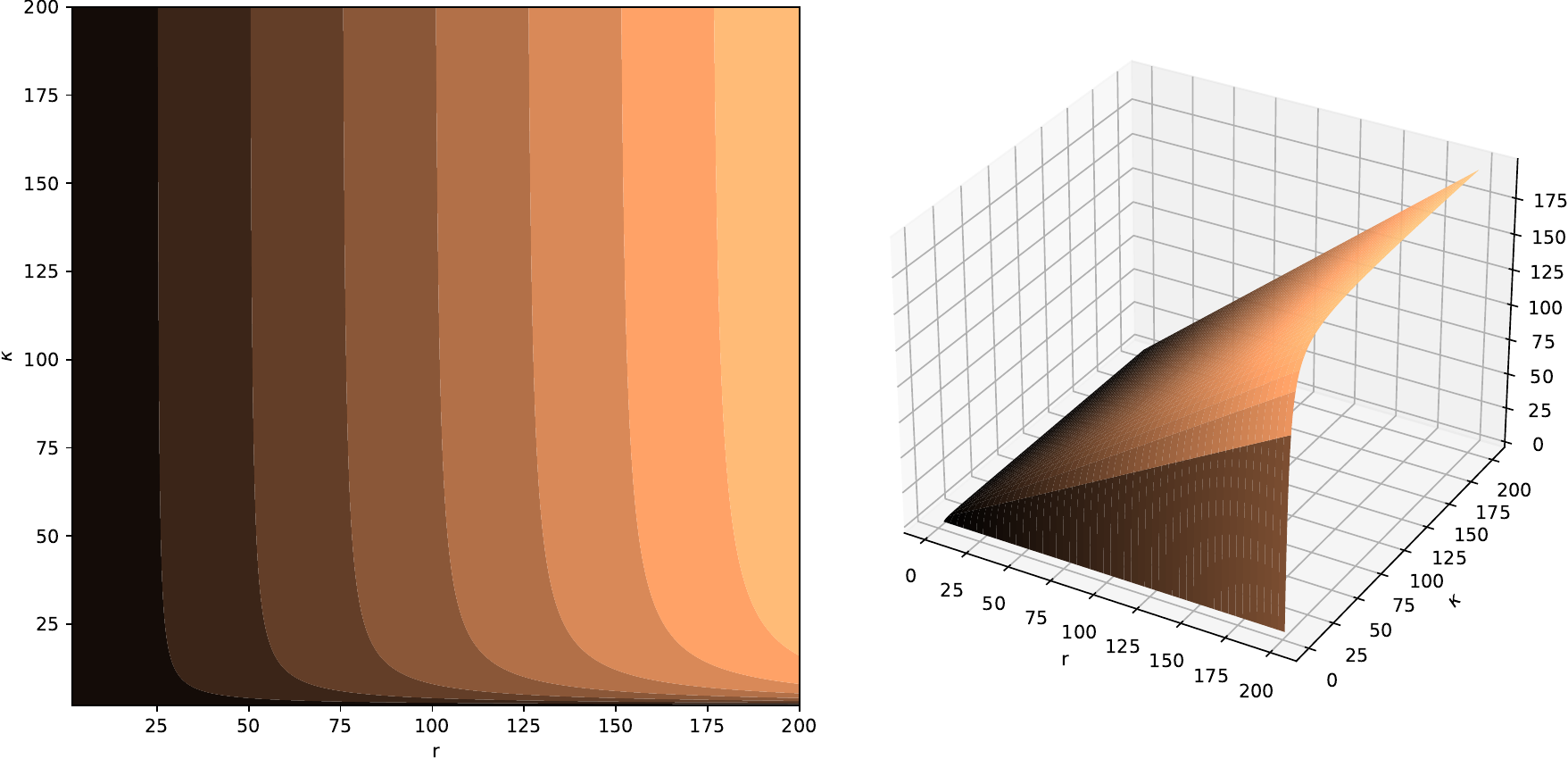}
		\end{adjustbox}
		\begin{adjustbox}{max width=0.6\textwidth}
			\includegraphics{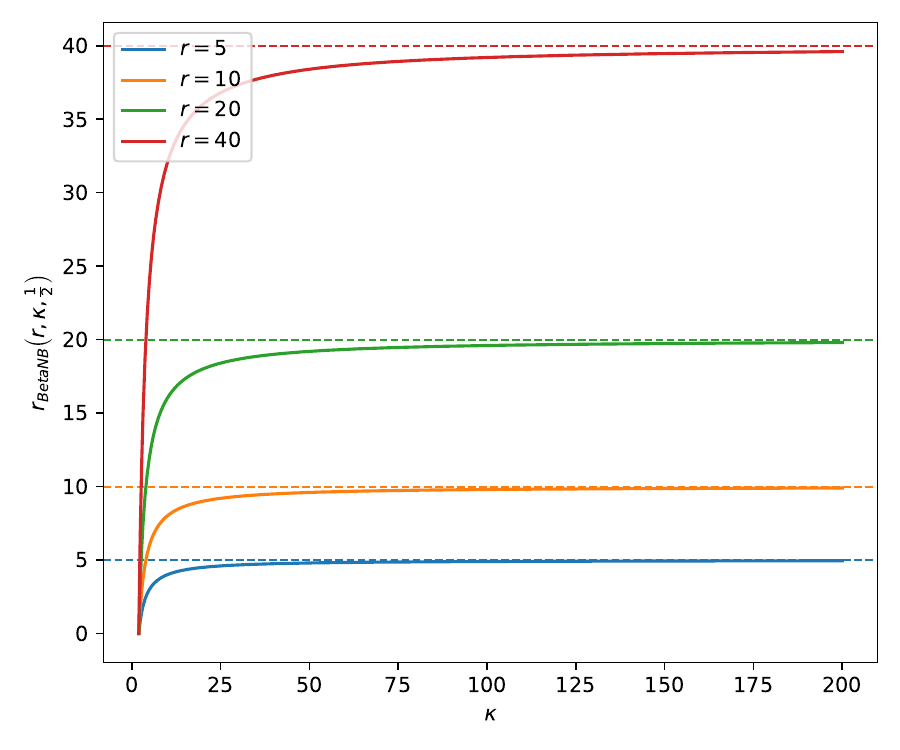}
		\end{adjustbox}
		\caption{Contour plot, 3D surface and 1D plot of $r_{\mathpzc{BetaNB}}(r, \kappa, \mu = \frac{1}{2})$. Notice how for high values of $\kappa$, the surface degenerates into what is effectively a plane that allows any combination of $r$ and $\kappa$. All 3 figures illustrate the same observation: reparametrization ceases to constrain parameter values as $\kappa$ grows.}
		\label{fig:reparam}
	\end{figure}
	Following the original definition, the $p$ variable is effectively linked to BAD and can be interpreted as a fraction of copies of a genome segment carrying the major allele of an SNP. This interpretation stands for the  binomial (Equation~\ref{eq:binomial_model}) or the NB model (Equation~\ref{eq:nbmodel}). 
	
	However,  BetaNB mean value is shifted for small values of $\kappa$ relatively to NB. For MCNB this interpretation is also misleading due to the different nature of $r$ parameter, which reflects rather a total read count rather than the read count supporting an alternative allele), see Table~\ref{table:moments}. Therefore, to make BetaNB and MCNB work adequately with BADs higher than 1 and maintain interpretability of the $r$ parameter, we transform it for NB and BetaNB so the expected values of the distributions agree with that of NB:
	$$r_{\mathpzc{MCNB}} = r \frac{1 - p^r}{1 - p},~ r_{\mathpzc{BetaNB}} = r \frac{(1-\mu)\kappa - 1}{\kappa (1 - \mu)}.$$
	$r_{\mathpzc{MCNB}}$ merely rescales $r$ parameter and does not constrain the parameter space. $r_{\mathpzc{BetaNB}}$, on the hand, links $r$ and $\kappa$ together that is more prominent for low values of the concentration parameter $\kappa$ and doesn't exist for high values $\kappa$ (which is expected as $\mathpzc{BetaNB} \rightarrow \mathpzc{NB}$ as $\kappa \rightarrow \infty$), see Figure~\ref{fig:reparam}.
	
	\section{Model parameters estimation}
	
	We use the maximum likelihood approach to obtain parameter estimates of the model distribution, i.e. we maximize a log-likelihood objective function with respect to its parameters vector $\theta$:
	\begin{equation}\label{eq:loglik}
		\mathcal{L}(\theta, X, Y, l) = \sum_{i=1}^n \ln f(y_i|\theta, x_i, l),
	\end{equation}
	where $X = \left\{x_{i}\right\}_{i=1}^n$,  $Y = \left\{y_{i}\right\}_{i=1}^n$ (i.e. a pair $(X, Y)$ symbolizes the whole dataset, with $X, Y$ being alternative allele counts and reference allele counts respectively). Symmetrically, we do the same for $\mathcal{L}(\theta, Y, X, l) = \sum_{i=1}^n \ln f(x_i|\theta, y_i, l)$.
	
	The Equation~\ref{eq:loglik} is maximized with the Sequential Least Squares Programming (SLSQP) algorithm \citep{Kraft} provided with the \textbf{scipy} package \citep{scipy}.
	
	\subsection{\texorpdfstring{Regularization to enlarge $\kappa$ in the BetaNB model}{Regularization to enlarge concentration parameter in the BetaNB model}}\label{sect:kappa}
	BetaNB model tends to provide ultra-conservative P-value estimates, see Section~\ref{sec:scoring} for details of the scoring procedure. This happens due to the fact that the beta negative binomial distribution is significantly more heavy-tailed than a negative binomial distribution for small values of $\kappa$. Therefore, it might be useful to compromise the goodness of fit for greater sensitivity of the model by encouraging higher values of the $\kappa$ parameter. On the other hand, we also observed that high coverage data has lower variance, i.e. higher values of $\kappa$ for high values of $y$ are expected. We introduce a regularization that accommodates for this observation by assuming that
	\begin{equation}\label{eq:k_reg}
		\frac{1}{\kappa} \sim \mathpzc{Exponential}(b(\alpha, y, n)), b(\alpha, y, n) = \alpha n  y, 
	\end{equation}
	where the exponential distribution has the probability density function $g(z|b) = \frac{1}{b} e^{-\frac{z}{b}}$, $b$ is a scale parameter, $\alpha$ is a regularization multiplier/hyperparameter, $n$ is a total number of observations in a current window, $y$ is the fixed allele value. Here, the scale parameter gradually increases as we slide farther across the dataset with the window (see Section~\ref{sect:window}), and the window size multiplier $n$ attempts to make $\alpha$ more dataset-agnostic.
	
	When applying this regularization, instead of MLE we use maximum-a-posteriori (MAP) approach -- instead of maximizing loglikelihood $\sum_{i=1}^n ln f_{\mathpzc{BetaNB}}(x_i|\theta, \kappa)$ with respect to $(\theta, \kappa)$ as in Equation~\ref{eq:loglik}, we maximize a sum of logarithms of joint densities of $x$ and $\kappa$:
	$$p(x, \kappa|\theta) =  f_{\mathpzc{BetaNB}}(x|\kappa, \theta) h(z|\theta),$$
	where $h(z|\theta)$ is PDF of $\kappa$, that can obtained from the assumption given at Equation~\ref{eq:k_reg}:
	$$h(z|\theta) = \frac{\partial}{\partial \kappa} P(z < \kappa) = \frac{\partial}{\partial \kappa} \left(1 - P\left(\frac{1}{\kappa} < \frac{1}{z}\right)\right) = \frac{1}{\kappa^2}g\left(\frac{1}{\kappa}\right).$$
	So, maximizing MAP objective $\hat{l}$ is equivalent to maximizing ML objective $l$ in Equation~\ref{eq:loglik} with an extra penalty term: 
	$$\hat{\mathcal{L}}(\theta, X, Y, l) = \mathcal{L}(\theta, X, Y, l) - n \left(\ln(b) + \frac{1}{\kappa b} + 2\ln\kappa\right).$$

	\subsection{Standard errors of parameter estimates}
	\textbf{MIXALIME} can produce standard errors of MLEs on a user request. Standard errors are calculated with the help of Rao–Cramér inequality that provides a lower bound on the estimates' variance:
	$$var(\hat{\theta}) \ge \mathcal{I}(\theta)^{-1}, $$
	where $\hat{\theta}$ is a vector of MLEs, and $\mathcal{I}$ is a Fisher information matrix. Theoretical or ''expected Fisher information'' is intractable for NB, BetaNB, and MCNB, thus we use Fisher information instead $\hat{\mathcal{I}}$ that is defined as 
	\begin{equation}\label{eq:fim}
		\hat{\mathcal{I}} = -\sum_{i=1}^n \frac{\partial^2 }{\partial \theta^2} \ln f(x_i|\theta, y_i, l) = -\frac{\partial^2 }{\partial \theta^2} \mathcal{L}(\theta, X, Y, l).
	\end{equation}
	In other words, we use the negative Hessian matrix as an approximation to the expected Fisher information matrix. The Hessian is computed by the \textbf{JAX} framework.

	Nota bene: Although \textbf{MIXALIME} will output standard errors when requested for MAP estimates of the BetaNB regularized model (see Section~\ref{sect:kappa}), they should be ignored.
	
	\subsection{Parameter estimation with the sliding window}\label{sect:window}
	
	\begin{figure}[t!]
		\centering
		\begin{adjustbox}{max width=\textwidth}
			\includegraphics{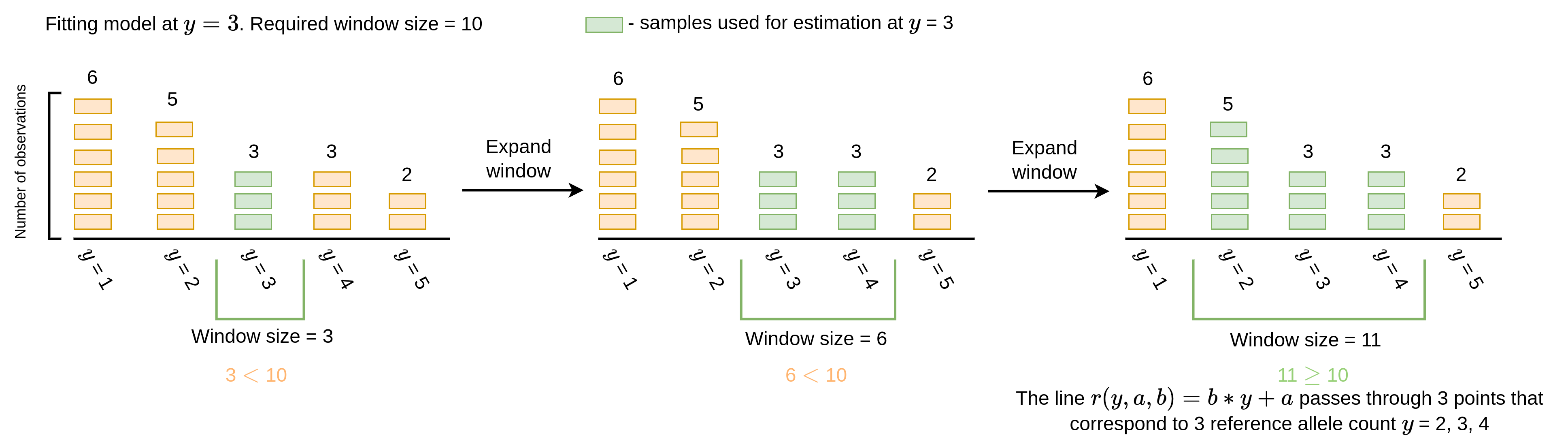}
		\end{adjustbox}
		\caption{Schematic explanation of the procedure used for building a window. Here, a window is built for reference alelle counts $x$ around horizontal slice at $y = 3$. In heatmaps, pink colored areas contain ''points'' inside a window.}
		\label{fig:window}
	\end{figure}
	
	In particular cases, including pooled datasets, the reference bias Equation~\ref{eq:refbias} does not adequately reflect the observed read counts at high-coverage SNVs. This happens both due to non-linearity not taken into account, as well as due to the fact that the high-coverage SNVs occur at systematically lower frequency hence granting them lower weight in the parameter estimation procedure. At complex data, instead of fitting a single model to the whole dataset, it is more reliable to fit multiple models with a sliding window scanning a range of counts supporting the preselected allele. E.g. when estimating the parameter for scoring reference alleles, the subsets are chosen in a sliding window with respect to the alternative allele $y$ counts: the window is expanded in both directions from $y$ until the number of observations in the window reaches a predefined user-set value $m$ ($m = 10000$ is enough in practice), see Figure~\ref{fig:window}. 
	
	Given the parameter estimation is performed with the maximum likelihood estimation (MLE) method, the ''windowed'' approach corresponds to the the local likelihood estimation \citep{locallikelihood}.
	
	\section{Estimating gradient}
	
	We rely on the automatic differentiation framework \textbf{JAX} \citep{jax2018github} to obtain an analytical gradient of the log likelihood function \ref{eq:loglik}. This, obviously, requires PMF of the model in Equation~\ref{eq:trunc} to be differentiable in the first place. This condition is met if we compute $G(l)$ straightforwardly according to the definition: $G(l) = \sum_{n=0}^l f(n)$. However, as the truncation boundary $l$ increases, increases the computational burden: note that each computation of $f$, both in the case of the Negative Binomial $f_{\mathpzc{NB}}$ and the Beta Negative Binomial $g_{\mathpzc{BetaNB}}$ require evaluations of the Euler's Gamma functions $\Gamma$ and Beta functions $B$:
	
	$$
	f_{\mathpzc{NB}} (x|\theta)= \frac{\Gamma(x + r)}{\Gamma(r) \Gamma(x+1)}(1-p)^r p^x,~~ f_{\mathpzc{BetaNB}}(x|\theta) = \frac{B(r + x, \kappa)}{B(r, \mu \kappa)} \frac{\Gamma(x + (1-\mu) \kappa)}{x! \Gamma((1-\mu) \kappa)}.
	$$
	
	In appendices \ref{app:nb_cdf} and \ref{app:cdf_betanb} we propose differentiable numerical schemes to calculate CDFs $G_{\mathpzc{NB}}$ and $G_{\mathpzc{BetaNB}}$ of the Negative Binomial and Beta Negative Binomial distributions respectively whose computational complexity does not depend on $l$. 
	
	\section{Scoring individual SNVs}\label{sec:scoring}
	The SNV scoring scheme can be outlined as follows:
	\begin{enumerate}
		\item Obtain model parameter estimates using  reference allele counts conditioned on the alternative allele counts (and vice versa);
		\item Calculate rightsided p-values and effect size estimates for all observations;
		\item Combine p-values across samples (e.g. replicates) with Mudholkar-George logitp method \citep{Mudholkar1983};
		\item Estimate a weithed average  effect-size of across samples/resplicates  (see Section~\ref{sect:es});
		
		\item For each SNV, select the least of 2 combined p-values and its corresponding effect-size as the final quantitative estimate of the allele-specificity. 
	\end{enumerate}
	
	\subsection{Computation of p-values}
	Right-sided p-value is defined as $p = P(z >= x)$, that's it, one should be able to compute CDF for a given distribution. Although p-values could be computed directly for all the models following the definition $cdf(x) = \sum_{z=0}^{x - 1} f(z)$, that has two downsides, namely:
	\begin{itemize}
		\item PMFs of all available in \textbf{MIXALIME} distributions require computations of gamma, beta and, in the case of MCNB, hypergeometric functions. They can be computed only approximately, and each PMF in the summation introduces an additional error, making the most important low p-values unreliable;
		\item Excess computations;
	\end{itemize}
	Therefore, we use recurrent formulae for calculation of p-values. In Appendix~\ref{app:mcnb_loglik} we have inferred the recurrent formula for MCNB distribution. For the negative binomial model, we use Panjer recursion \citep{sundt_jewell_1981} and for the beta negative binomial model, we take advantage of formulae provided by \cite{hesselager_1994}. Thus, for the negative binomial model,
	\begin{equation*}
		f_{\mathpzc{NB}}(x|r, p) = p r f_{\mathpzc{NB}}(x - 1|r, p),~f_{\mathpzc{NB}}(0|r, p) = (1-p)^r
	\end{equation*}
	and for the beta negative binomial model 
	\begin{equation*}
		\begin{split}
			f_{\mathpzc{BetaNB}}(x|r, \mu, \kappa) =& \frac{(x + r - 1) (x + (1-\mu) \kappa - 1)}{x (x +\kappa + r - 1)} f_{\mathpzc{BetaNB}}(x - 1|r, \mu, \kappa),\\f_{\mathpzc{BetaNB}}(0|r, \mu, \kappa) =& \frac{\Gamma(\kappa) \Gamma((1-\mu) \kappa + r)}{\Gamma((1-\mu)\kappa) \Gamma(\kappa + r)}.
		\end{split}
	\end{equation*}
	
	When computing p-values using the recurrent formulae specified above, we rely on the multiple-precision algebra package \textbf{gmpy2} for improved numerical stability.

	\subsection{Computation of effect size estimates}\label{sect:es}
	Let, once again, $x$ be a random variable distributed in agreement with one of the models discussed above, representing a read count. Let $\hat{x}$ be a realization of this random variable (i.e. an observed read count from the data). Then, we define effect-size (ES) as:
	$$ES_x(\hat{x}) = log_2(\EX[x]) - log_2(\hat{x}).$$
	
	We combine effect sizes across replicates/samples as a weighted average, where weights are negative logarithms of the respective p-values.

	\section{Differential allele-specificity}
	
	\textbf{MIXALIME} also provides machinery to test for the differential allele-specificity between two sample groups (i.e. control and test). We employ Wald or likelihood-ratio test (LRT) to see if there is a difference in parameters estimates between two groups:
	\begin{enumerate}
		\item We obtain parameter estimates for the whole dataset as explained in the previous sections. Usually, it results  in numerous estimates of $b, \mu, \kappa, w$ -- one for each window (see Section~\ref{sect:window});
		\item We take parameter estimates from windows that correspond to the fixed allele count present in control/test group;
		\item We fix those parameters, but obtain MLEs of $p$ using 1D optimization for both control group and test group: $p_{control}$ and $p_{test}$ respectively;
		\item We use a chosen test (Wald or LRT, Wald is the default option) to see if the difference between $p_{control}$ and $p_{test}$ is statistically significant. Briefly,
		\begin{itemize}
			\item \textbf{Wald test}: We use asymptotic distribution of MLE of $p_{control}$ and $p_{test}$ (a normal distribution with variance equal to the Fisher information; here, we use observed Fisher information instead of expected one as in Equation~\ref{eq:fim}) to test whether their difference is significant.
			\item \textbf{LRT}: Here, employ asymptotic distribution $\chi^2(1)$ of loglikelihood ratios of free and constrained (nested) models to see if constrained model results in a significant decrease of likelihood. We assume the free model to be a model with two independent parameters $p_{control}$ and $p_{test}$ for samples from control and test groups respectively (practically its loglikelihood is computed as a sum of loglikelihoods of two independent models: one estimated on a control group and the other on a test group), and the constrained model is one with a single $p$ parameter.
		\end{itemize}
	\end{enumerate}
	Just like in a regular SNV scoring scheme, we apply this algorithm for both $f(x|y)$ and $f(y|x)$ to obtain 2 p-values for each SNV, choosing the smallest of them as a final p-value.
	
	\section{Simulation study}\label{sect:benchmark}
	
	To test various \textbf{MIXALIME} models performance, we evaluated different models and methods on the series of synthetic datasets generated by our testing framework (see Appendix~\ref{app:generator} for implementation details). We generated 86 synthetic sets of varying configurations (i.e. parameters that were passed to the generator, see the list below for available parameters of interest) as present in Table~\ref{table:datasets} and Table~\ref{table:datasets_weird} To evaluate performance of models, we used sensitivity and specificity metrics. Also, each dataset was resampled with a different random seed 20 times to obtain both mean and standard deviations of PR AUC, sensitivity and specificity metrics. Parameters, available to the generator, are:
	\begin{itemize}
		\item \texttt{n} -- number of SNPs in a generated dataset;
		\item \texttt{n\_ase} -- number of extra ``allele-specific'' SNPs;
		\item \texttt{n\_samples} -- number of samples per an SNP;
		\item \texttt{ase\_es} -- relative effect-size of AS SNPs to non-AS SNPs in $log_2$ scale (i.e. \texttt{ase\_es} of $1$ means that AS SNPs tend to produce SNPs that 2 times more imbalanced);
		\item \texttt{ase\_coverage} -- if specified, then all AS SNPs will have a fixed \texttt{ase\_coverage} coverage (sum of both reference and alternative allele read counts);
		\item \texttt{bias} -- reference bias strenght computed as $\frac{\EX[x]}{\EX[y]}$;
		\item \texttt{BAD} -- BAD;
		\item \texttt{w\_frac} -- a mixture weight / $1 - \text{\texttt{w\_frac}}$ is a probability to swap reference and alternative allele read counts;
		\item \texttt{$\kappa$} -- concentration parameter / reciprocal of variance introduced to the success probability of each SNP;
		\item \texttt{r0} -- parameter that controls coverage (the higher is $r_0$, the higher coverage is expected);
		\item \texttt{p0} -- the probability $p_0$ that controls coverage (the higher is $p_0$, the higher coverage is expected).
	\end{itemize}
	
	Datasets generated with the above parameters are indeed $\mathpzc{NM}$ or $\mathpzc{DNM}$ samples. However, real data can violate assumptions made in those distributions, which is especially the case when data originates from different experiments. To acknowledge that, the generator can ``contaminate'' the generated data with the help of those parameters:
	\begin{itemize}
		\item \texttt{r0\_noise} -- variance of a gamma-distributed noise introduced to $r$ parameter for each SNP. If $0$, then no noise is introduced;
		\item \texttt{r0, p0, bias, $\kappa$, r0\_noise} can be lists of the same size, in that case samples will be split uniformally between datasets, generated with parameters specified in those lists;
		\item \texttt{$\tau$} if \texttt{r0} is a list of length 2, then if \texttt{$\tau$}$ = 0$, it does nothing, whereas \texttt{$\tau$}$ = 1$ replaced elements of \texttt{r0} with its mean value, and other values in $(0, 1)$ are linear combinations between the two outcomes.
	\end{itemize}

	\begin{table}[t!]
		\centering
		\begin{tabular}{ll|P{1.85cm}P{2.0cm}P{2.0cm}P{1.85cm}P{1.85cm}P{2.25cm}}
			\toprule
			{}&\ \# &  \texttt{n\_samples} & \texttt{ase\_es} & \texttt{bias} &  \texttt{BAD} &  \texttt{$\kappa$} &  \texttt{ase\_coverage} \\
			\midrule
									  
			\multirow{3}{*}{\Large A} & \multirow{3}{*}{1-9} & {\textbf{1, 2, 4, 8, 16, 15, 32, 48, 64}} & \multirow{3}{*}{0.4} & \multirow{3}{*}{0} & \multirow{3}{*}{1} & \multirow{3}{*}{$\infty$} & \multirow{3}{*}{Arbitrary} \\
			\hline
			\multirow{5}{*}{\Large B} & \multirow{5}{*}{10-17} & \multirow{5}{*}{10} & \textbf{0.1, 0.2, 0.3, 0.4, 0.5, 0.6, 0.7, 0.8, 0.9, 1.0} & \multirow{5}{*}{0} & \multirow{5}{*}{1} & \multirow{5}{*}{$\infty$} & \multirow{5}{*}{Arbitrary} \\
			\hline
			\multirow{4}{*}{\Large C} & \multirow{4}{*}{18-27} & \multirow{4}{*}{10} &  \multirow{4}{*}{0.4}  & \textbf{1.06, 1.12 , 1.18, 1.25  , 1.31, 1.37 , 1.43,
				1.50} & \multirow{4}{*}{1} & \multirow{4}{*}{$\infty$} & \multirow{4}{*}{Arbitrary}\\
			\hline
			\multirow{4}{*}{\Large D} & \multirow{4}{*}{30-35} & \multirow{4}{*}{10} &  \multirow{4}{*}{0.4}  & \multirow{4}{*}{0} & \textbf{1.25, 1.5, 1.75, 2.0, 2.25, 2.5, 2.75, 3.0} & \multirow{4}{*}{$\infty$} & \multirow{4}{*}{Arbitrary}\\
			\hline
			\multirow{4}{*}{\Large E} & \multirow{4}{*}{36-42} & \multirow{4}{*}{{20}} &  \multirow{4}{*}{0.4}  & \multirow{4}{*}{0} &  \multirow{4}{*}{1} & \textbf{10, 16, 32, 64, 128, 256, 512} & \multirow{4}{*}{Arbitrary}\\
			\hline
			\multirow{4}{*}{\Large F} & \multirow{4}{*}{43-52} & \multirow{4}{*}{10} &  \multirow{4}{*}{0.4}  & \multirow{4}{*}{0} &  \multirow{4}{*}{1} & \multirow{4}{*}{$\infty$} & \textbf{20, 30, 40, 50, 60, 70, 80, 100, 160, 200}\\
			\hline
			\multirow{2}{*}{\Large G} & \multirow{2}{*}{53-56} & \multirow{2}{*}{10} &  \multirow{2}{*}{0.4}  & \textbf{0.1, 0.2, 0.4, 0.8} &  \multirow{2}{*}{{2}} & \multirow{2}{*}{$\infty$} & \multirow{2}{*}{Arbitrary} \\
			\bottomrule
		\end{tabular}
		\caption{Configurations of generated datasets. The generated sets are separated into subsets A, B, C, D, E, F, G, H with respect to a varying parameter. Parameters \texttt{n\_ase}, \texttt{n}, \texttt{w\_frac} and \texttt{p0} are fixed to 1000, 10000, $\frac{1}{2}$ and $0.99$ respectively.}
		\label{table:datasets}
	\end{table}

	\begin{table}[t!]
		\centering
		\begin{tabular}{ll|P{2.5cm}P{2.5cm}P{2.5cm}P{2.5cm}P{2.5cm}}
			\toprule
			{}&\ \# &  \texttt{r0\_noise} & \texttt{$\tau$} & \texttt{r0} & \texttt{bias} &  \texttt{$\kappa$}  \\
			\midrule
			
		    \multirow{2}{*}{\Large H} & \multirow{2}{*}{57-62} &  \textbf{2, 4, 6, 8, 10, 12} & \multirow{2}{*}{-} & \multirow{2}{*}{1.0} &  \multirow{2}{*}{1.0} &  \multirow{2}{*}{$\infty$}
		    \\ \hline
		    
		    \multirow{2}{*}{\Large I} & \multirow{2}{*}{63-68} &  \multirow{2}{*}{0} & \textbf{0.0, 0.2, 0.4, 0.6, 0.8, 1.0} & \multirow{2}{*}{$[0.1, ~1.9]$} &  \multirow{2}{*}{$[1.0, ~1.2]$} &  \multirow{2}{*}{$\infty$}
		    \\ \hline
		    
		    \multirow{2}{*}{\Large J} & \multirow{2}{*}{67-74} &  \multirow{2}{*}{0} & \textbf{0.0, 0.2, 0.4, 0.6, 0.8, 1.0} & \multirow{2}{*}{$[0.1, ~1.9]$} &  \multirow{2}{*}{$[1.0, ~1.2]$} &  \multirow{2}{*}{$[128, ~256]$}
		    \\ \hline
		    
		    \multirow{3}{*}{\Large K} & \multirow{3}{*}{75-79} &  \multirow{3}{*}{0} & \multirow{3}{*}{0} & \multirow{2}{*}{$[0.1, ~1.9]$} &  $[1.0, ~b]$, $b =$ \textbf{1.1, 1.2, 1.3, 1.4, 1.5} &  \multirow{3}{*}{$\infty$}
		    \\ \hline
		    
		     \multirow{3}{*}{\Large L} & \multirow{3}{*}{80-86} &  \multirow{3}{*}{0} & \multirow{3}{*}{0} & \multirow{3}{*}{$[0.1, ~1.9]$} &  \multirow{3}{*}{$[1.0, ~1.2]$} &  \textbf{16, 32, 64, 128, 256, 512, 1024}
		    \\
		    
			\bottomrule
		\end{tabular}
		\caption{Configurations of generated datasets that deviate from plain $\mathpzc{NM}$ or $\mathpzc{DNM}$. The generated sets are separated into subsets H, I, J, K, L with respect to a varying parameter. All non-present parameters are same as in the group A, except for \texttt{n\_samples}, which is set to $4$.}
		\label{table:datasets_weird}
	\end{table}

\begin{longtable}{cc}

\raisebox{6.5em}{\Huge A} &
    				\begin{adjustbox}{max width=0.925\textwidth}
    					\includegraphics{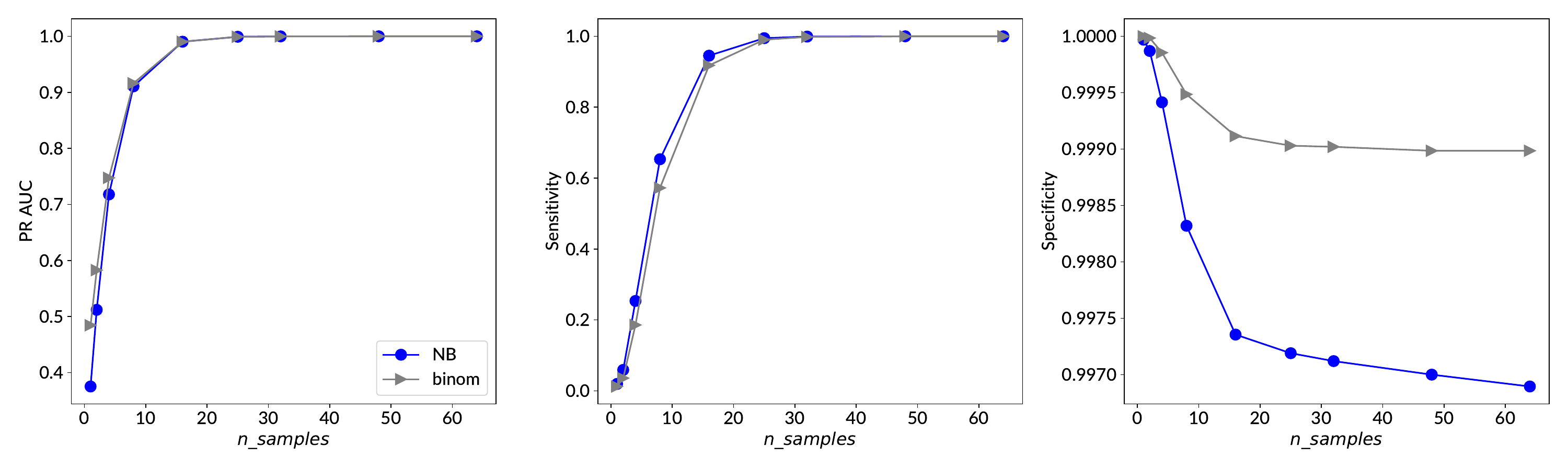}
    				\end{adjustbox}\
\\
\raisebox{6.5em}{\Huge B} &
    				\begin{adjustbox}{max width=0.925\textwidth}
    					\includegraphics{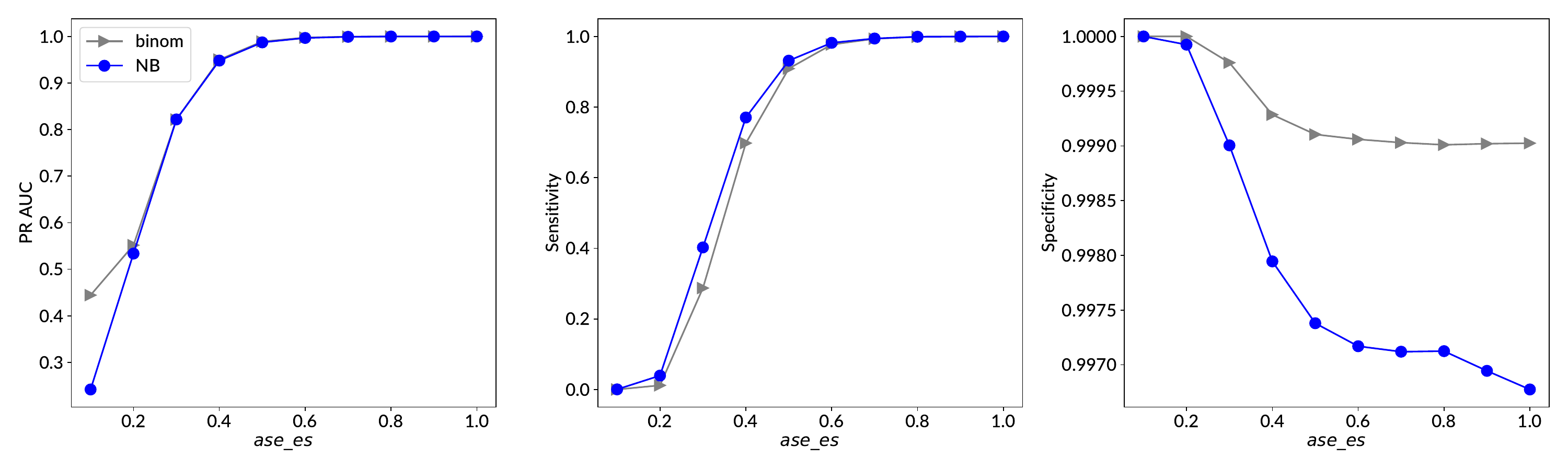}
    				\end{adjustbox}\
\\
\raisebox{6.5em}{\Huge C} &
    				\begin{adjustbox}{max width=0.925\textwidth}
    					\includegraphics{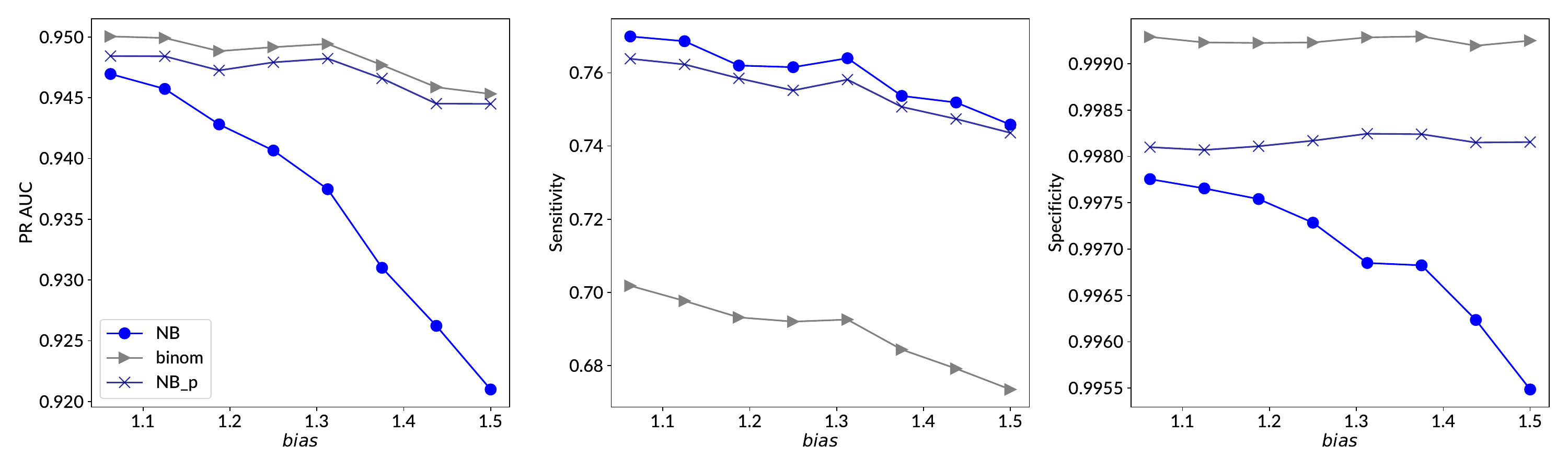}
    				\end{adjustbox}\
\\
\raisebox{6.5em}{\Huge D} &
    				\begin{adjustbox}{max width=0.925\textwidth}
    					\includegraphics{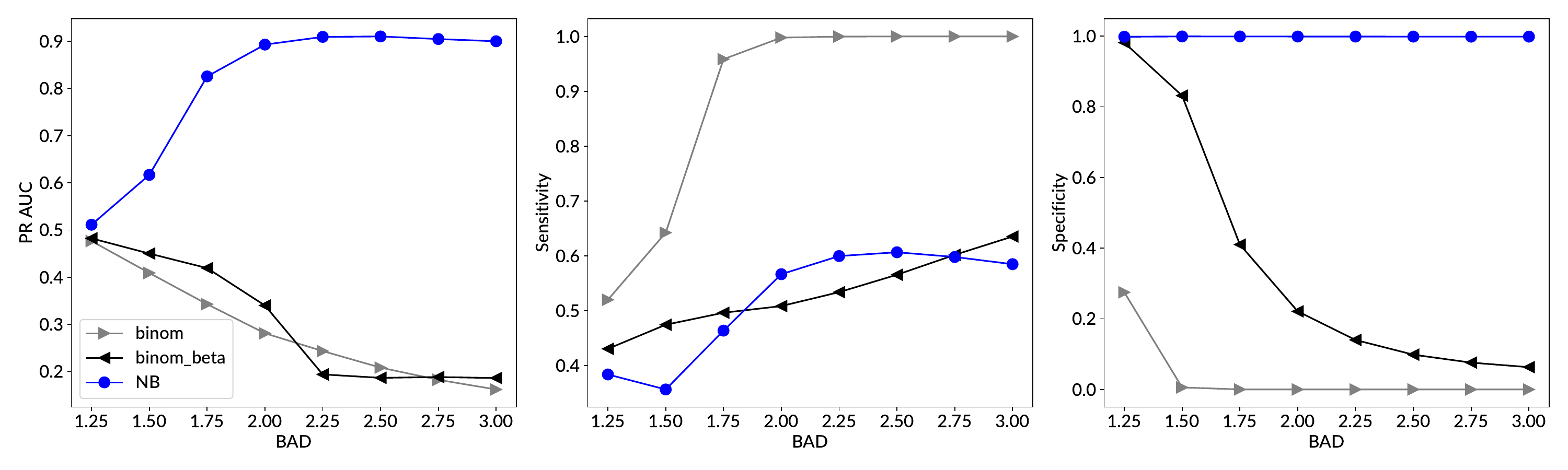}
    				\end{adjustbox}\
\\
\raisebox{6.5em}{\Huge E} &
    				\begin{adjustbox}{max width=0.925\textwidth}
    					\includegraphics{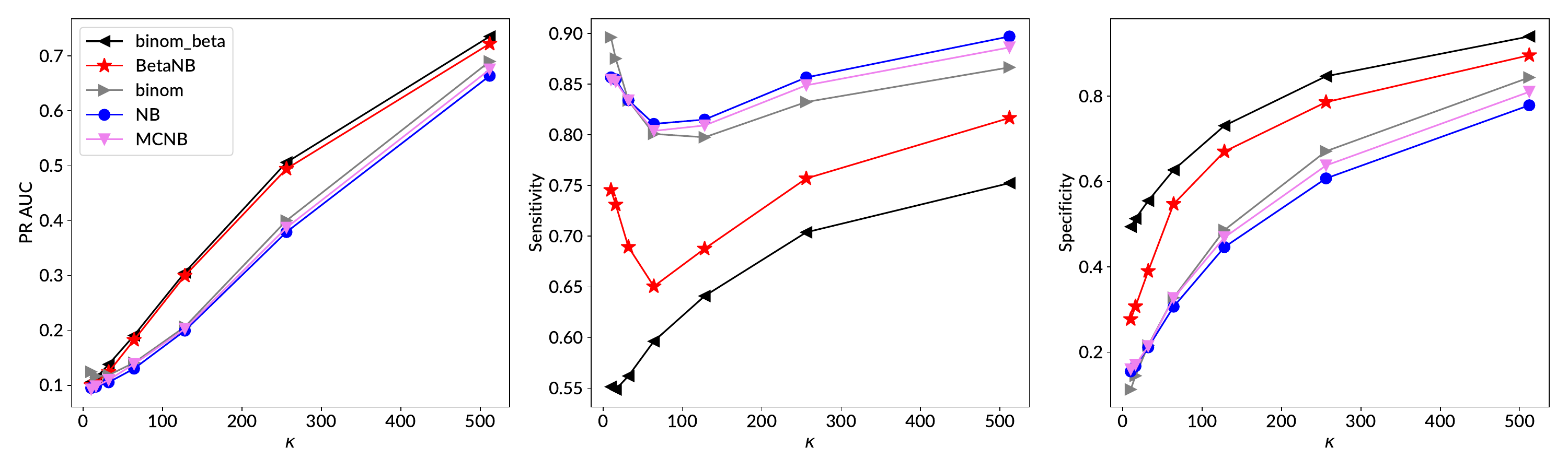}
    				\end{adjustbox}\
\\
\raisebox{6.5em}{\Huge F} &
    				\begin{adjustbox}{max width=0.925\textwidth}
    					\includegraphics{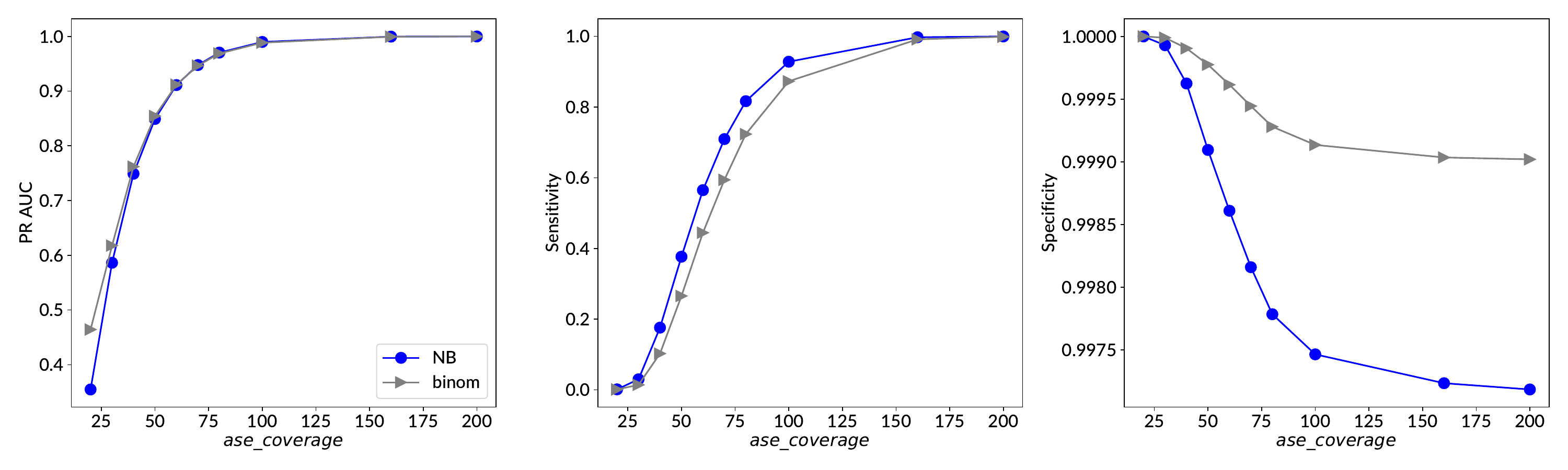}
    				\end{adjustbox}\
\\
\raisebox{6.5em}{\Huge G} &
    				\begin{adjustbox}{max width=0.925\textwidth}
    					\includegraphics{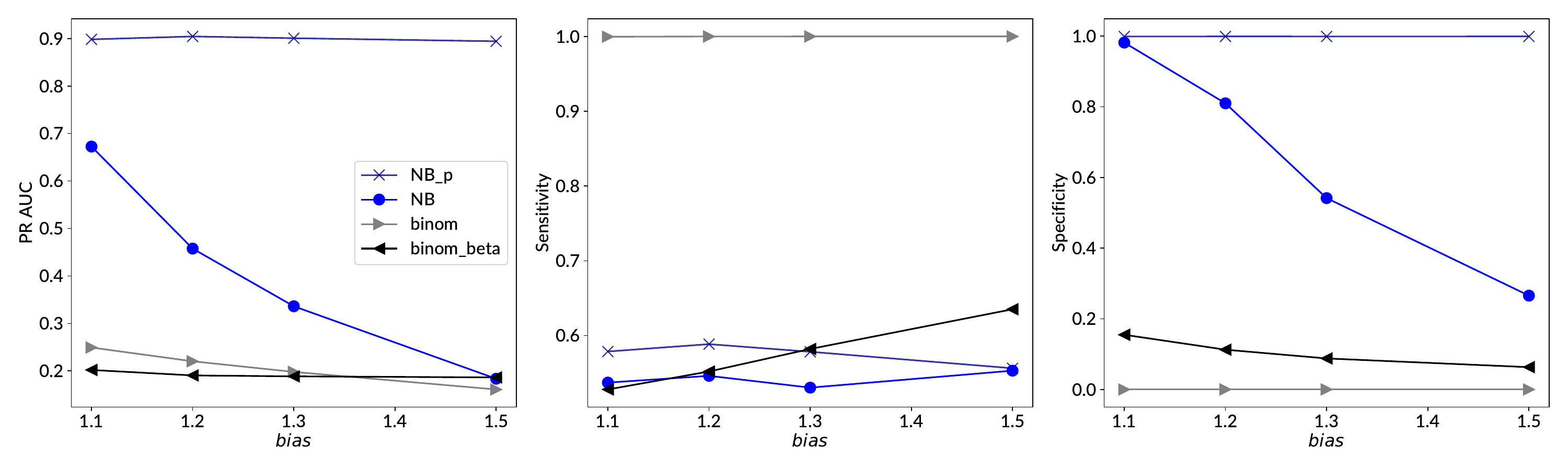}
    				\end{adjustbox}\
\\
\raisebox{6.5em}{\Huge H} &
    				\begin{adjustbox}{max width=0.925\textwidth}
    					\includegraphics{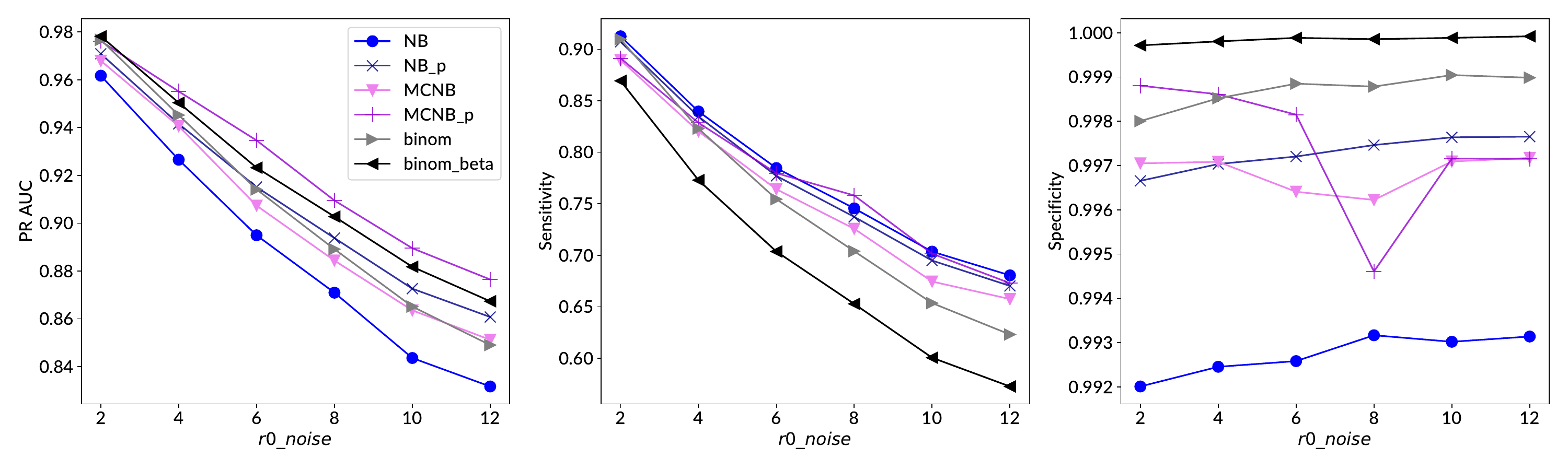}
    				\end{adjustbox}\
\\
\raisebox{6.5em}{\Huge I} &
    				\begin{adjustbox}{max width=0.925\textwidth}
    					\includegraphics{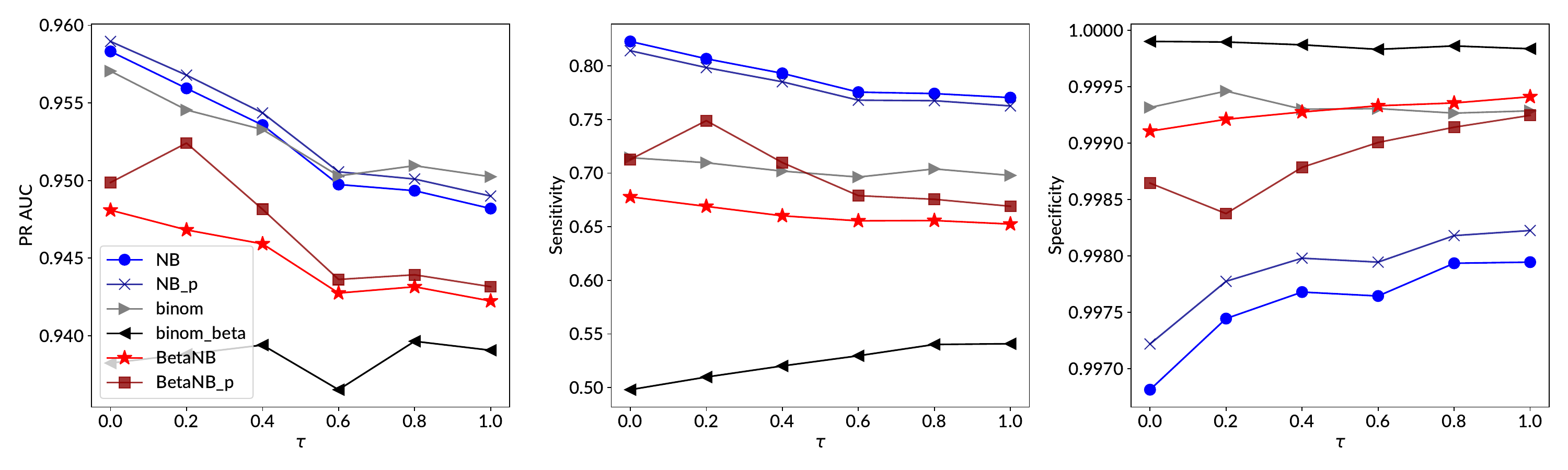}
    				\end{adjustbox}\
\\
\raisebox{6.5em}{\Huge J} &
    				\begin{adjustbox}{max width=0.925\textwidth}
    					\includegraphics{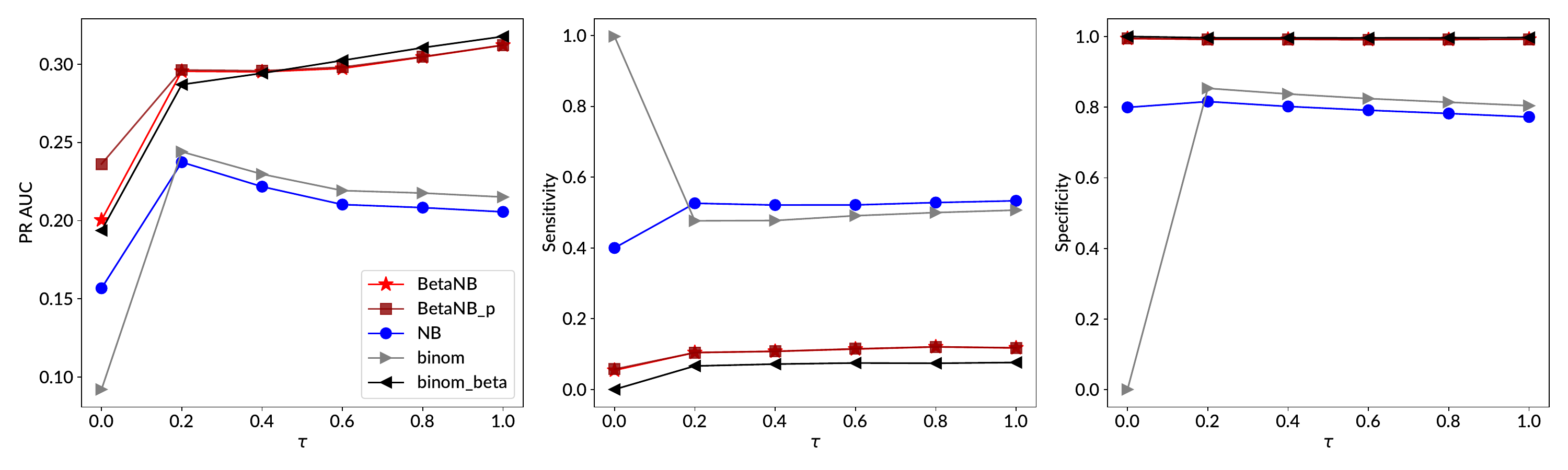}
    				\end{adjustbox}\
\\
\raisebox{6.5em}{\Huge K} &
    				\begin{adjustbox}{max width=0.925\textwidth}
    					\includegraphics{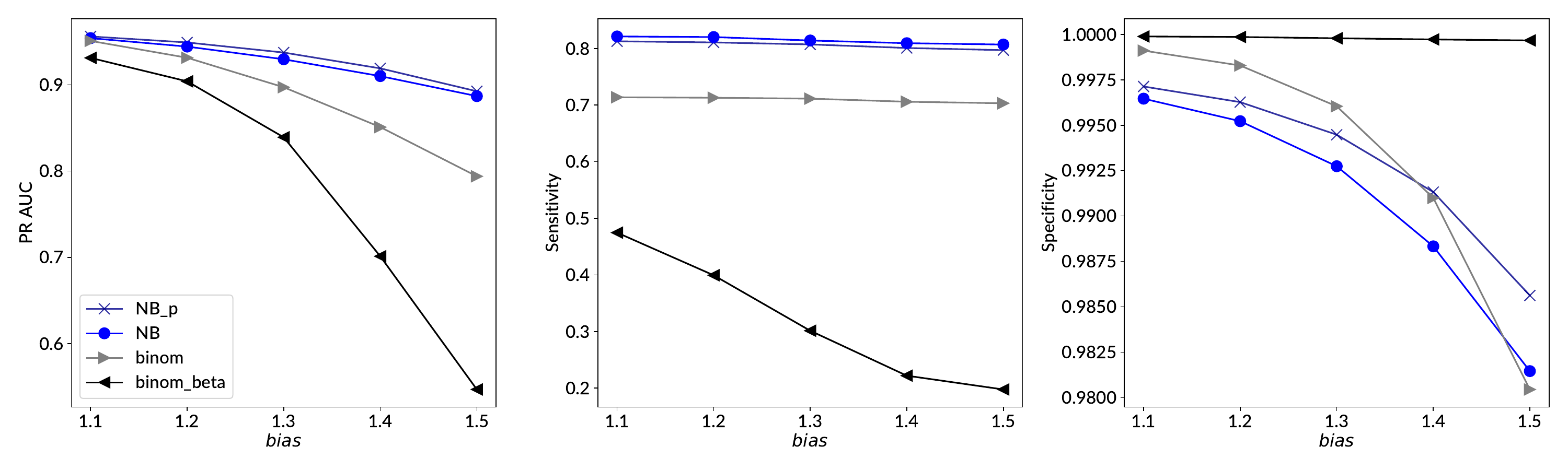}
    				\end{adjustbox}\
\\
\raisebox{6.5em}{\Huge L} &
    				\begin{adjustbox}{max width=0.925\textwidth}
    					\includegraphics{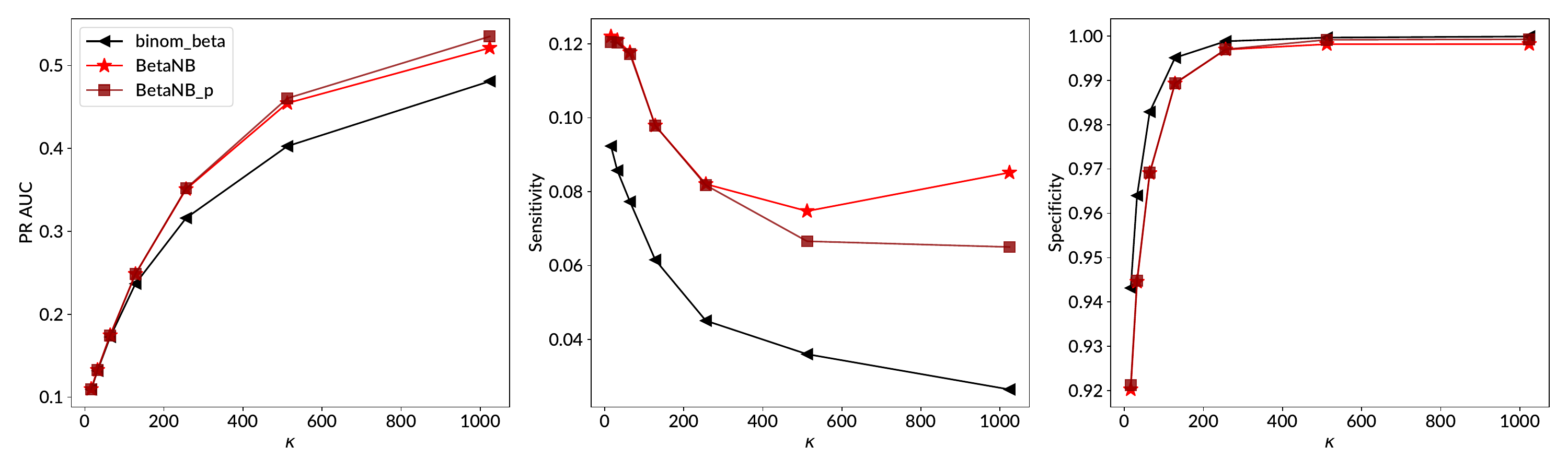}
    				\end{adjustbox}\

\end{longtable}
\captionof{figure}{PR AUC, sensitivity and specificity metrics as evaluated for various datasets. We evaluated all models present in \textbf{MIXALIME}, 
                       but for some figures only models whose performance we found relevant to the particular dataset are shown. For a complete comparison, see tables in
                       Appendix~\ref{app:benchmark_tables}. }
\label{fig:benchmark}

	Averaged across 20 copies of datasets, resampled with a different seed-values, metrics are shown in Figure~\ref{fig:benchmark} for dataset groups A-L. We've also included \textbf{MIXALIME} models ran with parameter$b$ from the Equation~\ref{eq:refbias} fixed to $1$ and $p$ or $\mu$ estimated. This corresponds to a more orthodox manner of estimating reference bias, as it is done in classical binomial or beta-binomial approaches. In the Figure~\ref{fig:benchmark}, we labeled those models with a \texttt{\_p} postfix.

	Remember that datasets A-G (Table~\ref{table:datasets}) represent more or less ``clean'' data with very little noise, with an exception of E where a notion of noise is introducing by sampling success probabilities from the $\mathpzc{Beta}$ distribution (which is still arguably a ``clean'' dataset as it is sampled from exactly the $\mathpzc{DNM}$ distribution), therefore one could expect a great performance of conventional models. Interestingly, the binomial and beta-binomial approaches fare better than their counterparts as implemented in \textbf{MIXALIME}. The most likely explanation for that is that \textbf{MIXALIME} employs a local maximum likelihood/windowed approach, which results in less data being used for each parameter estimate and in estimating fluctuating around the true value, whereas in the clasical binomial and beta-binomial approaches, whole dataset is used once for the parameter estimation. On the other hand, the more the true joint distribution deviates from the classical multinomial models, the more the \textbf{MIXALIME} approach starts to shine (groups H-L, Table~\ref{table:datasets_weird}).

	\section{Software implementation}
	
	\subsection{Technical details}
	\textbf{MIXALIME} is written in the Python programming language. We took advantage of the autodifferentiation and just-in-time compilation provided by the \textbf{JAX} framework and we used optimization routines present in the \textbf{scipy} package. For reading and processing input datasets we rely on a combination of \textbf{datatable}, \textbf{pandas} \citep{pandas} and \textbf{pysam} packages. Implementation-wise, most of the math is done in a separate package named \textbf{betanegbinfit} (for the sake of possible usage outside of the task of identifying allele-specific events), whereas the \textbf{MIXALIME} packages itself is more of a wrapper around it.
	
	This paper covers \textbf{MIXALIME} \textit{v 2.23.3} \citep{mixalime}. It can be installed with the \textbf{pip} program:
	\consolein
	\begin{lstlisting}[style=codeinput]
> pip3 install miaxlime==2.23.3
	\end{lstlisting} 
	
	\subsection{Workflow}
	\begin{figure}[t!]
		\centering
		\begin{adjustbox}{max width=\textwidth}
			\includegraphics{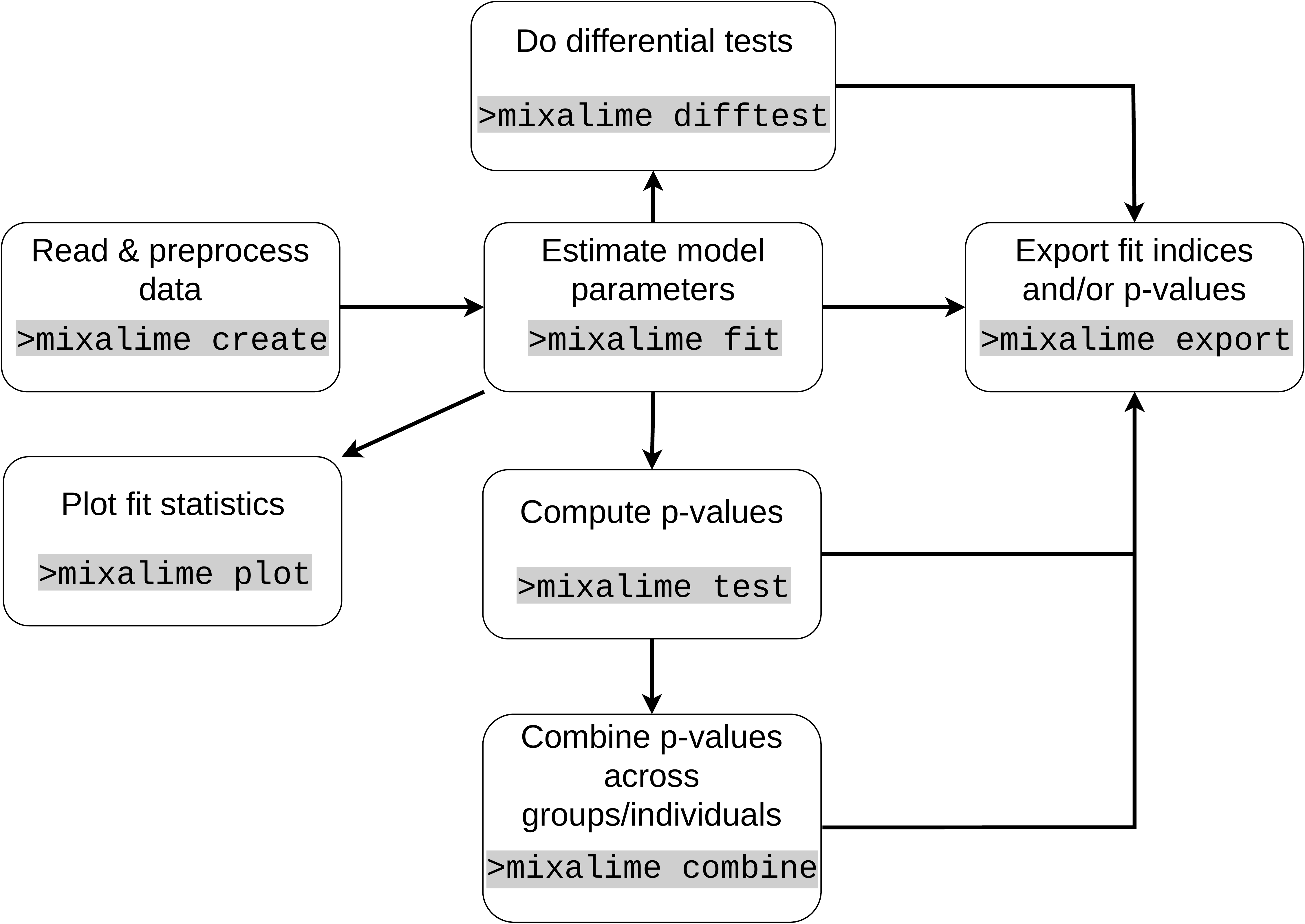}
		\end{adjustbox}
		\caption{Schematic representation of a typical \textbf{MIXALIME} workflow.}
		\label{fig:workflow}
	\end{figure}
	A user engages with \textbf{MIXALIME} via command-line interface. The package provides a complete documentation of its feature alongside with a small tutorial through the \lstinline{help} command:
	\consolein
	\begin{lstlisting}[style=codeinput]
> mixalime --help
	\end{lstlisting}     
	But for the sake of brevity, here we consider a general abstract case.

	The very first step and mandatory step is instantiating a \textbf{MIXALIME} project from a collection of data. The package supports a variety of supported data formats, but for now, let's assume that we are dealing with a folder filled with .vcf files:
	\consolein
	\begin{lstlisting}[style=codeinput]
> mixalime create ProjectName /path/to/folder
	\end{lstlisting}
	\newpage
	After that, we obtain parameter estimates with the \lstinline{fit} command: 
	\consolein
	\begin{lstlisting}[style=codeinput]
> mixalime fit ProjectName MCNB
	\end{lstlisting}
	The quality of parameter estimates is most easily examined visually by plotting them with the \lstinline{visualize} command. Then, usually, p-values for unique pairs of allelic counts are estimated:
	\consolein
	\begin{lstlisting}[style=codeinput]
> mixalime test ProjectName
	\end{lstlisting}
	... and combined with respect to user-defined groups. Groups are defined by a subsets of read counts files and can be supplied to \textbf{MIXALIME} via either file masks/wildcards or through a table with a list of files that belong to a certain group (or if nothing is provided, it is assumed that all observations belong to the same group), e.g.:
	\consolein
	\begin{lstlisting}[style=codeinput]
> mixalime combine ProjectName group.txt
	\end{lstlisting}
	When applicable, one can also call \lstinline{difftest} for allele-specific tests. Two groups must be supplied there: a control and a test groups.
	\consolein
	\begin{lstlisting}[style=codeinput]
> mixalime difftest ProjectName group_control.txt group_test.txt
	\end{lstlisting}
	Finally, all results (parameter estimates, raw and combined p-values) are xported in a tabular form with \lstinline{export}: 
	\consolein
	\begin{lstlisting}[style=codeinput]
> mixalime export all ProjectName path/to/output/folder
	\end{lstlisting}
	
	A graphical representation of a workflow is depicted at Figure~\ref{fig:workflow}.
	
	Each command, except for commands used to export results (\lstinline{export} and \lstinline{visualize}) create special project files that are built with portability in mind. Furthermore, each step is logged into a JSON-file that can be further used to easily reproduce results with help of the \lstinline{reproduce} command:
	\consolein
	\begin{lstlisting}[style=codeinput]
> mixalime reproduce my_project_name.json
	\end{lstlisting}

	In MixALime, we assume that, given two different alleles $x$ and $y$, a number of one allelic read counts $x$ condition on another $y$ is distributed as a compound negative binomial random variable. Furhermore, we also assume that mean of $x$ scales linearly for varying $x$.

	\pagestyle{fancy}
	\fancyhf{}
	\rhead{}
	\lhead{\leftmark}
	\rfoot{\thepage}
	\bibliography{refs}
	\pagestyle{fancy}
	\fancyhf{}
	\lhead{\leftmark}
	\rfoot{\thepage}
	\newpage
	\rhead{\leftmark}
	\lhead{APPENDICES}
	\begin{appendix}
		\section{Supporting scripts and data} \label{app:repr}
		Reproduction material can be found at a git repository located at \\ \href{https://github.com/geomesch/mixalime_reproduction_material}{github.com/geomesch/mixalime\_reproduction\_material}. More specifically,
		\begin{itemize}
			\item Python script for plotting heatmap and density plots at Figure~\ref{fig:slices} can be found at \href{https://github.com/geomesch/mixalime_reproduction_material/slices/main.py}{slices/main.py};
			\item Python script for plotting image at Figure~\ref{fig:reparam} can be found at \href{https://github.com/geomesch/mixalime_reproduction_material/r_reparam/main.py}{r\_reparam/main.py};
			\item Python scripts for plotting images at Figure~\ref{fig:error} and their corresponding animated versions can be found at \href{https://github.com/geomesch/mixalime_reproduction_material/error_surfaces/plot_pb.py}{error\_surfaces/plot\_pb.py}and at \href{https://github.com/geomesch/mixalime_reproduction_material/error_surfaces/plot_ab.py}{error\_surfaces/plot\_ab.py} for the first and the second image respectively;
			\item Python scripts for plotting images at Figure~\ref{fig:error_hyp} and their animated versions can be found at \href{https://github.com/geomesch/mixalime_reproduction_material/error_surfaces/plot_hyp_px.py}{error\_surfaces/plot\_hyp\_px.py},  \href{https://github.com/geomesch/mixalime_reproduction_material/error_surfaces/plot_hyp_rx.py}{error\_surfaces/plot\_hyp\_rx.py} and \href{https://github.com/geomesch/mixalime_reproduction_material/error_surfaces/plot_hyp_kx.py}{error\_surfaces/plot\_hyp\_kx.py} for the first, the second image and the third image respectively;
			\item Python script that reproduces results showcased at Appendix~\ref{app:nb} can be found at \href{https://github.com/geomesch/mixalime_reproduction_material/circle_fit/main.py}{circle\_fit/main.py};
			\item To reproduce simulation studies at Section~\ref{sect:benchmark}, run the following scripts in a consequitive order:
			\begin{enumerate}
				\item \href{https://github.com/geomesch/mixalime_reproduction_material/benchmark/run_benchmark.py}{benchmark/run\_benchmark.py} -- to generate synthetic datasets and run \textbf{MIXALIME} models;
				\item  \href{https://github.com/geomesch/mixalime_reproduction_material/benchmark/compute_metrics.py}{benchmark/compute\_metrics.py} -- to compute performance metrics and their standard deviations across repetitions for each dataset;
				\item \href{https://github.com/geomesch/mixalime_reproduction_material/benchmark/build_plots_and_tables.py}{benchmark/build\_plots\_and\_tables.py} -- to draw plots and build tables.
			\end{enumerate} 
			Then, tables in \textit{.tex} format as well as Figures~\ref{fig:benchmark},~\ref{fig:benchmark_full} can be found in the \textit{results/benchmark\_tabulars.tex, results/benchmark\_figures.tex} and \textit{results/benchmark\_figures\_full.tex} respectively.
		\end{itemize}
		\newpage
		\section{On the origins of NB distribution in allelic read counting}\label{app:nb_origins}
		
		To assess the statistical significance of the allelic imbalance, it is necessary to know PMF of $y|_{x}$ r.v. $g(y|x)$ and, conversely, PMF $f(x|y)$ of the $x|_{y}$ r.v. The major of this section is to prove that PMF follows the negative binomial law. Here we describe several alternative approaches leading to this conclusion.
		
		\subsection*{Inferring NB from assumptions imposed on other r.v.}
		The NB model in the basement of the {MIXALIME} framework is grounded in two key assumptions:
		
		\begin{enumerate}
			\item The number of read counts for alleles $x$ and $y$ conditioned on the total coverage $n = x + y$ is a binomial r.v.: $$x|_{n} \sim \mathpzc{Binom}(n, p),~~y|_{n} \sim \mathpzc{Binom}(n, 1-p),$$
			where $p$ is the probability of obtaining a read supporting a certain allele (e.g. $p = \frac{BAD}{BAD + 1}$). We shall denote marginal PMFs of $x, y$ as $f(x), g(y)$ and their conditionals will follow this notation (e.g. $f(x|y)$).
			\item The total coverage $n$ is a negative binomial r.v.:
			$$n \sim \mathpzc{NB}(r_0, p_0),$$
			were $r_0$ and $p_0$ are some parameters. This a common assumption in the analysis of high-throughput sequencing data, e.g. widely accepted for RNA-Seq. The marginal PMF of $n$ is denoted as $h(n)$.
			In practice, if directly estimated from the data, $r_0$ attains small values less than $1$ and $p_0 \rightarrow 1$. 
		\end{enumerate}
		
		To infer $g(y|x)$, the following lemma will be necessary.
		
		\begin{lemma}[Link between conditional distributions]
			Let $X, Y$ be random variables with a joint probability density function $f(x, y)$ and $N = X + Y$. Then, the following relation holds:
			\begin{equation}
				f(y|x) = h(n=x + y|x),
			\end{equation}
			where $f(x|y), h(n|y)$ are densities of $X$ and $N$, respectively, conditioned on $y$.
		\end{lemma}
		\begin{proof}
			Let's infer the joint distribution function of $(N, X)$: 
			\begin{equation*}
				\begin{split}
					P(N < n, X < x) = P(X + Y < n, X < x) = P(Y < n - X, X < x) = \int_{-\infty}^{x} \int_{-\infty}^{n - X} f(X, Y) dY dX
				\end{split}
			\end{equation*}
			According to the definition, the joint density of $(N, X)$ is:
			\begin{equation*}
				\begin{split}
					h(n, x) = \frac{\partial^2}{\partial x \partial n} P(N < n, X < x) = \frac{\partial^2}{\partial x \partial n} \int_{-\infty}^{x} \int_{-\infty}^{n - X} f(X, Y) dY dX = \frac{\partial}{\partial x} \int_{-\infty}^{x} \frac{\partial }{\partial n}\int_{-\infty}^{n - X} f(X, Y) dY dX = \\ =\frac{\partial}{\partial x} \int_{-\infty}^{x} f(X, n - X) dX = f(x, n - x).
				\end{split}
			\end{equation*}
			Hence, $f(x, n - x) = f(x, y) = h(n= x + y, y)$. Then, using the decomposition of joint density into a product of conditional density and marginal density,
			$$f(x, y) = f(y|x) f(x) = h(n, x) = h(n|x) f(x) \leftrightarrow f(y|x) = h(n=x + y|x).$$
		\end{proof} 
		
		The lemma combined with Bayes' theorem helps to infer $g(y|x)$:
		\begin{equation}\label{eq:cond_bayes}
			g(y|x) = h(n|x) =  \frac{\overbrace{f(x|n)}^{\mathpzc{Binom}(n, p)} \overbrace{h(n)}^{\mathpzc{NB}(r_0, p_0)}}{\underbrace{f(x)}_{\text{Yet unknown marginal}}}
		\end{equation} 
		We know the terms except for the marginal distribution $f(x)$.
		\begin{theorem}[Marginals]
			Let $x, y$ be random variables with known conditional distributions, and let the marginal distribution of $n = x + y$ be known as well:
			$$x|_{n} \sim \mathpzc{Binom}(n, p),~~y|_{n} \sim \mathpzc{Binom}(n, 1-p), n \sim \mathpzc{NB}(r_0, p_0).$$
			Then,
			\begin{equation} \label{eq:marginals}
				x \sim \mathpzc{NB}\left(r_0, \frac{p_0 p}{1 - (1 - p)p_0}\right),~y \sim \mathpzc{NB}\left(r_0,\frac{p_0 (1 - p)}{1 - p p_0} \right).
			\end{equation}
		\end{theorem}
		\begin{proof}
			We prove it by utilizing the joint distribution of $x$ and $n$ with PMF $h(x, n)$:
			\begin{equation}\label{eq:x_marginal}
				\begin{split}
					f(x) = \sum_{n=0}^\infty h(x, n) = \sum_{n=0}^\infty \overbrace{f(x|n)}^{\mathpzc{Binom}(n, p)} \overbrace{h(n)}^{\mathpzc{NB}(r_0, p_0)} = \sum_{n=x}^\infty f(x|n)h(n) = \sum_{n=x}^\infty \binom{n}{x} p^x (1-p)^{n-x} \binom{n+r_0-1}{r_0-1} {p_0}^n (1-p_0)^{r_0}  = \\ = \left( \left(\frac{p}{1-p}\right)^x (1 - p_0)^{r_0}) \frac{1}{x! (r_0 - 1)!} \right) \sum_{n=x}^\infty \left( \overbrace{p_0 (1 - p)}^{\hat{p}}\right)^n \frac{(n + r_0 - 1)!}{(n - x)!} = \\ =  \left( \left(\frac{p}{1-p}\right)^x (1 - p_0)^{r_0} \frac{1}{x! (r_0 - 1)!} \right) \sum_{i=0}^\infty {\hat{p}}^{i + x} \frac{(i + \overbrace{x + r_0 - 1}^\alpha)!}{i!} = \left( \left(\hat{p} \frac{p}{1-p}\right)^x (1 - p_0)^{r_0} \frac{1}{x! (r_0 - 1)!} \right) \sum_{i=0}^\infty {\hat{p}}^{i} \frac{(i + \alpha)!}{i!}
				\end{split}
			\end{equation}
			The series $\sum_{i=0}^\infty {\hat{p}}^{i} \frac{(i + \alpha)!}{i!}$ converge given that $0 \le \hat{p} < 1$, which is the case here, to $\beta(\hat{p}) = \frac{\alpha!}{(1 - \hat{p})^{\alpha + 1}}$. To prove this, consider Taylor series $\beta(\hat{p}) = \sum_{i=0} \hat{p}^i \frac{\partial^i}{\partial \hat{p}^i}\beta(0)$. What we really need to prove here is that $\frac{\partial^i}{\partial \hat{p}^i}\beta(0) = (i + \alpha)!$, which requires computing the $i$-th derivative
			$$\frac{\partial^i}{\partial \hat{p}^i}\beta(\hat{p}) = (1- \hat{p})^{-1 - a - i} (i + \alpha)!.$$
			It holds for the case of $i=0$. We assume that it holds for $i = m$, i.e.  and prove in accordance with the principle of the mathematical induction for $m +1$ that the derivative equation still holds:
			$$\frac{\partial^{m + 1}}{\partial \hat{p}^{m+1}}\beta(\hat{p}) =  (1- \hat{p})^{-2 - a - m} (m + \alpha)! (m + \alpha + 1) =  (1- \hat{p})^{-2 - a - m} (m + \alpha + 1)!. $$
			By plugging in $\hat{p} = 0$ we see that the relation still holds, therefore $\sum_{i=0}^\infty {\hat{p}}^{i} \frac{(i + \alpha)!}{i!} = \frac{\alpha!}{(1 - \hat{p})^{\alpha + 1}} = \frac{(x + r_0 - 1)!}{(1 - \hat{p})^{x + r}}$. Now, back to deriving the marginal PMF $f(x)$ at Equation~\ref{eq:x_marginal}:
			\begin{equation*}
				\begin{split}
					f(x) = \left( \left(\hat{p} \frac{p}{1-p}\right)^x (1 - p_0)^{r_0}) \frac{1}{x! (r_0 - 1)!} \right) \overbrace{\frac{(x + r_0 - 1)!}{(1 - \hat{p})^{x + r}}}^{\text{Here were the series}} =  \binom{x + r_0 - 1}{x}  \left( \frac{\hat{p} p}{(1-p)(1 - \hat{p})}\right)^x \left(\frac{1 - {p_0}}{1 -\hat{p}}\right)^{r_0}  = \\ \overset{\text{expand } \hat{p}}{=}
					\binom{x + r_0 - 1}{x}  \left( \frac{p_0 p}{1 - p_0 + p_0 p}\right)^x \left(\frac{1 - {p_0}}{1 -p_0 + p_0 p}\right)^{r_0}
				\end{split}
			\end{equation*}
			That's exactly PMF of $\mathpzc{NB}(r_0, p_m)$ where $p_m = \frac{p_0 p}{1 - p_0 + p_0 p}$. The same can be shown for the $y$ with the only difference being a different selection of the success probability of $1 - p$ for the r.v. $y|_{n}$.
		\end{proof}
		Now we have everything at hand to find the distribution of $y|_{x}$.
		\begin{theorem}[Conditional distributions]
			Let $x, y$ be random variables with known conditional distributions and a known marginal distribution of their sum $n = x + y$:
			$$x|_{n} \sim \mathpzc{Binom}(n, p),~~y|_{n} \sim \mathpzc{Binom}(n, 1-p), n \sim \mathpzc{NB}(r_0, p_0).$$
			Then,
			\begin{equation}\label{eq:conditionals}
			y|_{x} \sim \mathpzc{NB}(x + r_0, p_0 (1 - p)),~x|_{y} \sim \mathpzc{NB}(y + r_0, p_0 p)
			\end{equation}
		\end{theorem}
		\begin{proof}
			We know from the previous theorem that 
			$$x \sim \mathpzc{NB} \left(r_0, \frac{p_0 p}{1 - (1 - p)p_0}\right),~y \sim \mathpzc{NB} \left(r_0,\frac{p_0 (1 - p)}{1 - p p_0} \right).$$
			Then we can infer desired densities of $x|_{y}$ and $y|_{x}$ from the Bayes' theorem.

			Below we will use the collected variable $p_m = \frac{p_0 p}{1 - p_0 + p_0 p}$ from the proof of the theorem on marginals for brevity.
			\begin{equation*}
				\begin{split}
					g(n|x) =  \frac{\overbrace{f(x|n)}^{\mathpzc{Binom}(n, p)} \overbrace{h(n)}^{\mathpzc{NB}(r_0, p_0)}}{\underbrace{f(x)}_{\mathpzc{NB}(r_0, p_m)}} = \frac{\binom{n}{x} p^x (1-p)^{n-x} \binom{n+r_0-1}{r_0-1} {p_0}^n (1-p_0)^{r_0}}{\binom{x + r_0 - 1}{x}  {p_m }^x (1 - p_m)^{r_0}} = \\ \overset{\text{expand binomials}}{=} \left( \frac{n!}{x! (n - x)!} \frac{(n + r_0 - 1)!}{(r_0 - 1)! n!} \middle/ \left( \frac{(x + r_0 - 1)!}{x! (r_0 - 1)!}\right) \right) \left(\frac{p}{(1-p) p_m}\right)^x \left( (1-p) p_0 \right)^n \left(\frac{1 - p_0}{1-p_m} \right)^{r_0} = \\ \overset{\text{collect binomial}}{=} \binom{n + r_0 - 1}{n - x} \left(\frac{p}{(1-p) p_m}\right)^x \left( (1-p) p_0 \right)^n \left(\frac{1 - p_0}{1-p_m} \right)^{r_0} = \\ \overset{\text{expand } p_m}{=} \binom{n + r_0 - 1}{n - x} \left(\frac{1 - p_0 + p_0 p}{(1-p) p_0}\right)^x \left( (1-p) p_0 \right)^n \left(1 - p_0 + p_0 p \right)^{r_0} = \\ = \binom{n + r_0 - 1}{n - x} \left(1 -p_0 + p_0 p \right)^{r_0 + x} \left(p_0 (1 - p)\right)^{n - x} = g(y|x) = \binom{x + y + r_0 - 1}{y} \left(1 - (1 - p)p_0 \right)^{r_0 + x} \left(p_0 (1 - p)\right)^{y}
				\end{split}
			\end{equation*}
			
			The same can be shown for $x|_{y}$ with the only difference being a success probability $1 - p$.
		\end{proof}
		\subsection*{Building NB straight up from the logic of the experiment}
		
		Note that when performing allelic read counting at a genomic variant there are three possible outcomes: we get a read count from the reference allele, we get a read count from the alternative allele, or we get nothing (no more reads). The latter event can be thought of as a stop signal. I.e. here we are dealing with the following sequences:
		$$\left\{ref,~ ref,~ alt,~ ref,~ alt,~ ref,~ alt,~ alt,~ \textcolor{red}{\text{STOP}} \right\}$$
		Each of the events from $ref,~ alt,~ \textcolor{red}{\text{STOP}}$  has its own distinct probability of a ''success'': $p_r$, $p_a$ and $p_s$. Those quantities are bound to constraints:
		$$
		\begin{cases}
			p_r + p_a + p_s = 1 \\
			~\\
			\frac{p_r}{p_a} = BAD
		\end{cases} \leftrightarrows
		\begin{cases}
			p_a = \frac{1}{BAD + 1} (1 - p_s) = (1 - p) \overbrace{(1 - p_s)}^{p_0}  = p_0 (1 - p)\\
			~\\
			p_r = BAD p_a = \frac{BAD}{BAD+1} (1 - p_s) = p (1 - p_s) = p_0 p
		\end{cases}
		$$
		
		This three-variate ($ref, alt$ and the stop signal $\textcolor{red}{\text{STOP}}$) random variable follows the bivariate (with the stop-signal being interpreted as a failure) multinomial law $(x, y) \sim \mathpzc{NM}(r_0, p_0, p_0 (1 - p), p_0 p)$. Here, we allowed $r_0$ ''stop'' signs to appear before we stopped counting for the sake of generality. Now, we can use known negative multinomial properties to easily obtain the conditional r.v.:  $$y|x \sim \mathpzc{NB}(x + r_0, p_0 (1 - p)),$$
		which is exactly the same negative binomial distribution we've got in the previous subsection! 
		
		Here, we can think of conditioning negative multinomial variables as effectively ''collapsing'' the conditioned variables into ''failures''. Therefore, in the obtained negative binomial, the probability of success is deflated to accommodate for the greater number of ''failures'' (i.e. they now include both $ref$ and $\textcolor{red}{\text{STOP}}$ event).  
		
		\section{On linearity of the reference mapping bias} \label{app:nb}	
		\begin{figure}[H]
			\centering
			\begin{adjustbox}{max width=\textwidth}
				\includegraphics{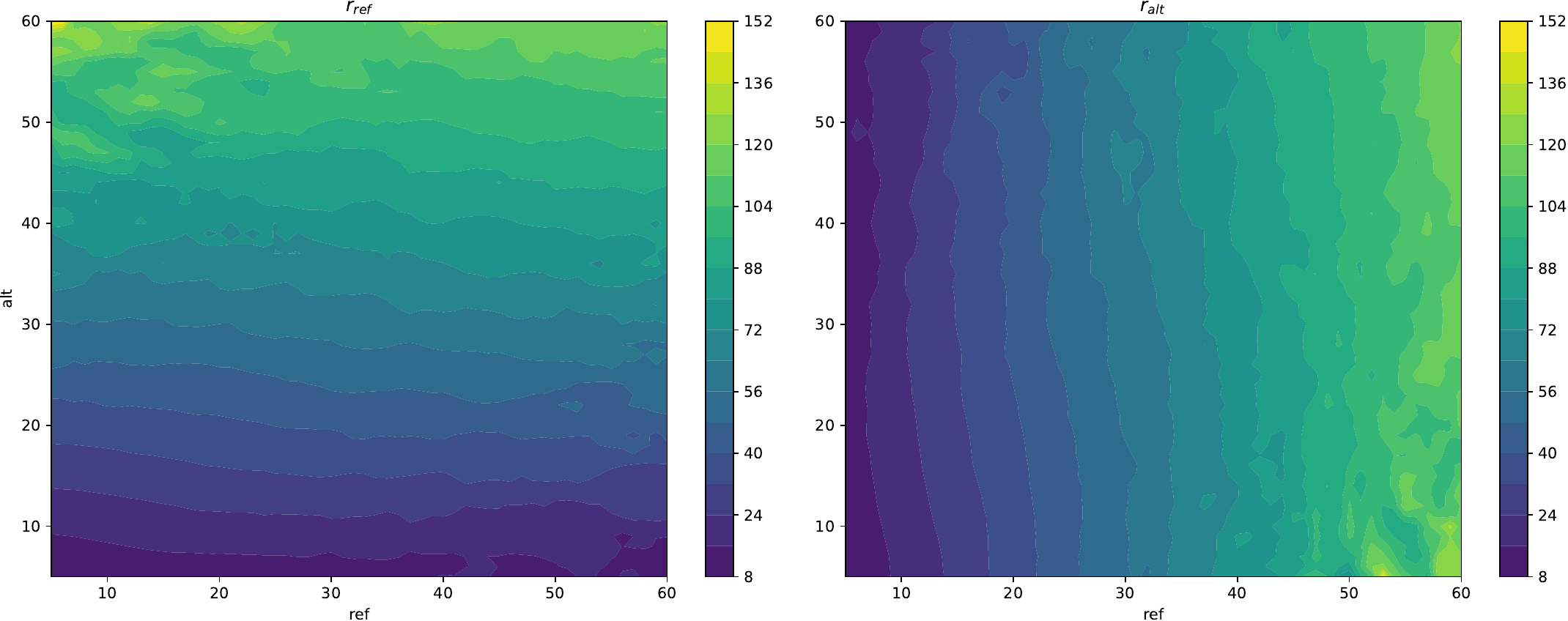}
			\end{adjustbox}
			\caption{Estimates of $r$ obtained from different windows.}
			\label{fig:dtbin}
		\end{figure}
		
		We examined DNAse-seq dataset for a cancerous cell-line K562 to see whether our idea of linear (or, at least, monotone) bias effect is any close to reality. For this task, we use a model different from those discussed in the paper, unusable for scoring SNVs, but helpful in terms of demonstrating the effect. We assume that read counts $x$ from an allele (be it a reference or an alternative one) are distributed as left-truncated at $a$ and right-truncated at $b$ binomial random variable (next denoted as $\mathpzc{DTBin}$):
		$$x \sim \mathpzc{DTBin}(r, a, b)$$
		The density function $x$ is:
		$$f_{\mathpzc{DTBin}}(x|r, a, b) = \frac{f_{\mathpzc{Bin}}(x|r)}{I_{1-p}(r-b, b + 1) - I_{1-p}(r-a, a + 1)} \mathbbm{1}_{a < x < b},$$
		where $f_{\mathpzc{\mathpzc{Bin}}}$ is a binomial density function and $I$ is a regularized incomplete beta function which computation we discuss in Appendix~\ref{app:nb_cdf}.
		
		We obtained estimates of $r$ by moving a two-dimensional window across a plane formed by reference and alternative allele read counts. The window is expanded until at least 4 unique entries/counts are inside of it. For each window position, truncation boundaries $a$ and $b$ are selected as minimal and maximal counts present in it.
		
		Figure~\ref{fig:dtbin} displays estimates of $r$ parameter at varying locations. It appears so, that the linearity hypothesis is sound, and we hope that any discrepancies are tackled by the local MLE/window model (Section~\ref{sect:window})
		\newpage
		\section{MCNB density function} \label{app:mcnb_proof}
		To prove that the Equation~\ref{eq:pre_mcnb} indeed marginalizes to the Equation~\ref{eq:mcnb}, remember the definition of a marginal distribution $f(y) = \sum_{x=-\infty}^\infty f(y|x) f(x)$ and apply it directly to our case:
		
		\begin{equation*}
			\begin{split}
				f_{\mathpzc{MCNB}}(y|r, p) = \sum_{i=1}^r f_{\mathpzc{NB}}(y|x, p) f_{ZTBin}(x|r, 1 - p) = \\ = \sum_{x=1}^r\left( (1-p)^x p^y \frac{\Gamma(y + x)}{\Gamma(y + 1) \Gamma(x)} \right) \left( \frac{(1-p)^x p^{r - x} \frac{\Gamma(r + 1)}{\Gamma(r - x + 1) \Gamma(x + 1)}}{1 - p^r}\right) = \\ = \frac{p^{r + y} \Gamma(r + 1)}{(1-p^r)\Gamma(y + 1) } \sum_{x=1}^r \frac{(1 - p)^{2x} p^{-x}\Gamma(y + x)}{\Gamma(x) \Gamma(r - x  + 1) \Gamma(x + 1)} = \frac{p^{r + y - 1} (1 - p)^2 \Gamma(r + 1)}{(1-p^r)\Gamma(y + 1) } \sum_{x=0}^{r-1} \underbrace{\frac{(1 - p)^{2x} p^{-x}\Gamma(y + x + 1)}{\Gamma(x + 1) \Gamma(r - x) \Gamma(x + 2)}}_{\alpha_x} = \\ =  \frac{p^{r + y - 1} (1 - p)^2 \Gamma(r + 1)}{(1-p^r)\Gamma(y + 1) } \sum_{x=0}^{r-1} \alpha_x = \frac{p^{r + y - 1} (1 - p)^2 \Gamma(r + 1)}{(1-p^r)\Gamma(y + 1) } \left( \underbrace{\sum_{x=0}^\infty \alpha_x}_{A} - \underbrace{\sum_{x=r}^\infty \alpha_x}_{B} \right)
			\end{split},
		\end{equation*}
		Conveniently, $A$ and $B$ can be transformed into hypergeometric series. For $A$ it is enough to take out the multiplier $\alpha_0 = \frac{\Gamma(y + 1)}{\Gamma(r)}$: $A = \alpha_0 \sum_{x=0}^\infty  \frac{\alpha_x}{\alpha_0} = \alpha_0 \sum_{x=0} \beta_i$. $\beta_i$ terms indeed suffice the hypergeometric function definition as $\beta_0 = 1$ and
		
		$$\frac{\beta_{i + 1}}{\beta_i} = -\frac{(x + y + 1) (x - r + 1)}{(x + 1) (x + 2)}\frac{(1 - p)^2}{p}$$
		Therefore, $$A = \frac{\Gamma(y + 1)}{\Gamma(r)} \, _2F_1(1 - r, y + 1, 2; -\frac{(1-p)^2}{p}) = r \frac{\Gamma(y + 1)}{\Gamma(r + 1)} \, _2F_1(1 - r, y + 1, 2; -\frac{(1-p)^2}{p}) $$
		As for the $B$, we also have to shift the infinite series left $r$ times to align it with the definition of the hypergeometric series: $$B = \sum_{x=r}^\infty \alpha_x = \sum_{x=0}^\infty \alpha_{x + r}  = (1 - p)^{2 r} p^{-r}\sum_{x=0}^\infty \frac{(1 - p)^{2x} p^{-x}\Gamma(y + x + r + 1)}{\Gamma(x + r + 1) \Gamma(- x) \Gamma(x + r + 2)}.$$
		Notice that $\Gamma(-x)$ in the denominator goes to $\pm \infty$, hence $B = 0$.
		$$ f_{\mathpzc{MCNB}}(y|r, p) = \frac{p^{r + y - 1} (1 - p)^2 \Gamma(r + 1)}{(1-p^r)\Gamma(y + 1) } A = \frac{r (p-1)^2 p^{r+y-1} \, _2F_1\left(1-r,y+1;2;-\frac{(p-1)^2}{p}\right)}{1 - p^r}.$$
		This concludes the proof.
		
		\newpage
		
		\section{Computation of the likelihood}\label{app:mcnb_loglik}
		Direct computation of Equation~\ref{eq:mcnb} implies computation of the hypergeometric function $\,_2F_1$. To do so rigorously is not an easy task in a general case \citep{hypgeom} and it takes a toll on the computational performance. Instead, here we infer the recurrent formula for MCNB density. To this end, we use the known consecutive neighbors relation for 
		\begin{equation}\label{eq:hyp_consneighb}
			\begin{split}
				\,_2F_1(1 - r, x + 1; 2; -\frac{(1 - p)^2}{p}) = \alpha_{x} \,_2F_1(1 - r, x; 2; -\frac{(1 - p)^2}{p}) + \beta_{x} \,_2F_1(1 - r, x - 1; 2; -\frac{(1 - p)^2}{p}),\\
				\alpha_{x} = \frac{p^2 (r+x-1)-2 p r+r+x-1}{(p-1) p x+x},~~~ \beta_{x} = \frac{p (2 - x)}{(p-1) p x+x}~~~~~~
			\end{split}.
		\end{equation}
		In turn, $f_{\mathpzc{MCNB}}$ for two consecutive neighbors is:
		\begin{equation}\label{eq:mcnb_consneighb}
			\begin{split}
				f_{\mathpzc{MCNB}}(x|r, p) =& C \, _2F_1\left(1-r,x+1;2;-\frac{(1-p)^2}{p}\right),~C = \frac{r (p-1)^2 p^{r+x-1}}{1 - p^r}\\
				f_{\mathpzc{MCNB}}(x - 1|r, p) =& \frac{C}{p} \, _2F_1\left(1-r,x;2;-\frac{(1-p)^2}{p}\right),\\
				f_{\mathpzc{MCNB}}(x - 2|r, p) =&\frac{C}{p^2} \, _2F_1\left(1-r,x+1;2;-\frac{(1-p)^2}{p}\right).
			\end{split}
		\end{equation}
		By combining Equation~\ref{eq:mcnb_consneighb} and Equation~\ref{eq:hyp_consneighb} we finally obtain the recurrent equation for the MCNB density:
		\begin{equation}\label{eq:mcnb_recurrent}
			f_{\mathpzc{MCNB}}(x|r, p) = \alpha_x p f_{\mathpzc{MCNB}}(x - 1|r, p) + \beta_x p^2 f_{\mathpzc{MCNB}}(x - 2|r, p),~x \ge 2.
		\end{equation}
		What's missing from the Equation~\ref{eq:mcnb_recurrent}, are initial conditions $f_{\mathpzc{MCNB}}(1|r, p)$ and $f_{\mathpzc{MCNB}}(0|r, p)$. See that hypergeometric function in $f_{\mathpzc{MCNB}}(1|r, p)$ attains form \begin{equation}\label{eq:1f0}
			\begin{split}
				\,_2F_1(1-r, 2; 2; -\frac{(1 - p)^2}{p}) = \,_1 F_0(1-r; -\frac{(1 - p)^2}{p}) = \sum_{i=0}^\infty \frac{\Gamma(1 - r + i)}{\Gamma(1 - r) i!} \left(-\frac{(1 - p)^2}{p}\right)^i =\\=  \sum_{i=0}^\infty \binom{i-r}{i} \left(-\frac{(1 - p)^2}{p}\right)^i  = \sum_{i=0}^\infty (-1)^i \binom{1-r}{i} \left(-\frac{(1 - p)^2}{p}\right)^i = \sum_{i=0}^\infty \binom{1-r}{i} \left(\frac{(1 - p)^2}{p}\right)^i  = \\ = \left(1 +\frac{(1 - p)^2}{p}\right)^{1 - r}.
			\end{split}
		\end{equation}
		As for $\,_2F_1(1-r, 1; 2; -\frac{(1 - p)^2}{p})$ in $f_{\mathpzc{MCNB}}(1|r, p)$, we shall take advantage of the other consecutive neighbor formula, this time with respect to the third parameter of the Gauss hypergeometric function:
		\begin{equation*}
			\begin{split}
				\,_2F_1(1-r, 1; c; -\frac{(1 - p)^2}{p}) = \alpha_c \,_2F_1(1-r, 1; c - 1; -\frac{(1 - p)^2}{p}) + \beta_c \,_2F_1(1-r, 1; c - 2; -\frac{(1 - p)^2}{p})\\
				\alpha_c = \frac{(c-1) (c (p (2 p-3)+2)+p (p (r-5)-2 r+8)+r-5)}{(c-2) (p-1)^2 (c+r-2)}, ~\beta_c = \frac{(1-c) ((p-1) p+1)}{(p-1)^2 (c+r-2)}.
			\end{split}
		\end{equation*}
		As $c \rightarrow 2$, the second term vanishes, whereas the hypergeometric series in the first term undergoes the same metamorphosis as in Equation~\ref{eq:1f0}. After simplification the hypergeometric function at $x = 0$ transforms into $\frac{p \left(\left(p+\frac{1}{p}-1\right)^r-1\right)}{(p-1)^2 r}$. 
		
		Finally, we substitute the obtained values for hypergeometric functions at $x = 1$, $x = 0$ into $f_{\mathpzc{MCNB}}(1|r, p)$ and $f_{\mathpzc{MCNB}}(0|r, p)$ respectively and obtain the initial conditions for the recurrence relation at Equation~\ref{eq:mcnb_recurrent}:
		\begin{equation}\label{eq:mcnb_init}
			f_{\mathpzc{MCNB}}(1|r, p) = \frac{r (p-1)^2 p^r \left(p+\frac{1}{p}-1\right)^{r-1}}{1 - p^r},~~ f_{\mathpzc{MCNB}}(0|r, p) = \frac{p^r \left(\left(p+\frac{1}{p}-1\right)^r-1\right)}{1 - p^r}.
		\end{equation}
		Here, extra caution should be taken when computing logarithms of $f_{\mathpzc{MCNB}}(0|r, p)$ due to the $\left(\left(p+\frac{1}{p}-1\right)^r-1\right)$ term. The $\left(p+\frac{1}{p}-1\right)^r$ part blows up for sufficiently large $r$, and the latter subtraction of $1$ makes it impossible to do calculations entirely in logarithmic space. A simple workaround is to omit the subtraction whenever $\left(p+\frac{1}{p}-1\right)^r >> 1$. To see why it works, consider a case when $\left(p+\frac{1}{p}-1\right)^r \approx 10^{400}$. This value can't be held in \texttt{float64}, and the result of straightforward computation will be \texttt{inf}. On other hand, even if it could be held (lower value of $10^{300}$ can, for instance, and the next statement will still hold) in that type, the results of $10^{400} - 1$ is still $10^{400}$ in any reasonable machine precision. Therefore, at those magnitudes, we are free to omit $-1$ altogether.

		Thus, for MCNB, we are able to completely avoid explicit computation of hypergeometric functions.
		\newpage
		\section{MCNB moments} \label{app:mcnb_props}
		In \textbf{MIXALIME} we need to know $\EX[x]$ to compute effect sizes. Formulae present in \ref{table:moments} were derived in a non-rigorous manner. They appear to be correct as they agree exactly with a numerically estimated moments. Here, we provide motivation by proving them for special integer values of $r$. But first, remember the MCNB density from Equation~\ref{eq:mcnb}:
		$$ f_{\mathpzc{MCNB}}(x|r, p) = \frac{r (p-1)^2 p^{r+x-1} \, _2F_1\left(1-r,x+1;2;-\frac{(1-p)^2}{p}\right)}{1 - p^r}.$$
		
		\subsection*{First moment / mean}
		The first moment is defined as 
		$$\EX_{\mathpzc{MCNB}}[x] = \sum_{i=0}^\infty i f_{\mathpzc{MCNB}}(i|r, p).$$
		For $r = 1$ $f_{\mathpzc{MCNB}}$ attains the form:
		$$f_{\mathpzc{MCNB}}(i|1, p) = \frac{(p-1)^2 p^{i} \, _2F_1\left(0,i+1;2;-\frac{(1-p)^2}{p}\right)}{1 - p}$$
		Remember, that $\, _2F_1(a, b;c;z) = \sum_{j=0}^\infty \frac{\Gamma(a + j) \Gamma(b + j) \Gamma(c)}{\Gamma(a) \Gamma(b) \Gamma(c + j) j!} z^j$. So, here 
		$$ _2F_1\left(0,i+1;2;-\frac{(1-p)^2}{p}\right) = \sum_{j=0}^\infty \frac{\Gamma(j) \Gamma(x + 1 + j)}{\Gamma(0) \Gamma(2 + j) j!} \left(-\frac{(1-p)^2}{p}\right)^j = 1.$$
		That's true because $\left| \frac{1}{\Gamma(0)}\right| \rightarrow 0$, hence all terms in a sum except the first one cancels to zero. Therefore,
		$$f_{\mathpzc{MCNB}}(i|1, p) = \frac{(1-p)^2 p^{i} \,}{1 - p} = (1-p) p^i,$$
		and the sum of interest is
		$$\EX_{\mathpzc{MCNB}}[x|r=1] = (1-p) \underbrace{\sum_{i=0}^\infty  i p^i}_{S(p)} = (1 - p )S(p).$$
		We deal with $S(p)$ by observing that
		$$\frac{\partial }{\partial p}\sum_{i=0}^\infty p^i = \sum_{i=0}^\infty i p^{i - 1} = \frac{1}{p} \underbrace{\sum_{i=0}^\infty i p^i}_{S(p)} = \frac{\partial }{\partial p} \left( \frac{1}{(1 - p)}\right) = \frac{1}{(1 -p)^2}.$$
		Then, 
		$$S(p) = \frac{p}{(1 - p)^2},$$
		and finally 
		$$\EX_{\mathpzc{MCNB}}[x|r=1] = \frac{p}{1-p}.$$
		For $r = 2$:
		
		$$ f_{\mathpzc{MCNB}}(i|2, p) = \frac{2 (p-1)^2 p^{1+x} \, _2F_1\left(-1,x+1;2;-\frac{(1-p)^2}{p}\right)}{1 - p^2},$$
		where $$\, _2F_1 = \sum_{j=0}^\infty \frac{\Gamma(-1 + j) \Gamma(x + 1 + j)}{\Gamma(-1) \Gamma(x + 1) \Gamma(2 + j)} \left( -\frac{(1-p)^2}{p}\right) = 1 + \frac{1 + x}{2} \frac{(1-p)^2}{p}.$$
		The equation holds as $\frac{\Gamma(-1 + j)}{\Gamma(-1)} = 0~ \forall ~j > 1$ and for $j = 1$ $\frac{\Gamma(0)}{\Gamma(-1)} = -1,~ \frac{\Gamma(x + 2)}{\Gamma(x + 1)} = x + 1$. After simplifications,
		
		$$\EX_{\mathpzc{MCNB}}(i|r = 2) = \sum_{i=0}^\infty i \frac{(1-p) p^i \left(p^2+(p-1)^2 i+1\right)}{p+1} = \frac{2 p}{1 -p^2}$$
		This can be proved a very much similar way as we did with $r = 1$ case, the only technical difference would be finding a sum with respect to the quadratic term $i^2 p^i$. We show how to find similar sums in the next section.
		
		\subsection*{Second moment}
		
		The second moment of the MCNB distribution was not present in Table~\ref{table:moments}, but we obtained it nevertheless to infer variance.It is equal to
		$$\EX_{\mathpzc{MCNB}}[x^2] = \sum_{i=0}^{\infty} i^2 f_{\mathpzc{MCNB}}(i|r, p) = \frac{p (p r (p (-r)+p+r)+r)}{(1-p) \left(1 - p^r\right)}.$$
		
		$$\EX_{\mathpzc{MCNB}}[x^2|r=1] = (1-p) \underbrace{\sum_{i=0}^\infty  i^2 p^i}_{S(p)} = (1 - p )S(p).$$
		
		We take a second derivative of a geometric series:
		$$\frac{\partial^2}{\partial p^2} \sum_{i=0}^\infty p^i = \frac{2}{(1-p)^3} = \sum_{i=0}^\infty i (i - 1) p^{i-2} = \frac{1}{p^2} \left( \underbrace{\sum_{i=0}^\infty i^2 p^i}_{S(p)} - \underbrace{\sum_{i=0}^\infty i p^i}_{E[x|r=1] / (1-p)}\right) =\frac{1}{p^2} \left(S(p)  - \frac{p}{(1 - p)^2}\right).$$
		In turn, 
		$$S(p) =  2 \frac{p^2} {(1-p)^3} + \frac{p}{(1-p)^2}=  \frac{p (p + 1)}{(1 - p)^3},$$
		and 
		$$\EX_{\mathpzc{MCNB}}[x^2|r=1] = \frac{p(p + 1)}{(1 - p^2)},$$
		which agrees with an assumed formula.
		
		Likewise, similar observations can be made for higher integer values of $r$.

		\subsection*{Variance}
		
		If we trust the formulae for $E[x]$ and $E[x^2]$, it is trivial to obtain variance estimate:
		
		$$var[x] = \EX[x^2] - \EX[x]^2$$
		One can verify that it agrees with $var[x]$ for MCNB present in Table~\ref{table:moments}.
		
		\subsection*{Higher values of $r$}
		The pattern for both moments appears to hold for higher values of integer $r$. Proving them manually becomes increasingly tiresome, but luckily that can be automated with symbolic algebra software such as \textbf{sympy}, \textbf{MAPLE} or \textbf{Mathematica}. For instance, in \textbf{Mathematica}, see: \newpage
		
		\consolein
		\begin{lstlisting}[style=codeinput, language=Mathematica]
(* MCNB density*)
f[x_, r_, p_] = 1/(1 - p^r) r (1 - p)^2 p^(-1 + r + x) Hypergeometric2F1[1 - r, 1 + x, 2, -((1 - p)^2/p)]
(* Let's see mean values for fist few integer r*)
Table[
FullSimplify[Sum[x * f[x, r, p], {x, 0, Infinity}]], {r, 1, 10}]
(* Same, but the second moment*)
Table[
FullSimplify[Sum[x^2 * f[x, r, p], {x, 0, Infinity}]], {r, 1, 10}]
		\end{lstlisting}
		
		The results obtained this way might require some simplifications and rearrangements done manually to obtain the nice and clean forms, though. See:
		\consoleout
		\begin{tcolorbox}[spartan,  colframe=black,  left=1mm,right=1mm,top=-2mm,bottom=1mm,  colback=codecolors]
			\begin{equation*}
				\begin{split}
					&\left\{-\frac{p}{p-1},-\frac{2 p}{p^2-1},-\frac{3 p}{p^3-1},-\frac{4 p}{p^4-1},-\frac{5 p}{p^5-1},-\frac{6 p}{p^6-1},-\frac{7 p}{p^7-1},-\frac{8 p}{p^8-1},-\frac{9 p}{p^9-1},-\frac{10 p}{p^{10}-1}\right\}\\
					&\left\{\frac{p (p+1)}{(p-1)^2},\frac{2}{(p-1)^2}+\frac{1}{p+1}+\frac{1}{p-1}-2,\frac{3 p (p (3-2 p)+1)}{(p-1)^2 \left(p^2+p+1\right)},\frac{4 p \left(-3 p^2+4 p+1\right)}{p^5-p^4-p+1},\frac{5 p (p (5-4 p)+1)}{p^6-p^5-p+1},\right.\\ &\left. \frac{6 p (p (6-5 p)+1)}{p^7-p^6-p+1},\frac{7 p (p (7-6 p)+1)}{p^8-p^7-p+1},\frac{8 p (p (8-7 p)+1)}{p^9-p^8-p+1},\frac{9 p (p (9-8 p)+1)}{(p-1) p^9-p+1},\frac{10 p (p (10-9 p)+1)}{p \left((p-1) p^9-1\right)+1}\right\}
				\end{split}
			\end{equation*}
		\end{tcolorbox}
		\subsection*{Numerical verification}
		To verify validity of the moments formulae, we used the following code in Python. The version of \textbf{betanegbinfit} package used here is \textit{v. 1.8.1}.
		\consolein
		\begin{lstlisting}[style=codeinput, language=python]
import betanegbinfit.distributions as dist
import numpy as np

n = 200
ps = np.linspace(0.1, 0.9, n)
rs = np.linspace(1, 50, n)
x = np.array([1000])
z = np.arange(x[-1])
res_1 = np.zeros((n, n))
res_2 = np.zeros((n, n))

def second_moment(r, p):
	return p * (p * r *(r + p - p * r) + r ) / ((1 - p) * (1 - p ** r))

for i in range(n):
	for j in range(n):
		r = np.array([rs[i]])
		p = ps[j]
		prob = np.exp(dist.MCNB.logprob_recurrent(x, r, p, z.max() + 1))
		m = np.sum(z * prob)
		m2 = np.sum(z ** 2 * prob)
		res_1[i, j] = np.abs(dist.MCNB.mean(r, p) - m)
		res_2[i, j] = np.abs(second_moment(r, p) - m2)
print('1st moment| Max absolute error:', res_1.max())
print('2nd moment| Max absolute error:', res_2.max())
		\end{lstlisting}
		
		\consoleout
		\begin{lstlisting}[style=codeoutput]
1st moment| Max absolute error: 3.566924533515703e-12
2nd moment| Max absolute error: 2.5556801119819283e-10
		\end{lstlisting}
		
		\newpage
		\section{\texorpdfstring{Computation of $G_{\mathpzc{NB}}$}{Computation of NB CDF}}\label{app:nb_cdf}
		It is known that $G_{\mathpzc{NB}}$ can be represented in terms of the Regularized Incomplete Beta Function $I$:
		
		\begin{equation}
			G_{\mathpzc{NB}}(x|\theta) = I_{p}(x + 1, r).
		\end{equation}
		
		Whereas there are methods to evaluate $I$ with a great precision, they represent it via piecewise function that is not differentiable \citep{Brown1994}. Instead, we use the continued fraction representation of $I$ \citep{Tretter1980}:
		
		\begin{equation}\label{eq:cont_incomplete_beta}
			I_{p}(x, r) =  \lim_{m\rightarrow\infty} C\left(s + \cfrac{r(1)}{q(1) + \cfrac{r(2)}{q(2) + \cfrac{r(3)}{\cfrac{\ddots}{\cfrac{}{}~~~~~~+\cfrac{r(m)}{q(m)}}}}}\right)=  C * \left(s + \mathbf{K}_{j=1}^\infty \frac{r(j)}{q(j)}\right) \approx C * \left(s + \mathbf{K}_{j=1}^m \frac{r(j)}{q(j)}\right) ,
		\end{equation}
		where 
		\begin{center}
			\begin{equation*}
				\begin{tabular}{lll}
					$	r(1) = \frac{pt(r-1)}{r(p+1)}$ & ~~~~&$r(n) = \frac{p^2 t^2 (n-1)(p + r + n - 2)(p + n - 1)(r - n)}{r^2(p + 2n - 3)(p + 2n - 2)^2(p + 2n - 1)}$ it $n > 1$ \\
					$q(n) = \frac{2(p t + 2r)n^2 + 2(p t + 2r)(p - 1)n _ pr(p-2-pt)}{r(p + 2n - 2)(p + 2n)}$ & & $t= \frac{r x}{p(1 - x)}$ \\
					$s = 1$ & & $C = \frac{x^p(1-x)^{r - 1}}{p B(p, r)}$ 
				\end{tabular}
			\end{equation*}
		\end{center}
		To compute the $n$-th convergent of a continued fraction $f_n = \mathbf{K}_{j=1}^m \frac{r(j)}{q{j}}$, we employ Lentz's recurrent formula \citep{Lentz}:
		\begin{equation}\label{eq:lentz}
			f_n = C_n D_n f_{n-1},~~C_n = q(n) + \frac{a(n)}{C_{n-1}}, ~~D_n = \frac{1}{q(n) + a(n) D_{n-1}},~ C_0 = s,~ D_0 = 0,
		\end{equation}
		where $n$ is such that it satisfies the stopping criterion $|C_n D_n - 1| < \epsilon$ for some small value of $\epsilon$ (e.g. $10^{-12}$ by default).
		
		\begin{figure}[H]
			\centering
			\begin{tabular}{ll}
				\raisebox{10.5em}{\Huge A} &
				\begin{adjustbox}{max width=0.96\textwidth}
					\includegraphics{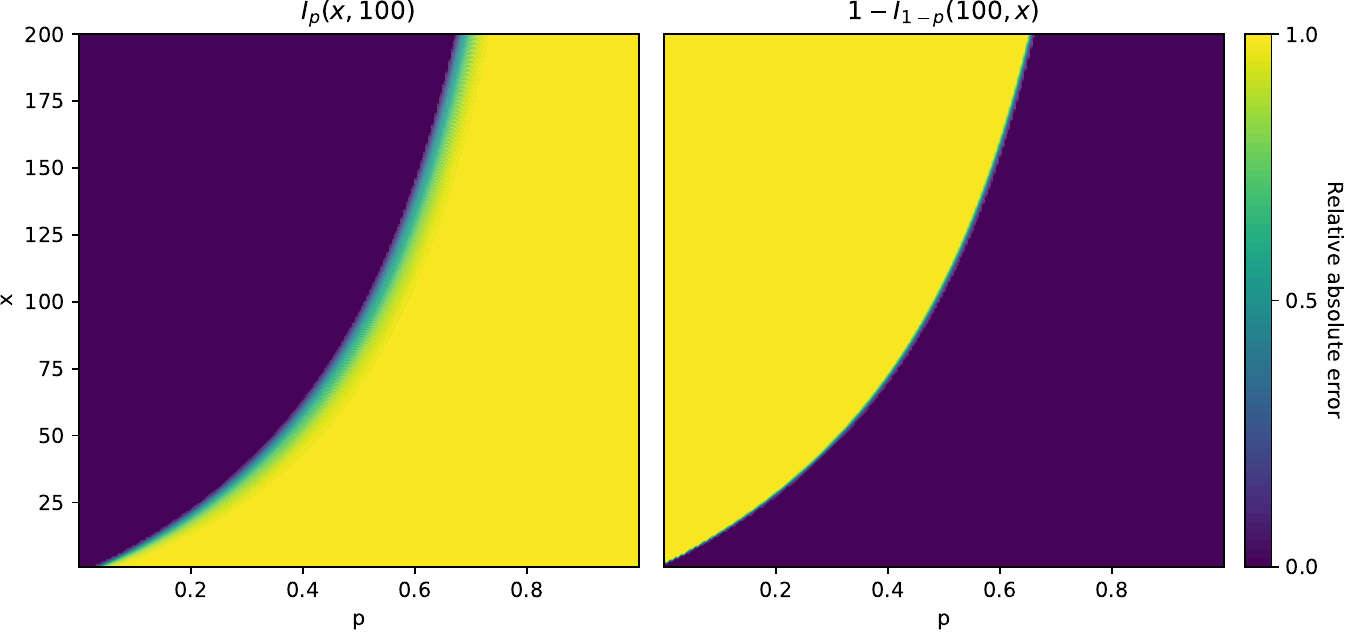}
				\end{adjustbox}\\
				\raisebox{10.5em}{\Huge B} &
				\begin{adjustbox}{max width=0.96\textwidth}
					\includegraphics{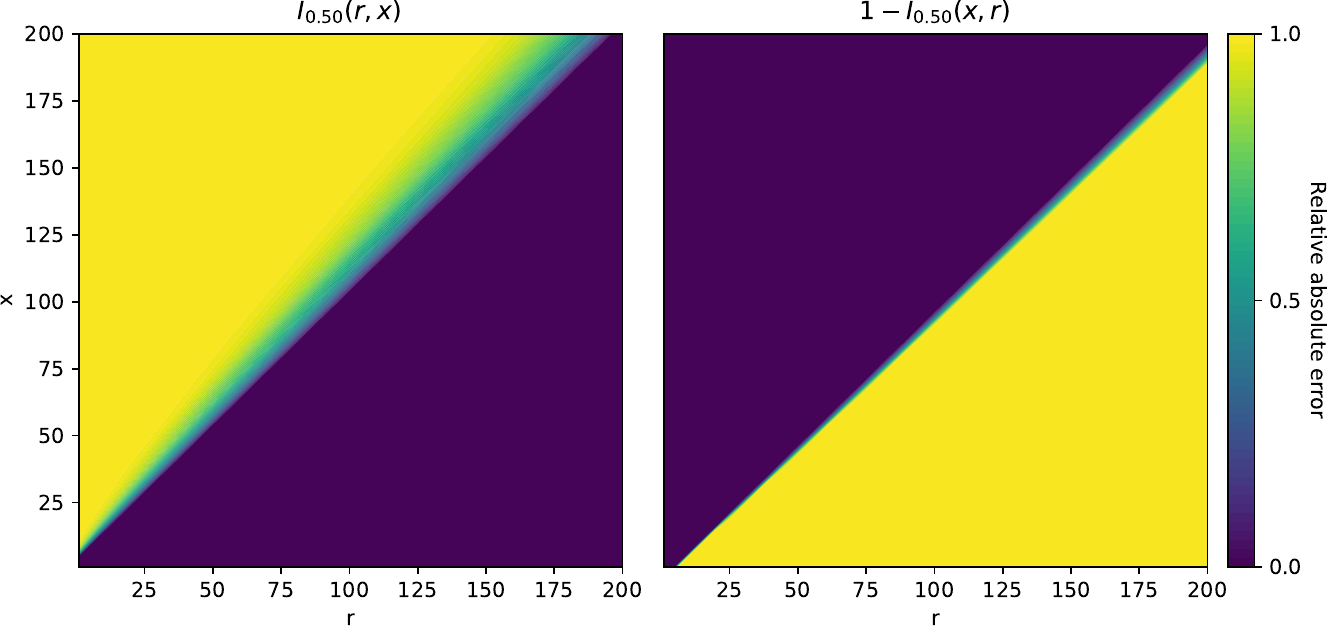}
				\end{adjustbox}
			\end{tabular}
			\caption{Relative error surfaces for the regularized incomplete beta function $I_p(x, r)$. \textbf{A}. Fixed $r = 100$, varying $x, p$. \textbf{B}. Fixed $p=\frac{1}{2}$, varying $x, r$. Animated versions of \textbf{A} and \textbf{B} with varying $x$ and $p$ respectively are available in the supplementary data.}
			\label{fig:error}
		\end{figure}
		
		Finally, we examined relative error surfaces between the proposed approximation (Equation~\ref{eq:cont_incomplete_beta}) and the ground-truth values and it turns out that the approximation systemically fails in half of the function argument space ($p, r, x$). However, it is possible to exploit the well-known property of the regularized incomplete beta function \citep{Abramowitz}
		\begin{equation}\label{eq:betainc_symmetry}
			I_p(x, r) = 1 - I_{1 - p}(r, x).
		\end{equation}
		The relative error surface of the right side of the Equation~\ref{eq:betainc_symmetry} $1 - I_{1 - p}(r, x)$ appears to ''complement'' the relative error surface of the left side, see Figure~\ref{fig:error}. Hence, to accurately (for our purposes) compute Equation~\ref{eq:cont_incomplete_beta}, we employ the following trick:
		\begin{equation}\label{eq:cdf_nb}
			G_{\mathpzc{NB}}(x|\theta) = \begin{cases}
				I_p(x + 1, r) & if~r \le \frac{1-p}{p} x\\
				1 - I_{1-p}(r, x + 1) & if~r > \frac{1-p}{p} x
			\end{cases}\, .
		\end{equation}
		The decision rule was established empirically by examining the error surfaces in a wide parameter range.
		\newpage 
		\section{\texorpdfstring{Computation of $G_{\mathpzc{BetaNB}}$}{Computation of BetaNB CDF}}\label{app:cdf_betanb}
		
		In case of beta-negative binomial distribution, CDF is
		$
		G_{\mathpzc{BetaNB}}(n| \theta) = \sum_{i=0}^n f_{\mathpzc{BetaNB}}(i| \theta) = \sum_{i=0}^n \frac{B(r + i, \kappa)}{B(r, \mu \kappa)} \frac{\Gamma(i + (1-\mu) \kappa)}{i! \Gamma((1-\mu) \kappa)} 
		$. It is possible to represent this definite sum via continuous functions to avoid the time of computations being dependent on $n$:
		\begin{equation}\label{eq:bnb_cdf}
			\begin{split}
				G_{\mathpzc{BetaNB}}(x|\theta) =  1-\frac{ B(r+\kappa  \mu ,\kappa(1 - \mu) + x +1) \, _3F_2(1,{r+ x +1}, {\kappa(1 - \mu) + x +1}; {x +2}, {r+\kappa + x +1};1)}{(x + 1) B(r, x + 1) B(\kappa  \mu ,\kappa (1 - \mu) )} = \\ = 1 - C \, _3F_2(1,\overbrace{r+ x +1}^{a_1}, \overbrace{\kappa(1 - \mu) + x +1}^{a_2}; \overbrace{x +2}^{b_1}, \overbrace{r+\kappa + x +1}^{b_2};1),
			\end{split}
		\end{equation}
		where ${}_3F_2$ is a generalized hypergeometric function, defined as
		${}_p F_q(a_1, a_2, \dots a_p; b_1, b_2 \dots b_q;  x)  = \sum_{i=0}^\infty \beta_i = \sum_{i=0}^\infty \left({\prod_{j=1}^{q} \frac{\Gamma(a_j + i)}{\Gamma(a_j)}}\right)/ \left({{\prod_{j=1}^{p} \frac{\Gamma(b_j + i)}{\Gamma(b_j)}}}\right) \frac{z^i}{i!}$. 
		Proving it is fairly straightforward by observing, first, that $\frac{b_{i+1}}{b_i} = \frac{\prod_{j=1}^p (a_j + i)}{(i + 1)\prod_{j=1}^q (b_j + i)} z$. Then, note that $G_{\mathpzc{BetaNB}}(x|\theta)$ can be represented as
		
		$$G_{\mathpzc{BetaNB}}(x|\theta) =  \sum_{i=0}^x f_{\mathpzc{BetaNB}}(i| \theta) = 1 -  \sum_{i=x + 1}^\infty f_{\mathpzc{BetaNB}}(i| \theta).$$
		By examining the latter sum, it becomes clear that it is a hypergeometric series:
		\begin{equation*}
			\begin{split}
				\sum_{i=x + 1}^\infty g_{\mathpzc{BetaNB}}(i| \theta) = \sum_{i=x + 1}^\infty \frac{B(r + i, \kappa)}{B(r, \mu \kappa)} \frac{\Gamma(i + (1-\mu) \kappa)}{i! \Gamma((1-\mu) \kappa)} = \\ = \frac{1}{B(r, \mu \kappa) \Gamma((1 - \mu)\kappa)}\sum_{i=0}^\infty \frac{B(r + x + i + 1, \kappa)\Gamma(x + i + (1-\mu) \kappa + 1)}{(i + x + 1)!} = \\ =
				\frac{1}{B(r, \mu \kappa) \Gamma((1 - \mu)\kappa)}\sum_{i=0}^\infty \frac{B(r + x + i + 1, \kappa)\Gamma(x + i + (1-\mu) \kappa + 1)}{\Gamma(i + x + 2)} = \\ =
				\frac{B(r + x + 1, \kappa) \Gamma(x + (1 - \mu) \kappa + 1)}{B(r, \mu \kappa) \Gamma((1 - \mu)\kappa) \Gamma(x + 2)}\sum_{i=0}^\infty \frac{B(r + x + i + 1, \kappa)\Gamma(x + i + (1-\mu) \kappa + 1) \Gamma(x + 2)}{B(r + x + 1, \kappa) \Gamma(x + (1 - \mu) \kappa + 1)\Gamma(i + x + 2)} =  \\ =
				\frac{B(r + x + 1, \kappa) \Gamma(x + (1 - \mu) \kappa + 1)}{B(r, \mu \kappa) \Gamma((1 - \mu)\kappa)  \Gamma(x + 2)}\sum_{i=0}^\infty \beta_i,
			\end{split}
		\end{equation*}
		where $\beta_i$ suffices $\beta_0 = 1$ and $\frac{\beta_{i + 1}}{\beta_i} = \frac{(i+r+x+1) (i+(1 - \mu)\kappa+x+1)}{(i+x+2) (i+\kappa+r+x+1)} * \frac{i + 1}{i + 1}$, hence $$\sum_{i=0}^\infty \beta_i = {}_3 F_2 (1,{r+ x +1}, {\kappa(1 - \mu) + x +1}; {x +2}, {r+\kappa + x +1};1)).$$
		We rearrange the multiplier in front of the sum by using the property $\Gamma(x + 2) = (x + 1)\Gamma(x + 1)$ and use the definition of the Beta function $B(a, b) = \frac{\Gamma(a) \Gamma(b)}{\Gamma(a + b)}$ :
		\begin{equation*}
			\begin{split}
				\frac{B(r + x + 1, \kappa) \Gamma(x + (1 - \mu) \kappa + 1)}{B(r, \mu \kappa) \Gamma((1 - \mu)\kappa)  \Gamma(x + 2)} = 
				\frac{B(r + x + 1, \kappa) \Gamma(x + (1 - \mu) \kappa + 1)}{B(r, \mu \kappa) \Gamma((1 - \mu)\kappa)  \Gamma(x + 1) (x + 1)} = \\ = \frac{1}{x + 1}  \frac{\Gamma(r + x + 1) \Gamma(\kappa)}{\Gamma(r + x + \kappa + 1)}  \frac{\Gamma(r + \mu \kappa)} {\Gamma(r) \Gamma(\mu \kappa)}  \frac{\Gamma(x + (1 - \mu) \kappa + 1)}{\Gamma((1 - \mu) \kappa) \Gamma(x + 1)} = \\
				=  \frac{1}{x + 1} \frac{\Gamma(r + \kappa \mu) \Gamma(x + (1 - \mu)\kappa + 1)}{\Gamma(r + x + \kappa + 1)} \frac{\Gamma(\kappa)}{\Gamma(\mu \kappa) \Gamma((1-\mu)\kappa)} \frac{\Gamma(r + x + 1)}{\Gamma(r) \Gamma(x + 1)} = \\ =
				\frac{1}{x + 1} \frac{B(r + \kappa \mu, x + (1 - \mu)\kappa + 1)}{B(\mu \kappa, (1-\mu) \kappa) B(r, x + 1)}
			\end{split}
		\end{equation*}
		and we finally obtain the Equation~\ref{eq:bnb_cdf}. Here, we rearranged the  gamma functions to obtain different beta functions to make sure that both arguments of beta functions have comparabler magnitude on average allowing for a greater numerical stability.
		
		The $_3 F_2$ term is problematic, as there is no \textbf{Python} package that can compute $_3 F_2$ using CPU-friendly arithmetic (e.g., \textbf{mpmath} uses arbitrary precision arithmetic to compute $_p F_q$ in a straightforward fashion using the definition, but that is seriously downgrading performance), let it alone to provide a differentiable function under the \textbf{JAX} framework.
		Fortunately, a continued fraction representation similar to that at Equation~\ref{eq:cont_incomplete_beta} exists \citep{hg_fractions}:
		
		\begin{equation}\label{eq:contf_hq} 
			_3 F_2(1, a_1, a_2; b_1, b_2; 1) \approx s + \mathbf{K}_{j=0}^{m} \frac{r(j - 1)}{q(j - 1)},~ r(n) = r_i\left(\frac{n - i}{3}\right), q(n) = q_i\left(\frac{n - i}{3}\right), i = n ~ mod ~ 3,\end{equation}
		
		where 
		
		\begin{center}
			\begin{equation*}
				\begin{tabular}{ll}
					$r_0(n) = -\frac{(n + b_1 - a_2 - 1)(n + b_2 - a_2 - 1)}{(2n - 1)(2n + a_2 - 1)}$& $r_1(n) = -\frac{(n + b_1 - 1)(n + b_2 - 1)}{2n (2n + a_1 - 1)} $ \\
					$r_0(0) = 1$, $r_1(0) = -1$ & $r_2(n) = -\frac{(n + b_1 - a_1)(n + b_2 - a_1)}{(2n + a_1)(2n + a_2)}
					$ \\ & \\
					\hline
					& \\
					$q_0(n) = \frac{(3n + b_1 - 1)(3n + b_2 - 1) - 2n (2n + a_2)}{2n (2n + a_1 - 1)}$ &  $q_1(n) = \frac{(3n + b_1) (3n + b_2) - (2n + 1) (2n + a_1)}{(2n + a_1) (2n + a_2)}$\\
					$q_0(n) = 1$ & $q_2(n) = \frac{(3n + b_1 + 1) (3n + b_2 + 1) - (2n + a_1 + 1) (2n + a_2 + 1)}{(2n + 1) (2n + a_2 + 1)}$  \\
					$s = 0$
				\end{tabular}
			\end{equation*}
		\end{center}
		
		Likewise, we use Lentz's method to efficiently estimate the continued fraction in Equation~\ref{eq:contf_hq}. However, a slight change of the algorithm is required: $C_1$ in the Equation~\ref{eq:lentz} blows up if $s = 0$ (the first term in a continued fraction representation), which is the case with $_3F_2$ here, therefore we replace $s$ with an arbitrarily chosen value of $10^{-30}$ that is subtracted from the result  thus caping accuracy of the approximation for low values of $_3F_2$. An alternative would be to use the algorithm induced by the Euler-Wallis equations, or the Steed's algorithm, but in practice,  those approaches often failed to converge, which is less favorable than reduced precision fat low values.
		
		There is one final step we have to consider. The algorithm in Equation~\ref{eq:contf_hq} for ${}_3 F_2$ exhibits similar problems as $I_p(r, x)$. Fortunately, it turns out that there is a property of $G_{\mathpzc{BetaNB}}$ that mimics the Equation~\ref{eq:betainc_symmetry}:
		
		\begin{equation}\label{eq:hyp_symmetry}
			G_{\mathpzc{BetaNB}}(x|r, \mu, \kappa) = 1 - G_{\mathpzc{BetaNB}}(r - 1|x + 1, 1 - \mu, \kappa).
		\end{equation}
		
\begin{figure}[h]
			\centering
			\begin{tabular}{ll}
				\raisebox{9.5em}{\Huge A} &
				\begin{adjustbox}{max width=0.9\textwidth}
					\includegraphics{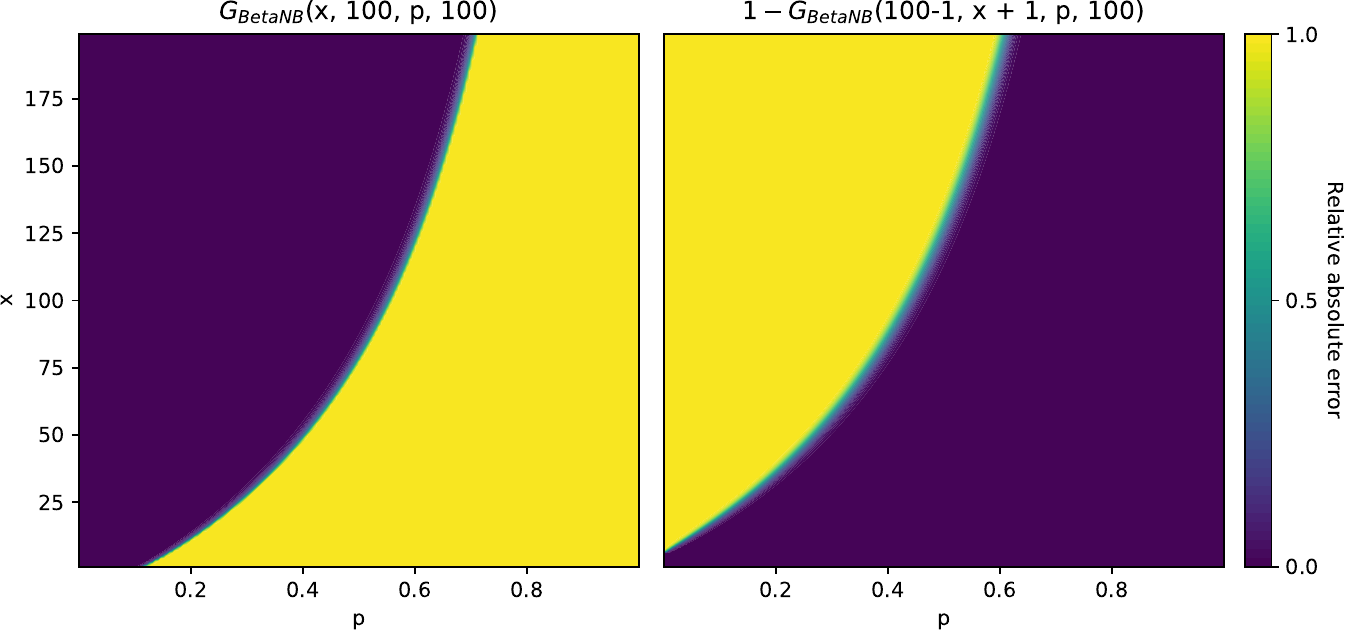}
				\end{adjustbox}\\
				\raisebox{9.5em}{\Huge B} &
				\begin{adjustbox}{max width=0.95\textwidth}
					\includegraphics{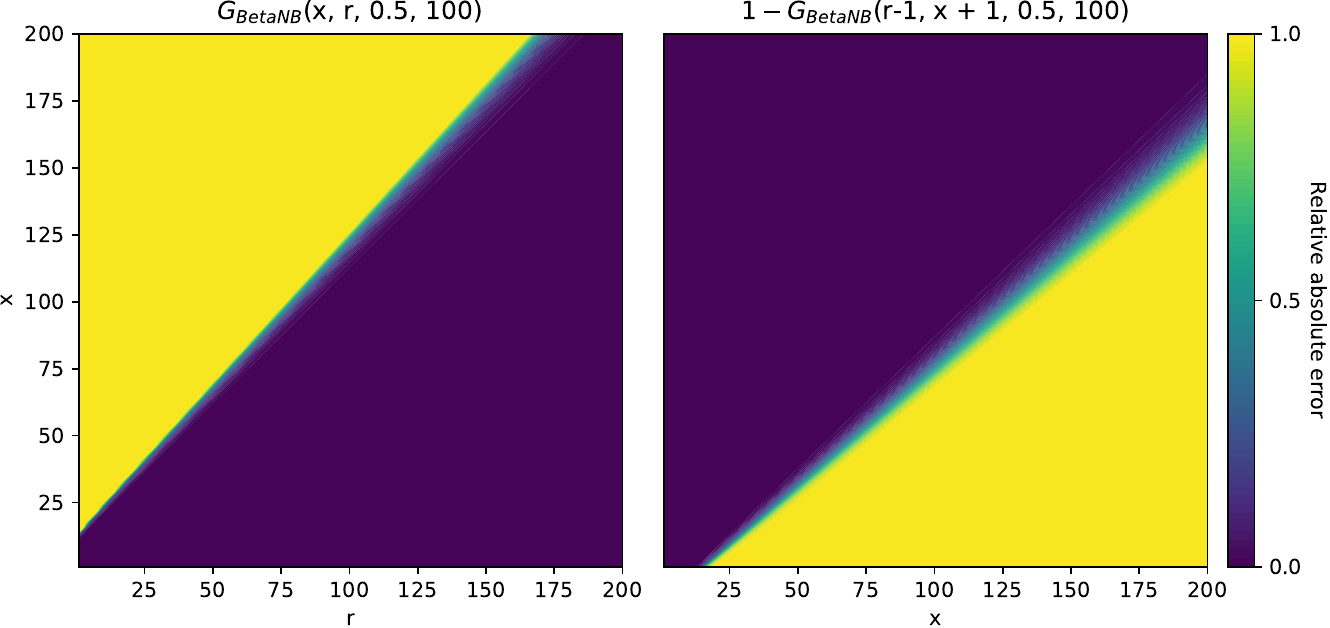}
				\end{adjustbox}
				\\
				\raisebox{9.5em}{\Huge C} &
				\begin{adjustbox}{max width=0.9\textwidth}
					\includegraphics{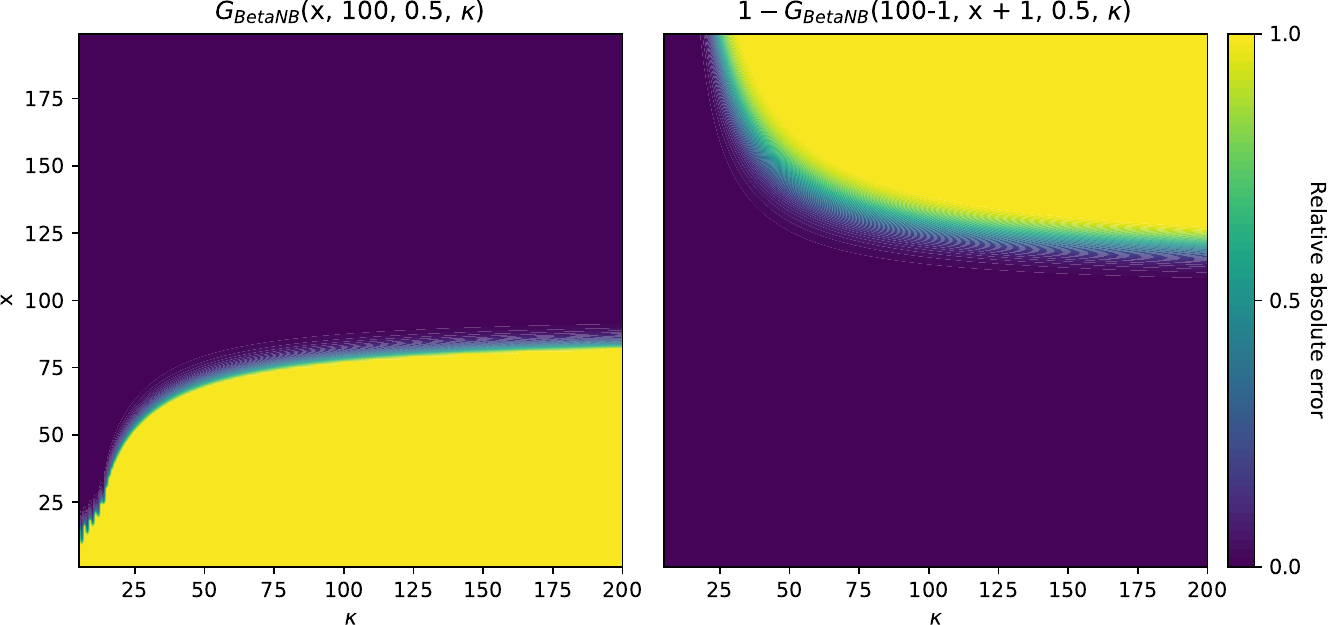}
				\end{adjustbox}
			\end{tabular}
			\caption{Relative error surfaces for the cumulative distribution functions of BetaNB $G_{\mathpzc{BNB}}$. \textbf{A}. Fixed $p = 100, \kappa = 50$, varying $r, p$. \textbf{B}. Fixed $p=\frac{1}{2}, \kappa = 50$, varying $x, r$. \textbf{C}. Fixed $r= 100,p = \frac{1}{2}$, varying $x, \kappa$. An animated version of \textbf{B} with the varying concentration parameter $\kappa$ is available in supplementary materials. In the animation, it should be evident that high-error zones converge on the diagonal, mimicking error zones of the regularized incomplete beta function.}
			\label{fig:error_hyp}
		\end{figure}

		To the best of our knowledge, the property was not described in literature as of today. It appeared to be non-trivial to prove analytically, but as shown in Figure~\ref{fig:error_hyp}, it holds numerically. There, higher values of the concentration parameter $\kappa$ increase an area of the ''high-error zone'', however it is mirrored by similar low-error zones of $I_{p}(x, r)$, hence a similar decision rule as in Equation~\ref{eq:cdf_nb} applicable:
		\begin{equation}
			G_{\mathpzc{BetaNB}}(x|\theta) = \begin{cases}
				1-\frac{ B(r+\kappa  \mu ,\kappa(1 - \mu) + x +1) \, _3F_2(1,{r+ x +1}, {\kappa(1 - \mu) + x +1}; {x +2}, {r+\kappa + x +1};1)}{(x + 1) B(r, x + 1) B(\kappa  \mu ,\kappa (1 - \mu) )} & if~r \le \frac{1-p}{p} x\\
				\frac{ B(x + \kappa  \mu + 1,\kappa(1 - \mu) + r) \, _3F_2(1,{r + x +1}, {\kappa(1 - \mu) + r}; {r + 1}, {r + \kappa + x + 1};1)}{r B(r, x + 1) B(\kappa  \mu ,\kappa (1 - \mu) )} & if~r > \frac{1-p}{p} x
			\end{cases}\,
		\end{equation}
		\afterpage{\clearpage}
		\newpage
		\section{Syntehetic dataset generator}\label{app:generator}
		\subsection*{General sampling scheme}
		We assume data $(x, y)$ to be sampled from either Negative multinomial distribution $\mathpzc{NM}(r_0, p_0, p_1, p_2)$ (where $p_0 + p_1 + p_2 = 1$) as outlined in Appendix~\ref{app:nb_origins} or Dirichlet negative multinomial distribution $\mathpzc{DMN}(r_0, \alpha_0, \alpha_1, \alpha_2)$ if we seek to introduce noise to $p_k$ probabilities. In either case, sampling is done by first sampling $x$ from the marginal $P(X)$ and then sampling $y$ from the conditional distribution $P(Y|X)$. We've derived them previously for $\mathcal{NM}$ in Appendix~\ref{app:nb_origins}, see Equation~\ref{eq:marginals} and Equation~\ref{eq:conditionals}, and it is easy to derive them for $\mathcal{DNM}$.
		
		As we deal with data that is at least zero-truncated, multivariate multinomial distributions have to be truncated as well. It is achieved with rejection sampling (note, that marginal of a truncated multivariate distribution is not just a truncated marginal and the sampling is non-trivial: $P(X|Y > left) \neq P(X)$).
		\begin{algorithm}
			\caption{Sampling an SNP}
			\begin{algorithmic}
				\Function{SampleSNP}{$r_0, p_0, p, \kappa, num\_samples, left, r\_noise$} \Comment{$p$ is a success probability for the reference allele, $p_0$ controls coverage/sequence length, $\kappa$ is a concentration parameter, $r\_noise$ is mean of an exponential noise to be added to $r_0$ }
				\Require $left > 0$
				\If{$r\_noise \ge 0$}
				\State $r_0 \gets \mathpzc{Gamma}(\alpha=\frac{r_0^2}{r\_noise}, \beta=\frac{r_0}{r\_noise})$
				\EndIf
				\State $p_m \gets \text{\texttt{getMarginalP}}(p_0, p, \kappa)$ \Comment{$p_m$ and $p_c$ are transformed success probabilities}
				\State $p_c \gets \text{\texttt{getConditionalP}}(p_0, p, \kappa)$
				\State $X \gets \text{\texttt{zeros}}(num\_samples)$
				\State $Y \gets \text{\texttt{zeros}}(num\_samples)$
				\While{$\text{\texttt{sum}}(ind \gets (X \le left) ~|| ~(Y \le left))$}
				\State $X[ind] \gets \mathpzc{NB}(r_0, p_m)$ \Comment{$P(X)$}
				\State $Y[ind] \gets \mathpzc{NB}(r_0 + X[ind], p_c)$ \Comment{$P(Y|X)$}
				\EndWhile
				\State \Return $X, Y$
				\EndFunction
			\end{algorithmic}
		\end{algorithm}
		Functions \texttt{getMarginalP}, \texttt{getConditionalP}.
		\subsection*{Negative multinomial distribution}
		For $\mathcal{NM}$, we use parameters as inferred in Equation~\ref{eq:marginals} and Equation~\ref{eq:conditionals}:
		\begin{equation}\label{eq:sample_pair}
			x \sim \mathpzc{NB}\left(r_0, \frac{p_0 p}{1 - (1 - p)p_0}\right),~y|_{x} \sim \mathpzc{NB}(x + r_0, p_0 (1 - p)).
		\end{equation}
		Hence,
		\begin{algorithm}
			\caption{Getting success probabilities for $\mathpzc{NM}$}
			\begin{algorithmic}
				\Function{getMarginalP}{$p_0, p, \kappa$} 
				\State $p_m \gets  \frac{p_0 p}{1 - (1 - p)p_0}$
				\State \Return $p_m$
				\EndFunction
				
				\Function{getConditionalP}{$p_0, p, \kappa, X$} 
				\State $p_c \gets p_0 (1 - p)$
				\State \Return $p_c$
				\EndFunction
			\end{algorithmic}
		\end{algorithm}

		\subsection*{Dirichlet negative multinomial distribution}
		If $$(x, y) \sim \mathpzc{DMN}(r_0, \alpha_0, \alpha_1, \alpha_2),$$
		then marginal and conditional distributions are 
		\begin{equation}\label{eq:sample_pair_d}
		x \sim \mathpzc{BetaNB}(r_0, \alpha_1, \alpha_0), ~y|_x \sim \mathpzc{BetaNB}(r_0 + x, \alpha_2, \alpha_0 + \alpha_1)
		\end{equation}
		We choose such values of $\alpha_0, \alpha_1, \alpha_2$ that means of $x, y|_x$ from Equation~\ref{eq:sample_pair} and Equation~\ref{eq:sample_pair_d} agree:
		\begin{equation} \label{eq:dirichlet_alphas}
			\begin{cases*}
				\alpha_0 = \frac{1 - p_0}{p_0} \kappa \\
				\alpha_1 = p \kappa \\
				\alpha_2 = (1-p) \kappa
			\end{cases*}.
		\end{equation}
		Remember that $x \sim \mathpzc{BetaN}(r_0, \alpha, \beta)$ is same as $p \sim \mathpzc{Beta}(\alpha, \beta),~x \sim \mathpzc{NB}(r_0, p)$, therefore we can sample some $p_m, p_c$ from $\mathpzc{Beta}$ inside of \texttt{getMarginalP} and \texttt{getConditionalP} and still retain sampling from $\mathpzc{NB}$ insode of the \texttt{while} loop.
		\begin{algorithm}
			\caption{Getting success probabilities for $\mathpzc{DNM}$}
			\begin{algorithmic}
				\Function{getMarginalP}{$p_0, p, \kappa$} 
				\State $\alpha_0 \gets \frac{1 - p_0}{p_0} \kappa$
				\State $\alpha_1 \gets p \kappa $
				\State $p_m \gets  \mathpzc{TruncatedBeta}(\alpha_1, \alpha_0)$
				\State \Return $p_m$
				\EndFunction
				
				\Function{getConditionalP}{$p_0, p, \kappa, X$} 
				\State $\alpha_0 \gets \frac{1 - p_0}{p_0} \kappa$
				\State $\alpha_1 \gets p \kappa $
				\State $\alpha_2 \gets (1-p) \kappa$
				\State $p_c \gets \mathpzc{TruncatedBeta}(\alpha_2, \alpha_0 + \alpha_1)$
				\State \Return $p_c$
				\EndFunction
			\end{algorithmic}
		\end{algorithm}\\
		
		Sampling from plain $\mathpzc{Beta}$ leads to occasional extremely high (e.g. $0.9999999$) or low values, which, in turn, might result in rejection sampling taking forever, while also producing $(x, y)$ pairs of unrealistically high coverage. Therefore, sampling from a truncated version of $\mathpzc{Beta}$ might be appealing. However, if we were to select even seemingly liberal bounds such as $10^{-5}$ and $1-10^{-5}$, this introduces bias into the model. More specifically, it results in a mean of sampled beta to deviate from the assumption that means of $x, y|_x$ from Equation~\ref{eq:sample_pair} and Equation~\ref{eq:sample_pair_d} should agree, which we hoped to impose by Equation~\ref{eq:dirichlet_alphas}. To counteract this issue, we first select an arbitrary right truncation bound (we use $1-10^{-5}$ by default) and then we estimate the left truncation bound by minimizing the difference between means of truncated and non-truncated variables. That's it, the minimized function is
		\begin{equation}
			f(left|right, \alpha, \beta) = \left(\frac{\alpha}{\alpha + \beta} - \frac{\widehat{I}_{left}(1 + \alpha, \beta) - \widehat{I}_{right}(1 + \alpha, \beta)}{\widehat{I}_{left}(\alpha, \beta) - \widehat{I}_{right}(\alpha, \beta)}\right)^2,
		\end{equation} 
		where $\widehat{I}_{z}(\alpha, \beta)$ is an incomplete beta function that can be computed from the regularized incomplete beta function $I_{z}(\alpha, \beta)$ from Appendix~\ref{app:nb_cdf}: $\widehat{I}_{z}(\alpha, \beta) = 	B(\alpha, \beta ) {I}_{z}(\alpha, \beta)$.
		
		\subsection*{Introducing allele-specificity and biases}
		Both bias and allele-specificity are introduced by altering the $p$ probability. In the case of AS SNPs, changes in $p$ are introduced by solving the logit equation $logit(p_{new}) - logit(p) = \gamma$ for $p_{new}$ (similar formula used to computed effect-sizes in Section~\ref{sect:es}), where $\gamma$ is an effect size and the logit function uses base-2:
		$$p_{new} = \frac{p}{p + (1 - p) 2^{-\gamma}}.$$
		For example,for a dataset with with $BAD = 1$ and no reference bias, plugging in $\gamma = 1$ in the equation above in the $p_{new} = \frac{2}{3}$ instead of $\frac{1}{2}$ for AS SNPs.
		
		As for the reference bias, we use slightly different formula: we solve the equation $\frac{\EX[x]}{\EX[y]} = \gamma$ for $p$, which is just $$p_{new} = \frac{ \gamma * BAD}{1 + \gamma * BAD}.$$
		
		\subsection*{Sampling for a fixed coverage}
		This is easier as we have to sample only $x|_{n}$ and $y$ then is given as $n-x$. For $\mathpzc{NM}$, $x|_{n} \sim \mathpzc{Binom}(n, p)$ and for $\mathpzc{DNM}$ $x|_{n} \sim \mathpzc{BetaBinom}(n, \alpha_1, \alpha_2)$, where $\alpha_1$ and $\alpha_2$ are given from the Equation~\ref{eq:dirichlet_alphas}.
		
		\newpage 
		
		\section{Simulation studies tabulars}\label{app:benchmark_tables}
	
		\subsubsection*{Group A}
\begin{table}[H]
\centering
\begin{adjustbox}{max width=\linewidth}
\begin{tabular}{lllllllll}
\toprule
 & NB & NB\_p & BetaNB & BetaNB\_p & MCNB & MCNB\_p & binom & binom\_beta \\
$n\_samples$ &  &  &  &  &  &  &  &  \\
\midrule
1 & 0.3753 $\pm$ 0.0154 & 0.3750 $\pm$ 0.0154 & 0.3757 $\pm$ 0.0188 & 0.3769 $\pm$ 0.0159 & 0.3444 $\pm$ 0.0447 & 0.3195 $\pm$ 0.0144 & 0.4845 $\pm$ 0.0195 & 0.5130 $\pm$ 0.0212 \\
2 & 0.5120 $\pm$ 0.0130 & 0.5135 $\pm$ 0.0129 & 0.5042 $\pm$ 0.0441 & 0.5136 $\pm$ 0.0125 & 0.4602 $\pm$ 0.0245 & 0.4728 $\pm$ 0.0152 & 0.5829 $\pm$ 0.0142 & 0.5818 $\pm$ 0.0134 \\
4 & 0.7179 $\pm$ 0.0100 & 0.7199 $\pm$ 0.0100 & 0.7137 $\pm$ 0.0121 & 0.7156 $\pm$ 0.0103 & 0.6941 $\pm$ 0.0120 & 0.7007 $\pm$ 0.0104 & 0.7475 $\pm$ 0.0092 & 0.7369 $\pm$ 0.0087 \\
8 & 0.9107 $\pm$ 0.0063 & 0.9118 $\pm$ 0.0062 & 0.9045 $\pm$ 0.0065 & 0.9048 $\pm$ 0.0065 & 0.9063 $\pm$ 0.0074 & 0.9071 $\pm$ 0.0069 & 0.9162 $\pm$ 0.0062 & 0.9049 $\pm$ 0.0064 \\
16 & 0.9904 $\pm$ 0.0019 & 0.9905 $\pm$ 0.0019 & 0.9877 $\pm$ 0.0021 & 0.9876 $\pm$ 0.0021 & 0.9907 $\pm$ 0.0017 & 0.9898 $\pm$ 0.0031 & 0.9903 $\pm$ 0.0020 & 0.9852 $\pm$ 0.0021 \\
25 & 0.9992 $\pm$ 0.0004 & 0.9992 $\pm$ 0.0004 & 0.9986 $\pm$ 0.0006 & 0.9986 $\pm$ 0.0006 & 0.9993 $\pm$ 0.0003 & 0.9992 $\pm$ 0.0004 & 0.9992 $\pm$ 0.0004 & 0.9979 $\pm$ 0.0009 \\
32 & 0.9999 $\pm$ 0.0001 & 0.9999 $\pm$ 0.0001 & 0.9998 $\pm$ 0.0002 & 0.9998 $\pm$ 0.0002 & 0.9999 $\pm$ 0.0001 & 0.9999 $\pm$ 0.0001 & 0.9999 $\pm$ 0.0001 & 0.9995 $\pm$ 0.0004 \\
48 & 1.0000 $\pm$ 0.0000 & 1.0000 $\pm$ 0.0000 & 1.0000 $\pm$ 0.0000 & 1.0000 $\pm$ 0.0000 & 1.0000 $\pm$ 0.0000 & 1.0000 $\pm$ 0.0000 & 1.0000 $\pm$ 0.0000 & 1.0000 $\pm$ 0.0000 \\
64 & 1.0000 $\pm$ 0.0000 & 1.0000 $\pm$ 0.0000 & 1.0000 $\pm$ 0.0000 & 1.0000 $\pm$ 0.0000 & 1.0000 $\pm$ 0.0000 & 1.0000 $\pm$ 0.0000 & 1.0000 $\pm$ 0.0000 & 1.0000 $\pm$ 0.0000 \\
\bottomrule
\end{tabular}

\end{adjustbox}
\caption{Group A, PR AUC, values after $\pm$ are standard deviations.}
\end{table}

\begin{table}[H]
\centering
\begin{adjustbox}{max width=\linewidth}
\begin{tabular}{lllllllll}
\toprule
 & BetaNB & BetaNB\_p & MCNB & MCNB\_p & NB & NB\_p & binom & binom\_beta \\
$n\_samples$ &  &  &  &  &  &  &  &  \\
\midrule
1 & 0.0160 $\pm$ 0.0084 & 0.0158 $\pm$ 0.0087 & 0.0040 $\pm$ 0.0020 & 0.0084 $\pm$ 0.0053 & 0.0196 $\pm$ 0.0082 & 0.0202 $\pm$ 0.0081 & 0.0124 $\pm$ 0.0069 & 0.0012 $\pm$ 0.0017 \\
2 & 0.0403 $\pm$ 0.0115 & 0.0410 $\pm$ 0.0134 & 0.0143 $\pm$ 0.0066 & 0.0312 $\pm$ 0.0112 & 0.0591 $\pm$ 0.0132 & 0.0600 $\pm$ 0.0134 & 0.0359 $\pm$ 0.0117 & 0.0057 $\pm$ 0.0046 \\
4 & 0.1854 $\pm$ 0.0187 & 0.1990 $\pm$ 0.0270 & 0.1324 $\pm$ 0.0166 & 0.1861 $\pm$ 0.0205 & 0.2536 $\pm$ 0.0185 & 0.2503 $\pm$ 0.0192 & 0.1859 $\pm$ 0.0204 & 0.0962 $\pm$ 0.0139 \\
8 & 0.5380 $\pm$ 0.0169 & 0.5554 $\pm$ 0.0299 & 0.5509 $\pm$ 0.0178 & 0.5896 $\pm$ 0.0193 & 0.6534 $\pm$ 0.0172 & 0.6462 $\pm$ 0.0182 & 0.5725 $\pm$ 0.0180 & 0.4209 $\pm$ 0.0182 \\
16 & 0.8800 $\pm$ 0.0104 & 0.8839 $\pm$ 0.0150 & 0.9239 $\pm$ 0.0117 & 0.9284 $\pm$ 0.0106 & 0.9454 $\pm$ 0.0093 & 0.9421 $\pm$ 0.0091 & 0.9180 $\pm$ 0.0106 & 0.8062 $\pm$ 0.0135 \\
25 & 0.9744 $\pm$ 0.0042 & 0.9747 $\pm$ 0.0045 & 0.9921 $\pm$ 0.0026 & 0.9922 $\pm$ 0.0025 & 0.9948 $\pm$ 0.0016 & 0.9937 $\pm$ 0.0019 & 0.9904 $\pm$ 0.0023 & 0.9473 $\pm$ 0.0050 \\
32 & 0.9931 $\pm$ 0.0021 & 0.9932 $\pm$ 0.0021 & 0.9990 $\pm$ 0.0008 & 0.9989 $\pm$ 0.0006 & 0.9992 $\pm$ 0.0007 & 0.9991 $\pm$ 0.0007 & 0.9982 $\pm$ 0.0010 & 0.9821 $\pm$ 0.0049 \\
48 & 0.9999 $\pm$ 0.0004 & 0.9999 $\pm$ 0.0003 & 1.0000 $\pm$ 0.0000 & 1.0000 $\pm$ 0.0000 & 1.0000 $\pm$ 0.0000 & 1.0000 $\pm$ 0.0000 & 1.0000 $\pm$ 0.0000 & 0.9988 $\pm$ 0.0011 \\
64 & 1.0000 $\pm$ 0.0000 & 1.0000 $\pm$ 0.0000 & 1.0000 $\pm$ 0.0000 & 1.0000 $\pm$ 0.0000 & 1.0000 $\pm$ 0.0000 & 1.0000 $\pm$ 0.0000 & 1.0000 $\pm$ 0.0000 & 1.0000 $\pm$ 0.0000 \\
\bottomrule
\end{tabular}

\end{adjustbox}
\caption{Group A, Sensitivity, values after $\pm$ are standard deviations.}
\end{table}

\begin{table}[H]
\centering
\begin{adjustbox}{max width=\linewidth}
\begin{tabular}{lllllllll}
\toprule
 & BetaNB & BetaNB\_p & MCNB & MCNB\_p & NB & NB\_p & binom & binom\_beta \\
$n\_samples$ &  &  &  &  &  &  &  &  \\
\midrule
1 & 1.0000 $\pm$ 0.0000 & 1.0000 $\pm$ 0.0001 & 1.0000 $\pm$ 0.0000 & 1.0000 $\pm$ 0.0000 & 1.0000 $\pm$ 0.0001 & 1.0000 $\pm$ 0.0001 & 1.0000 $\pm$ 0.0000 & 1.0000 $\pm$ 0.0000 \\
2 & 0.9999 $\pm$ 0.0001 & 0.9999 $\pm$ 0.0001 & 1.0000 $\pm$ 0.0000 & 0.9999 $\pm$ 0.0001 & 0.9999 $\pm$ 0.0001 & 0.9999 $\pm$ 0.0001 & 1.0000 $\pm$ 0.0000 & 1.0000 $\pm$ 0.0000 \\
4 & 0.9998 $\pm$ 0.0002 & 0.9997 $\pm$ 0.0002 & 0.9999 $\pm$ 0.0001 & 0.9997 $\pm$ 0.0002 & 0.9994 $\pm$ 0.0002 & 0.9994 $\pm$ 0.0002 & 0.9999 $\pm$ 0.0001 & 1.0000 $\pm$ 0.0001 \\
8 & 0.9994 $\pm$ 0.0002 & 0.9993 $\pm$ 0.0003 & 0.9994 $\pm$ 0.0002 & 0.9989 $\pm$ 0.0003 & 0.9983 $\pm$ 0.0004 & 0.9985 $\pm$ 0.0003 & 0.9995 $\pm$ 0.0002 & 0.9999 $\pm$ 0.0001 \\
16 & 0.9995 $\pm$ 0.0003 & 0.9994 $\pm$ 0.0003 & 0.9988 $\pm$ 0.0003 & 0.9983 $\pm$ 0.0012 & 0.9974 $\pm$ 0.0006 & 0.9979 $\pm$ 0.0004 & 0.9991 $\pm$ 0.0003 & 0.9998 $\pm$ 0.0001 \\
25 & 0.9997 $\pm$ 0.0002 & 0.9996 $\pm$ 0.0003 & 0.9987 $\pm$ 0.0003 & 0.9986 $\pm$ 0.0004 & 0.9972 $\pm$ 0.0006 & 0.9979 $\pm$ 0.0005 & 0.9990 $\pm$ 0.0003 & 0.9999 $\pm$ 0.0001 \\
32 & 0.9997 $\pm$ 0.0002 & 0.9997 $\pm$ 0.0002 & 0.9987 $\pm$ 0.0003 & 0.9987 $\pm$ 0.0004 & 0.9971 $\pm$ 0.0006 & 0.9979 $\pm$ 0.0004 & 0.9990 $\pm$ 0.0003 & 0.9999 $\pm$ 0.0001 \\
48 & 0.9998 $\pm$ 0.0001 & 0.9998 $\pm$ 0.0001 & 0.9987 $\pm$ 0.0003 & 0.9988 $\pm$ 0.0005 & 0.9970 $\pm$ 0.0006 & 0.9980 $\pm$ 0.0004 & 0.9990 $\pm$ 0.0003 & 0.9999 $\pm$ 0.0001 \\
64 & 0.9999 $\pm$ 0.0001 & 0.9999 $\pm$ 0.0001 & 0.9987 $\pm$ 0.0003 & 0.9989 $\pm$ 0.0005 & 0.9969 $\pm$ 0.0006 & 0.9981 $\pm$ 0.0004 & 0.9990 $\pm$ 0.0003 & 1.0000 $\pm$ 0.0001 \\
\bottomrule
\end{tabular}

\end{adjustbox}
\caption{Group A, Specificity, values after $\pm$ are standard deviations.}
\end{table}
\subsubsection*{Group B}
\begin{table}[H]
\centering
\begin{adjustbox}{max width=\linewidth}
\begin{tabular}{lllllllll}
\toprule
 & NB & NB\_p & BetaNB & BetaNB\_p & MCNB & MCNB\_p & binom & binom\_beta \\
$ase\_es$ &  &  &  &  &  &  &  &  \\
\midrule
0.1 & 0.2419 $\pm$ 0.0278 & 0.2402 $\pm$ 0.0284 & 0.2418 $\pm$ 0.0296 & 0.2413 $\pm$ 0.0296 & 0.4175 $\pm$ 0.1391 & 0.2036 $\pm$ 0.0348 & 0.4444 $\pm$ 0.1308 & 0.5334 $\pm$ 0.0545 \\
0.2 & 0.5337 $\pm$ 0.0135 & 0.5345 $\pm$ 0.0136 & 0.5292 $\pm$ 0.0129 & 0.5287 $\pm$ 0.0133 & 0.5086 $\pm$ 0.0170 & 0.5140 $\pm$ 0.0163 & 0.5518 $\pm$ 0.0117 & 0.5453 $\pm$ 0.0141 \\
0.3 & 0.8218 $\pm$ 0.0080 & 0.8226 $\pm$ 0.0079 & 0.8118 $\pm$ 0.0085 & 0.8138 $\pm$ 0.0078 & 0.8117 $\pm$ 0.0080 & 0.8144 $\pm$ 0.0076 & 0.8215 $\pm$ 0.0076 & 0.8044 $\pm$ 0.0083 \\
0.4 & 0.9482 $\pm$ 0.0048 & 0.9490 $\pm$ 0.0048 & 0.9422 $\pm$ 0.0051 & 0.9432 $\pm$ 0.0055 & 0.9466 $\pm$ 0.0046 & 0.9463 $\pm$ 0.0042 & 0.9502 $\pm$ 0.0048 & 0.9391 $\pm$ 0.0054 \\
0.5 & 0.9873 $\pm$ 0.0019 & 0.9877 $\pm$ 0.0019 & 0.9854 $\pm$ 0.0023 & 0.9853 $\pm$ 0.0023 & 0.9879 $\pm$ 0.0017 & 0.9875 $\pm$ 0.0018 & 0.9891 $\pm$ 0.0018 & 0.9845 $\pm$ 0.0025 \\
0.6 & 0.9968 $\pm$ 0.0008 & 0.9969 $\pm$ 0.0007 & 0.9959 $\pm$ 0.0012 & 0.9962 $\pm$ 0.0008 & 0.9972 $\pm$ 0.0006 & 0.9971 $\pm$ 0.0006 & 0.9975 $\pm$ 0.0007 & 0.9961 $\pm$ 0.0008 \\
\bottomrule
\end{tabular}

\end{adjustbox}
\caption{Group B, PR AUC, values after $\pm$ are standard deviations.}
\end{table}

\begin{table}[H]
\centering
\begin{adjustbox}{max width=\linewidth}
\begin{tabular}{lllllllll}
\toprule
 & BetaNB & BetaNB\_p & MCNB & MCNB\_p & NB & NB\_p & binom & binom\_beta \\
$ase\_es$ &  &  &  &  &  &  &  &  \\
\midrule
0.1 & 0.0004 $\pm$ 0.0006 & 0.0004 $\pm$ 0.0006 & 0.0001 $\pm$ 0.0002 & 0.0003 $\pm$ 0.0006 & 0.0004 $\pm$ 0.0006 & 0.0005 $\pm$ 0.0006 & 0.0001 $\pm$ 0.0005 & 0.0001 $\pm$ 0.0002 \\
0.2 & 0.0227 $\pm$ 0.0090 & 0.0234 $\pm$ 0.0100 & 0.0068 $\pm$ 0.0045 & 0.0241 $\pm$ 0.0088 & 0.0394 $\pm$ 0.0101 & 0.0394 $\pm$ 0.0097 & 0.0118 $\pm$ 0.0061 & 0.0014 $\pm$ 0.0015 \\
0.3 & 0.3002 $\pm$ 0.0219 & 0.3261 $\pm$ 0.0300 & 0.2740 $\pm$ 0.0225 & 0.3381 $\pm$ 0.0198 & 0.4025 $\pm$ 0.0187 & 0.3961 $\pm$ 0.0190 & 0.2874 $\pm$ 0.0218 & 0.1699 $\pm$ 0.0245 \\
0.4 & 0.6525 $\pm$ 0.0200 & 0.6691 $\pm$ 0.0401 & 0.6903 $\pm$ 0.0185 & 0.7173 $\pm$ 0.0176 & 0.7704 $\pm$ 0.0177 & 0.7627 $\pm$ 0.0171 & 0.6978 $\pm$ 0.0207 & 0.5408 $\pm$ 0.0184 \\
0.5 & 0.8611 $\pm$ 0.0108 & 0.8648 $\pm$ 0.0165 & 0.9029 $\pm$ 0.0094 & 0.9093 $\pm$ 0.0083 & 0.9314 $\pm$ 0.0068 & 0.9269 $\pm$ 0.0068 & 0.9090 $\pm$ 0.0082 & 0.8071 $\pm$ 0.0127 \\
0.6 & 0.9457 $\pm$ 0.0075 & 0.9470 $\pm$ 0.0066 & 0.9746 $\pm$ 0.0047 & 0.9752 $\pm$ 0.0049 & 0.9820 $\pm$ 0.0039 & 0.9801 $\pm$ 0.0041 & 0.9770 $\pm$ 0.0045 & 0.9267 $\pm$ 0.0069 \\
\bottomrule
\end{tabular}

\end{adjustbox}
\caption{Group B, Sensitivity, values after $\pm$ are standard deviations.}
\end{table}

\begin{table}[H]
\centering
\begin{adjustbox}{max width=\linewidth}
\begin{tabular}{lllllllll}
\toprule
 & BetaNB & BetaNB\_p & MCNB & MCNB\_p & NB & NB\_p & binom & binom\_beta \\
$ase\_es$ &  &  &  &  &  &  &  &  \\
\midrule
0.1 & 1.0000 $\pm$ 0.0000 & 1.0000 $\pm$ 0.0000 & 1.0000 $\pm$ 0.0000 & 1.0000 $\pm$ 0.0000 & 1.0000 $\pm$ 0.0000 & 1.0000 $\pm$ 0.0000 & 1.0000 $\pm$ 0.0000 & 1.0000 $\pm$ 0.0000 \\
0.2 & 1.0000 $\pm$ 0.0001 & 1.0000 $\pm$ 0.0001 & 1.0000 $\pm$ 0.0000 & 0.9999 $\pm$ 0.0001 & 0.9999 $\pm$ 0.0001 & 0.9999 $\pm$ 0.0001 & 1.0000 $\pm$ 0.0000 & 1.0000 $\pm$ 0.0000 \\
0.3 & 0.9996 $\pm$ 0.0002 & 0.9995 $\pm$ 0.0003 & 0.9997 $\pm$ 0.0002 & 0.9994 $\pm$ 0.0003 & 0.9990 $\pm$ 0.0003 & 0.9991 $\pm$ 0.0003 & 0.9998 $\pm$ 0.0002 & 0.9999 $\pm$ 0.0001 \\
0.4 & 0.9994 $\pm$ 0.0003 & 0.9992 $\pm$ 0.0005 & 0.9991 $\pm$ 0.0003 & 0.9987 $\pm$ 0.0004 & 0.9979 $\pm$ 0.0004 & 0.9982 $\pm$ 0.0004 & 0.9993 $\pm$ 0.0003 & 0.9998 $\pm$ 0.0001 \\
0.5 & 0.9995 $\pm$ 0.0003 & 0.9994 $\pm$ 0.0003 & 0.9989 $\pm$ 0.0003 & 0.9985 $\pm$ 0.0005 & 0.9974 $\pm$ 0.0006 & 0.9979 $\pm$ 0.0005 & 0.9991 $\pm$ 0.0003 & 0.9998 $\pm$ 0.0001 \\
0.6 & 0.9997 $\pm$ 0.0003 & 0.9997 $\pm$ 0.0002 & 0.9988 $\pm$ 0.0004 & 0.9987 $\pm$ 0.0003 & 0.9972 $\pm$ 0.0006 & 0.9979 $\pm$ 0.0005 & 0.9991 $\pm$ 0.0003 & 0.9999 $\pm$ 0.0001 \\
\bottomrule
\end{tabular}

\end{adjustbox}
\caption{Group B, Specificity, values after $\pm$ are standard deviations.}
\end{table}
\subsubsection*{Group C}
\begin{table}[H]
\centering
\begin{adjustbox}{max width=\linewidth}
\begin{tabular}{lllllllll}
\toprule
 & NB & NB\_p & BetaNB & BetaNB\_p & MCNB & MCNB\_p & binom & binom\_beta \\
$bias$ &  &  &  &  &  &  &  &  \\
\midrule
1.0625 & 0.9470 $\pm$ 0.0051 & 0.9484 $\pm$ 0.0048 & 0.9419 $\pm$ 0.0049 & 0.9419 $\pm$ 0.0051 & 0.9445 $\pm$ 0.0053 & 0.9424 $\pm$ 0.0087 & 0.9500 $\pm$ 0.0050 & 0.9386 $\pm$ 0.0057 \\
1.1250 & 0.9457 $\pm$ 0.0043 & 0.9484 $\pm$ 0.0040 & 0.9417 $\pm$ 0.0047 & 0.9421 $\pm$ 0.0045 & 0.9426 $\pm$ 0.0077 & 0.9456 $\pm$ 0.0040 & 0.9499 $\pm$ 0.0041 & 0.9381 $\pm$ 0.0053 \\
1.1875 & 0.9428 $\pm$ 0.0044 & 0.9472 $\pm$ 0.0037 & 0.9405 $\pm$ 0.0045 & 0.9412 $\pm$ 0.0053 & 0.9415 $\pm$ 0.0044 & 0.9430 $\pm$ 0.0073 & 0.9488 $\pm$ 0.0038 & 0.9372 $\pm$ 0.0048 \\
1.2500 & 0.9407 $\pm$ 0.0047 & 0.9479 $\pm$ 0.0038 & 0.9399 $\pm$ 0.0072 & 0.9419 $\pm$ 0.0041 & 0.9393 $\pm$ 0.0055 & 0.9451 $\pm$ 0.0042 & 0.9492 $\pm$ 0.0039 & 0.9377 $\pm$ 0.0050 \\
1.3125 & 0.9375 $\pm$ 0.0054 & 0.9482 $\pm$ 0.0053 & 0.9403 $\pm$ 0.0061 & 0.9420 $\pm$ 0.0055 & 0.9363 $\pm$ 0.0076 & 0.9420 $\pm$ 0.0060 & 0.9494 $\pm$ 0.0049 & 0.9387 $\pm$ 0.0062 \\
1.3750 & 0.9310 $\pm$ 0.0056 & 0.9466 $\pm$ 0.0052 & 0.9401 $\pm$ 0.0051 & 0.9401 $\pm$ 0.0057 & 0.9323 $\pm$ 0.0057 & 0.9428 $\pm$ 0.0065 & 0.9477 $\pm$ 0.0049 & 0.9362 $\pm$ 0.0051 \\
1.4375 & 0.9262 $\pm$ 0.0070 & 0.9445 $\pm$ 0.0058 & 0.9368 $\pm$ 0.0101 & 0.9378 $\pm$ 0.0061 & 0.9261 $\pm$ 0.0086 & 0.9425 $\pm$ 0.0059 & 0.9459 $\pm$ 0.0056 & 0.9343 $\pm$ 0.0060 \\
1.5000 & 0.9210 $\pm$ 0.0068 & 0.9445 $\pm$ 0.0054 & 0.9377 $\pm$ 0.0055 & 0.9378 $\pm$ 0.0062 & 0.9258 $\pm$ 0.0077 & 0.9400 $\pm$ 0.0064 & 0.9453 $\pm$ 0.0054 & 0.9340 $\pm$ 0.0058 \\
\bottomrule
\end{tabular}

\end{adjustbox}
\caption{Group C, PR AUC, values after $\pm$ are standard deviations.}
\end{table}

\begin{table}[H]
\centering
\begin{adjustbox}{max width=\linewidth}
\begin{tabular}{lllllllll}
\toprule
 & BetaNB & BetaNB\_p & MCNB & MCNB\_p & NB & NB\_p & binom & binom\_beta \\
$bias$ &  &  &  &  &  &  &  &  \\
\midrule
1.0625 & 0.6540 $\pm$ 0.0211 & 0.6613 $\pm$ 0.0256 & 0.6918 $\pm$ 0.0196 & 0.7222 $\pm$ 0.0189 & 0.7699 $\pm$ 0.0161 & 0.7638 $\pm$ 0.0174 & 0.7018 $\pm$ 0.0207 & 0.5451 $\pm$ 0.0254 \\
1.1250 & 0.6526 $\pm$ 0.0208 & 0.6701 $\pm$ 0.0280 & 0.6958 $\pm$ 0.0159 & 0.7152 $\pm$ 0.0190 & 0.7686 $\pm$ 0.0185 & 0.7623 $\pm$ 0.0193 & 0.6976 $\pm$ 0.0202 & 0.5384 $\pm$ 0.0228 \\
1.1875 & 0.6449 $\pm$ 0.0171 & 0.6758 $\pm$ 0.0465 & 0.6845 $\pm$ 0.0173 & 0.7170 $\pm$ 0.0137 & 0.7620 $\pm$ 0.0161 & 0.7585 $\pm$ 0.0162 & 0.6931 $\pm$ 0.0157 & 0.5341 $\pm$ 0.0160 \\
1.2500 & 0.6379 $\pm$ 0.0265 & 0.6602 $\pm$ 0.0358 & 0.6824 $\pm$ 0.0189 & 0.7183 $\pm$ 0.0168 & 0.7615 $\pm$ 0.0189 & 0.7552 $\pm$ 0.0174 & 0.6920 $\pm$ 0.0195 & 0.5331 $\pm$ 0.0210 \\
1.3125 & 0.6349 $\pm$ 0.0232 & 0.6629 $\pm$ 0.0371 & 0.6935 $\pm$ 0.0152 & 0.7229 $\pm$ 0.0165 & 0.7640 $\pm$ 0.0156 & 0.7581 $\pm$ 0.0142 & 0.6925 $\pm$ 0.0177 & 0.5313 $\pm$ 0.0206 \\
1.3750 & 0.6279 $\pm$ 0.0167 & 0.6608 $\pm$ 0.0270 & 0.6693 $\pm$ 0.0158 & 0.7105 $\pm$ 0.0161 & 0.7537 $\pm$ 0.0114 & 0.7507 $\pm$ 0.0148 & 0.6844 $\pm$ 0.0156 & 0.5235 $\pm$ 0.0219 \\
1.4375 & 0.6201 $\pm$ 0.0229 & 0.6475 $\pm$ 0.0241 & 0.6743 $\pm$ 0.0188 & 0.7048 $\pm$ 0.0203 & 0.7519 $\pm$ 0.0165 & 0.7474 $\pm$ 0.0187 & 0.6792 $\pm$ 0.0191 & 0.5225 $\pm$ 0.0236 \\
1.5000 & 0.6136 $\pm$ 0.0158 & 0.6437 $\pm$ 0.0292 & 0.6725 $\pm$ 0.0171 & 0.7017 $\pm$ 0.0176 & 0.7458 $\pm$ 0.0150 & 0.7436 $\pm$ 0.0163 & 0.6735 $\pm$ 0.0139 & 0.5146 $\pm$ 0.0181 \\
\bottomrule
\end{tabular}

\end{adjustbox}
\caption{Group C, Sensitivity, values after $\pm$ are standard deviations.}
\end{table}

\begin{table}[H]
\centering
\begin{adjustbox}{max width=\linewidth}
\begin{tabular}{lllllllll}
\toprule
 & BetaNB & BetaNB\_p & MCNB & MCNB\_p & NB & NB\_p & binom & binom\_beta \\
$bias$ &  &  &  &  &  &  &  &  \\
\midrule
1.0625 & 0.9993 $\pm$ 0.0002 & 0.9993 $\pm$ 0.0003 & 0.9988 $\pm$ 0.0004 & 0.9980 $\pm$ 0.0019 & 0.9978 $\pm$ 0.0004 & 0.9981 $\pm$ 0.0004 & 0.9993 $\pm$ 0.0002 & 0.9998 $\pm$ 0.0002 \\
1.1250 & 0.9993 $\pm$ 0.0002 & 0.9991 $\pm$ 0.0004 & 0.9987 $\pm$ 0.0007 & 0.9986 $\pm$ 0.0004 & 0.9977 $\pm$ 0.0004 & 0.9981 $\pm$ 0.0004 & 0.9992 $\pm$ 0.0002 & 0.9998 $\pm$ 0.0002 \\
1.1875 & 0.9993 $\pm$ 0.0002 & 0.9990 $\pm$ 0.0006 & 0.9988 $\pm$ 0.0003 & 0.9983 $\pm$ 0.0008 & 0.9975 $\pm$ 0.0005 & 0.9981 $\pm$ 0.0004 & 0.9992 $\pm$ 0.0003 & 0.9998 $\pm$ 0.0001 \\
1.2500 & 0.9993 $\pm$ 0.0003 & 0.9991 $\pm$ 0.0005 & 0.9987 $\pm$ 0.0005 & 0.9985 $\pm$ 0.0004 & 0.9973 $\pm$ 0.0006 & 0.9982 $\pm$ 0.0004 & 0.9992 $\pm$ 0.0002 & 0.9998 $\pm$ 0.0002 \\
1.3125 & 0.9994 $\pm$ 0.0002 & 0.9992 $\pm$ 0.0004 & 0.9984 $\pm$ 0.0005 & 0.9982 $\pm$ 0.0006 & 0.9969 $\pm$ 0.0006 & 0.9982 $\pm$ 0.0005 & 0.9993 $\pm$ 0.0003 & 0.9998 $\pm$ 0.0001 \\
1.3750 & 0.9995 $\pm$ 0.0002 & 0.9991 $\pm$ 0.0004 & 0.9987 $\pm$ 0.0004 & 0.9985 $\pm$ 0.0006 & 0.9968 $\pm$ 0.0007 & 0.9982 $\pm$ 0.0005 & 0.9993 $\pm$ 0.0002 & 0.9999 $\pm$ 0.0001 \\
1.4375 & 0.9993 $\pm$ 0.0004 & 0.9991 $\pm$ 0.0004 & 0.9983 $\pm$ 0.0005 & 0.9986 $\pm$ 0.0004 & 0.9962 $\pm$ 0.0006 & 0.9981 $\pm$ 0.0004 & 0.9992 $\pm$ 0.0004 & 0.9998 $\pm$ 0.0002 \\
1.5000 & 0.9994 $\pm$ 0.0002 & 0.9992 $\pm$ 0.0003 & 0.9981 $\pm$ 0.0007 & 0.9984 $\pm$ 0.0005 & 0.9955 $\pm$ 0.0006 & 0.9982 $\pm$ 0.0004 & 0.9993 $\pm$ 0.0003 & 0.9998 $\pm$ 0.0002 \\
\bottomrule
\end{tabular}

\end{adjustbox}
\caption{Group C, Specificity, values after $\pm$ are standard deviations.}
\end{table}
\subsubsection*{Group D}
\begin{table}[H]
\centering
\begin{adjustbox}{max width=\linewidth}
\begin{tabular}{lllllllll}
\toprule
 & NB & NB\_p & BetaNB & BetaNB\_p & MCNB & MCNB\_p & binom & binom\_beta \\
BAD &  &  &  &  &  &  &  &  \\
\midrule
1.2 & 0.5112 $\pm$ 0.0171 & 0.5131 $\pm$ 0.0175 & 0.4986 $\pm$ 0.0216 & 0.5102 $\pm$ 0.0177 & 0.5105 $\pm$ 0.0176 & 0.5100 $\pm$ 0.0175 & 0.4766 $\pm$ 0.0171 & 0.4823 $\pm$ 0.0170 \\
1.5 & 0.6168 $\pm$ 0.0154 & 0.6102 $\pm$ 0.0187 & 0.6199 $\pm$ 0.0155 & 0.5801 $\pm$ 0.0230 & 0.5391 $\pm$ 0.0156 & 0.5493 $\pm$ 0.0176 & 0.4090 $\pm$ 0.0173 & 0.4501 $\pm$ 0.0169 \\
1.8 & 0.8255 $\pm$ 0.0098 & 0.8111 $\pm$ 0.0109 & 0.7969 $\pm$ 0.0095 & 0.7827 $\pm$ 0.0113 & 0.7412 $\pm$ 0.0148 & 0.7502 $\pm$ 0.0145 & 0.3428 $\pm$ 0.0146 & 0.4193 $\pm$ 0.0177 \\
2.0 & 0.8929 $\pm$ 0.0048 & 0.8817 $\pm$ 0.0058 & 0.8762 $\pm$ 0.0060 & 0.8645 $\pm$ 0.0068 & 0.8487 $\pm$ 0.0081 & 0.8512 $\pm$ 0.0089 & 0.2811 $\pm$ 0.0133 & 0.3400 $\pm$ 0.0406 \\
2.2 & 0.9093 $\pm$ 0.0073 & 0.9001 $\pm$ 0.0075 & 0.8980 $\pm$ 0.0073 & 0.8887 $\pm$ 0.0081 & 0.8834 $\pm$ 0.0103 & 0.8833 $\pm$ 0.0091 & 0.2434 $\pm$ 0.0128 & 0.1939 $\pm$ 0.0084 \\
2.5 & 0.9102 $\pm$ 0.0060 & 0.9029 $\pm$ 0.0066 & 0.8997 $\pm$ 0.0063 & 0.8924 $\pm$ 0.0104 & 0.8901 $\pm$ 0.0095 & 0.8906 $\pm$ 0.0073 & 0.2084 $\pm$ 0.0099 & 0.1867 $\pm$ 0.0063 \\
2.8 & 0.9047 $\pm$ 0.0068 & 0.8986 $\pm$ 0.0067 & 0.8726 $\pm$ 0.0778 & 0.8899 $\pm$ 0.0067 & 0.8895 $\pm$ 0.0071 & 0.8886 $\pm$ 0.0077 & 0.1825 $\pm$ 0.0120 & 0.1884 $\pm$ 0.0059 \\
3.0 & 0.9000 $\pm$ 0.0075 & 0.8942 $\pm$ 0.0078 & 0.8766 $\pm$ 0.0242 & 0.8857 $\pm$ 0.0075 & 0.8825 $\pm$ 0.0101 & 0.8837 $\pm$ 0.0083 & 0.1621 $\pm$ 0.0120 & 0.1863 $\pm$ 0.0058 \\
\bottomrule
\end{tabular}

\end{adjustbox}
\caption{Group D, PR AUC, values after $\pm$ are standard deviations.}
\end{table}

\begin{table}[H]
\centering
\begin{adjustbox}{max width=\linewidth}
\begin{tabular}{lllllllll}
\toprule
 & BetaNB & BetaNB\_p & MCNB & MCNB\_p & NB & NB\_p & binom & binom\_beta \\
BAD &  &  &  &  &  &  &  &  \\
\midrule
1.2 & 0.3654 $\pm$ 0.0323 & 0.3538 $\pm$ 0.0182 & 0.3403 $\pm$ 0.0170 & 0.3878 $\pm$ 0.0197 & 0.3836 $\pm$ 0.0193 & 0.3667 $\pm$ 0.0173 & 0.5199 $\pm$ 0.0154 & 0.4304 $\pm$ 0.0190 \\
1.5 & 0.3029 $\pm$ 0.0218 & 0.3289 $\pm$ 0.0269 & 0.3653 $\pm$ 0.0184 & 0.3567 $\pm$ 0.0158 & 0.3563 $\pm$ 0.0192 & 0.3565 $\pm$ 0.0181 & 0.6421 $\pm$ 0.0151 & 0.4743 $\pm$ 0.0147 \\
1.8 & 0.3253 $\pm$ 0.0170 & 0.3402 $\pm$ 0.0355 & 0.3741 $\pm$ 0.0205 & 0.3801 $\pm$ 0.0155 & 0.4636 $\pm$ 0.0166 & 0.4524 $\pm$ 0.0144 & 0.9586 $\pm$ 0.0079 & 0.4962 $\pm$ 0.0110 \\
2.0 & 0.4110 $\pm$ 0.0223 & 0.4068 $\pm$ 0.0269 & 0.4372 $\pm$ 0.0185 & 0.4627 $\pm$ 0.0178 & 0.5664 $\pm$ 0.0141 & 0.5504 $\pm$ 0.0148 & 0.9979 $\pm$ 0.0011 & 0.5083 $\pm$ 0.0149 \\
2.2 & 0.4549 $\pm$ 0.0185 & 0.4382 $\pm$ 0.0164 & 0.4655 $\pm$ 0.0155 & 0.5060 $\pm$ 0.0161 & 0.5996 $\pm$ 0.0134 & 0.5788 $\pm$ 0.0132 & 0.9998 $\pm$ 0.0004 & 0.5337 $\pm$ 0.0155 \\
2.5 & 0.4641 $\pm$ 0.0129 & 0.4569 $\pm$ 0.0287 & 0.4709 $\pm$ 0.0194 & 0.5181 $\pm$ 0.0148 & 0.6064 $\pm$ 0.0165 & 0.5855 $\pm$ 0.0167 & 1.0000 $\pm$ 0.0000 & 0.5655 $\pm$ 0.0115 \\
2.8 & 0.4394 $\pm$ 0.0585 & 0.4489 $\pm$ 0.0192 & 0.4483 $\pm$ 0.0198 & 0.5129 $\pm$ 0.0150 & 0.5979 $\pm$ 0.0184 & 0.5745 $\pm$ 0.0174 & 1.0000 $\pm$ 0.0000 & 0.6018 $\pm$ 0.0160 \\
3.0 & 0.4342 $\pm$ 0.0529 & 0.4376 $\pm$ 0.0217 & 0.4362 $\pm$ 0.0243 & 0.4992 $\pm$ 0.0203 & 0.5847 $\pm$ 0.0203 & 0.5563 $\pm$ 0.0189 & 1.0000 $\pm$ 0.0000 & 0.6352 $\pm$ 0.0112 \\
\bottomrule
\end{tabular}

\end{adjustbox}
\caption{Group D, Sensitivity, values after $\pm$ are standard deviations.}
\end{table}

\begin{table}[H]
\centering
\begin{adjustbox}{max width=\linewidth}
\begin{tabular}{lllllllll}
\toprule
 & BetaNB & BetaNB\_p & MCNB & MCNB\_p & NB & NB\_p & binom & binom\_beta \\
BAD &  &  &  &  &  &  &  &  \\
\midrule
1.2 & 0.9943 $\pm$ 0.0130 & 0.9993 $\pm$ 0.0003 & 0.9995 $\pm$ 0.0002 & 0.9980 $\pm$ 0.0006 & 0.9985 $\pm$ 0.0004 & 0.9990 $\pm$ 0.0003 & 0.2751 $\pm$ 0.0046 & 0.9818 $\pm$ 0.0016 \\
1.5 & 0.9997 $\pm$ 0.0002 & 0.9995 $\pm$ 0.0002 & 0.9980 $\pm$ 0.0010 & 0.9991 $\pm$ 0.0003 & 0.9993 $\pm$ 0.0003 & 0.9993 $\pm$ 0.0003 & 0.0058 $\pm$ 0.0008 & 0.8317 $\pm$ 0.0137 \\
1.8 & 0.9998 $\pm$ 0.0001 & 0.9997 $\pm$ 0.0002 & 0.9991 $\pm$ 0.0004 & 0.9992 $\pm$ 0.0004 & 0.9992 $\pm$ 0.0002 & 0.9991 $\pm$ 0.0002 & 0.0001 $\pm$ 0.0001 & 0.4101 $\pm$ 0.0099 \\
2.0 & 0.9997 $\pm$ 0.0002 & 0.9997 $\pm$ 0.0002 & 0.9993 $\pm$ 0.0004 & 0.9991 $\pm$ 0.0004 & 0.9991 $\pm$ 0.0002 & 0.9990 $\pm$ 0.0003 & 0.0000 $\pm$ 0.0000 & 0.2211 $\pm$ 0.0057 \\
2.2 & 0.9997 $\pm$ 0.0002 & 0.9997 $\pm$ 0.0002 & 0.9995 $\pm$ 0.0003 & 0.9992 $\pm$ 0.0003 & 0.9990 $\pm$ 0.0004 & 0.9989 $\pm$ 0.0004 & 0.0000 $\pm$ 0.0000 & 0.1401 $\pm$ 0.0060 \\
2.5 & 0.9996 $\pm$ 0.0002 & 0.9996 $\pm$ 0.0003 & 0.9994 $\pm$ 0.0003 & 0.9992 $\pm$ 0.0002 & 0.9989 $\pm$ 0.0004 & 0.9989 $\pm$ 0.0004 & 0.0000 $\pm$ 0.0000 & 0.0985 $\pm$ 0.0033 \\
2.8 & 0.9996 $\pm$ 0.0002 & 0.9995 $\pm$ 0.0003 & 0.9995 $\pm$ 0.0002 & 0.9991 $\pm$ 0.0002 & 0.9987 $\pm$ 0.0004 & 0.9988 $\pm$ 0.0002 & 0.0000 $\pm$ 0.0000 & 0.0760 $\pm$ 0.0039 \\
3.0 & 0.9996 $\pm$ 0.0002 & 0.9996 $\pm$ 0.0002 & 0.9995 $\pm$ 0.0003 & 0.9992 $\pm$ 0.0003 & 0.9988 $\pm$ 0.0005 & 0.9989 $\pm$ 0.0004 & 0.0000 $\pm$ 0.0000 & 0.0631 $\pm$ 0.0032 \\
\bottomrule
\end{tabular}

\end{adjustbox}
\caption{Group D, Specificity, values after $\pm$ are standard deviations.}
\end{table}
\subsubsection*{Group E}
\begin{table}[H]
\centering
\begin{adjustbox}{max width=\linewidth}
\begin{tabular}{lllllllll}
\toprule
 & NB & NB\_p & BetaNB & BetaNB\_p & MCNB & MCNB\_p & binom & binom\_beta \\
$\kappa$ &  &  &  &  &  &  &  &  \\
\midrule
10 & 0.0950 $\pm$ 0.0036 & 0.0964 $\pm$ 0.0034 & 0.0999 $\pm$ 0.0041 & 0.1034 $\pm$ 0.0060 & 0.0923 $\pm$ 0.0025 & 0.0972 $\pm$ 0.0032 & 0.1245 $\pm$ 0.0059 & 0.1061 $\pm$ 0.0050 \\
16 & 0.0973 $\pm$ 0.0034 & 0.1003 $\pm$ 0.0036 & 0.1049 $\pm$ 0.0040 & 0.0979 $\pm$ 0.0037 & 0.0979 $\pm$ 0.0032 & 0.1022 $\pm$ 0.0037 & 0.1163 $\pm$ 0.0057 & 0.1133 $\pm$ 0.0050 \\
32 & 0.1054 $\pm$ 0.0038 & 0.1123 $\pm$ 0.0043 & 0.1241 $\pm$ 0.0048 & 0.1199 $\pm$ 0.0046 & 0.1097 $\pm$ 0.0042 & 0.1155 $\pm$ 0.0040 & 0.1185 $\pm$ 0.0051 & 0.1380 $\pm$ 0.0063 \\
64 & 0.1303 $\pm$ 0.0056 & 0.1393 $\pm$ 0.0059 & 0.1825 $\pm$ 0.0097 & 0.1821 $\pm$ 0.0097 & 0.1387 $\pm$ 0.0058 & 0.1418 $\pm$ 0.0059 & 0.1410 $\pm$ 0.0063 & 0.1907 $\pm$ 0.0107 \\
128 & 0.1994 $\pm$ 0.0063 & 0.2039 $\pm$ 0.0068 & 0.2995 $\pm$ 0.0098 & 0.2996 $\pm$ 0.0099 & 0.2027 $\pm$ 0.0064 & 0.2090 $\pm$ 0.0069 & 0.2066 $\pm$ 0.0069 & 0.3056 $\pm$ 0.0100 \\
256 & 0.3792 $\pm$ 0.0157 & 0.3847 $\pm$ 0.0162 & 0.4943 $\pm$ 0.0178 & 0.4887 $\pm$ 0.0319 & 0.3871 $\pm$ 0.0157 & 0.3969 $\pm$ 0.0163 & 0.4001 $\pm$ 0.0164 & 0.5061 $\pm$ 0.0181 \\
512 & 0.6639 $\pm$ 0.0155 & 0.6683 $\pm$ 0.0153 & 0.7222 $\pm$ 0.0147 & 0.7017 $\pm$ 0.0308 & 0.6753 $\pm$ 0.0152 & 0.6840 $\pm$ 0.0149 & 0.6901 $\pm$ 0.0149 & 0.7356 $\pm$ 0.0138 \\
\bottomrule
\end{tabular}

\end{adjustbox}
\caption{Group E, PR AUC, values after $\pm$ are standard deviations.}
\end{table}

\begin{table}[H]
\centering
\begin{adjustbox}{max width=\linewidth}
\begin{tabular}{lllllllll}
\toprule
 & BetaNB & BetaNB\_p & MCNB & MCNB\_p & NB & NB\_p & binom & binom\_beta \\
$\kappa$ &  &  &  &  &  &  &  &  \\
\midrule
10 & 0.7455 $\pm$ 0.0124 & 0.4065 $\pm$ 0.1084 & 0.8542 $\pm$ 0.0109 & 0.7557 $\pm$ 0.0128 & 0.8567 $\pm$ 0.0110 & 0.7983 $\pm$ 0.0110 & 0.8962 $\pm$ 0.0098 & 0.5515 $\pm$ 0.0164 \\
16 & 0.7311 $\pm$ 0.0072 & 0.7282 $\pm$ 0.0348 & 0.8525 $\pm$ 0.0094 & 0.7699 $\pm$ 0.0103 & 0.8548 $\pm$ 0.0091 & 0.8035 $\pm$ 0.0099 & 0.8752 $\pm$ 0.0084 & 0.5488 $\pm$ 0.0092 \\
32 & 0.6894 $\pm$ 0.0136 & 0.6961 $\pm$ 0.0133 & 0.8339 $\pm$ 0.0110 & 0.7643 $\pm$ 0.0125 & 0.8341 $\pm$ 0.0117 & 0.7905 $\pm$ 0.0118 & 0.8338 $\pm$ 0.0128 & 0.5621 $\pm$ 0.0116 \\
64 & 0.6505 $\pm$ 0.0143 & 0.6519 $\pm$ 0.0148 & 0.8039 $\pm$ 0.0144 & 0.7676 $\pm$ 0.0120 & 0.8108 $\pm$ 0.0131 & 0.7790 $\pm$ 0.0126 & 0.8009 $\pm$ 0.0125 & 0.5963 $\pm$ 0.0128 \\
128 & 0.6877 $\pm$ 0.0122 & 0.6878 $\pm$ 0.0123 & 0.8093 $\pm$ 0.0128 & 0.7928 $\pm$ 0.0127 & 0.8151 $\pm$ 0.0107 & 0.7968 $\pm$ 0.0120 & 0.7976 $\pm$ 0.0117 & 0.6410 $\pm$ 0.0127 \\
256 & 0.7570 $\pm$ 0.0147 & 0.7591 $\pm$ 0.0179 & 0.8488 $\pm$ 0.0140 & 0.8410 $\pm$ 0.0149 & 0.8566 $\pm$ 0.0121 & 0.8467 $\pm$ 0.0140 & 0.8324 $\pm$ 0.0153 & 0.7038 $\pm$ 0.0145 \\
512 & 0.8167 $\pm$ 0.0103 & 0.8317 $\pm$ 0.0258 & 0.8862 $\pm$ 0.0071 & 0.8835 $\pm$ 0.0083 & 0.8971 $\pm$ 0.0082 & 0.8901 $\pm$ 0.0087 & 0.8665 $\pm$ 0.0081 & 0.7524 $\pm$ 0.0149 \\
\bottomrule
\end{tabular}

\end{adjustbox}
\caption{Group E, Sensitivity, values after $\pm$ are standard deviations.}
\end{table}

\begin{table}[H]
\centering
\begin{adjustbox}{max width=\linewidth}
\begin{tabular}{lllllllll}
\toprule
 & BetaNB & BetaNB\_p & MCNB & MCNB\_p & NB & NB\_p & binom & binom\_beta \\
$\kappa$ &  &  &  &  &  &  &  &  \\
\midrule
10 & 0.2773 $\pm$ 0.0047 & 0.6390 $\pm$ 0.1117 & 0.1597 $\pm$ 0.0042 & 0.2699 $\pm$ 0.0038 & 0.1550 $\pm$ 0.0035 & 0.2206 $\pm$ 0.0041 & 0.1126 $\pm$ 0.0034 & 0.4942 $\pm$ 0.0048 \\
16 & 0.3073 $\pm$ 0.0052 & 0.3087 $\pm$ 0.0295 & 0.1704 $\pm$ 0.0036 & 0.2671 $\pm$ 0.0076 & 0.1673 $\pm$ 0.0035 & 0.2251 $\pm$ 0.0039 & 0.1446 $\pm$ 0.0037 & 0.5131 $\pm$ 0.0047 \\
32 & 0.3901 $\pm$ 0.0061 & 0.3822 $\pm$ 0.0053 & 0.2140 $\pm$ 0.0042 & 0.3059 $\pm$ 0.0047 & 0.2113 $\pm$ 0.0043 & 0.2673 $\pm$ 0.0045 & 0.2156 $\pm$ 0.0042 & 0.5549 $\pm$ 0.0061 \\
64 & 0.5474 $\pm$ 0.0100 & 0.5450 $\pm$ 0.0102 & 0.3269 $\pm$ 0.0082 & 0.3846 $\pm$ 0.0067 & 0.3066 $\pm$ 0.0036 & 0.3607 $\pm$ 0.0048 & 0.3278 $\pm$ 0.0045 & 0.6272 $\pm$ 0.0040 \\
128 & 0.6703 $\pm$ 0.0055 & 0.6705 $\pm$ 0.0047 & 0.4690 $\pm$ 0.0057 & 0.5050 $\pm$ 0.0079 & 0.4463 $\pm$ 0.0056 & 0.4835 $\pm$ 0.0058 & 0.4864 $\pm$ 0.0066 & 0.7309 $\pm$ 0.0047 \\
256 & 0.7862 $\pm$ 0.0039 & 0.7818 $\pm$ 0.0192 & 0.6377 $\pm$ 0.0069 & 0.6566 $\pm$ 0.0072 & 0.6075 $\pm$ 0.0056 & 0.6334 $\pm$ 0.0064 & 0.6711 $\pm$ 0.0068 & 0.8468 $\pm$ 0.0033 \\
512 & 0.8961 $\pm$ 0.0029 & 0.8756 $\pm$ 0.0298 & 0.8104 $\pm$ 0.0041 & 0.8182 $\pm$ 0.0045 & 0.7789 $\pm$ 0.0049 & 0.7941 $\pm$ 0.0048 & 0.8441 $\pm$ 0.0036 & 0.9402 $\pm$ 0.0031 \\
\bottomrule
\end{tabular}

\end{adjustbox}
\caption{Group E, Specificity, values after $\pm$ are standard deviations.}
\end{table}
\subsubsection*{Group F}
\begin{table}[H]
\centering
\begin{adjustbox}{max width=\linewidth}
\begin{tabular}{lllllllll}
\toprule
 & NB & NB\_p & BetaNB & BetaNB\_p & MCNB & MCNB\_p & binom & binom\_beta \\
$ase\_coverage$ &  &  &  &  &  &  &  &  \\
\midrule
20 & 0.3547 $\pm$ 0.0134 & 0.3520 $\pm$ 0.0136 & 0.3540 $\pm$ 0.0149 & 0.3516 $\pm$ 0.0135 & 0.3731 $\pm$ 0.0344 & 0.3097 $\pm$ 0.0229 & 0.4643 $\pm$ 0.0239 & 0.5125 $\pm$ 0.0193 \\
30 & 0.5863 $\pm$ 0.0124 & 0.5862 $\pm$ 0.0123 & 0.5864 $\pm$ 0.0128 & 0.5866 $\pm$ 0.0121 & 0.5288 $\pm$ 0.0189 & 0.5617 $\pm$ 0.0169 & 0.6177 $\pm$ 0.0104 & 0.6407 $\pm$ 0.0090 \\
40 & 0.7490 $\pm$ 0.0111 & 0.7482 $\pm$ 0.0112 & 0.7496 $\pm$ 0.0111 & 0.7487 $\pm$ 0.0111 & 0.6898 $\pm$ 0.0163 & 0.7209 $\pm$ 0.0156 & 0.7620 $\pm$ 0.0115 & 0.7739 $\pm$ 0.0117 \\
50 & 0.8491 $\pm$ 0.0095 & 0.8486 $\pm$ 0.0095 & 0.8502 $\pm$ 0.0097 & 0.8488 $\pm$ 0.0095 & 0.8022 $\pm$ 0.0130 & 0.8242 $\pm$ 0.0128 & 0.8545 $\pm$ 0.0098 & 0.8606 $\pm$ 0.0100 \\
60 & 0.9111 $\pm$ 0.0080 & 0.9110 $\pm$ 0.0086 & 0.9105 $\pm$ 0.0091 & 0.9111 $\pm$ 0.0086 & 0.8780 $\pm$ 0.0109 & 0.8965 $\pm$ 0.0104 & 0.9118 $\pm$ 0.0081 & 0.9150 $\pm$ 0.0082 \\
70 & 0.9477 $\pm$ 0.0047 & 0.9475 $\pm$ 0.0052 & 0.9471 $\pm$ 0.0058 & 0.9477 $\pm$ 0.0051 & 0.9254 $\pm$ 0.0062 & 0.9380 $\pm$ 0.0065 & 0.9466 $\pm$ 0.0045 & 0.9480 $\pm$ 0.0047 \\
\bottomrule
\end{tabular}

\end{adjustbox}
\caption{Group F, PR AUC, values after $\pm$ are standard deviations.}
\end{table}

\begin{table}[H]
\centering
\begin{adjustbox}{max width=\linewidth}
\begin{tabular}{lllllllll}
\toprule
 & BetaNB & BetaNB\_p & MCNB & MCNB\_p & NB & NB\_p & binom & binom\_beta \\
$ase\_coverage$ &  &  &  &  &  &  &  &  \\
\midrule
20 & 0.0013 $\pm$ 0.0012 & 0.0013 $\pm$ 0.0012 & 0.0010 $\pm$ 0.0015 & 0.0015 $\pm$ 0.0043 & 0.0013 $\pm$ 0.0012 & 0.0014 $\pm$ 0.0014 & 0.0010 $\pm$ 0.0011 & 0.0008 $\pm$ 0.0010 \\
30 & 0.0298 $\pm$ 0.0122 & 0.0335 $\pm$ 0.0124 & 0.0080 $\pm$ 0.0091 & 0.0386 $\pm$ 0.0192 & 0.0298 $\pm$ 0.0122 & 0.0345 $\pm$ 0.0128 & 0.0141 $\pm$ 0.0081 & 0.0098 $\pm$ 0.0072 \\
40 & 0.1765 $\pm$ 0.0143 & 0.1840 $\pm$ 0.0152 & 0.0405 $\pm$ 0.0155 & 0.1649 $\pm$ 0.0180 & 0.1759 $\pm$ 0.0151 & 0.1842 $\pm$ 0.0148 & 0.1026 $\pm$ 0.0141 & 0.0704 $\pm$ 0.0134 \\
50 & 0.3789 $\pm$ 0.0197 & 0.3846 $\pm$ 0.0191 & 0.1584 $\pm$ 0.0186 & 0.3317 $\pm$ 0.0223 & 0.3766 $\pm$ 0.0187 & 0.3857 $\pm$ 0.0183 & 0.2655 $\pm$ 0.0215 & 0.2209 $\pm$ 0.0190 \\
60 & 0.5633 $\pm$ 0.0251 & 0.5745 $\pm$ 0.0238 & 0.3341 $\pm$ 0.0211 & 0.5275 $\pm$ 0.0224 & 0.5648 $\pm$ 0.0229 & 0.5750 $\pm$ 0.0242 & 0.4444 $\pm$ 0.0188 & 0.3878 $\pm$ 0.0188 \\
70 & 0.7063 $\pm$ 0.0233 & 0.7172 $\pm$ 0.0198 & 0.5096 $\pm$ 0.0188 & 0.6794 $\pm$ 0.0202 & 0.7093 $\pm$ 0.0181 & 0.7177 $\pm$ 0.0195 & 0.5939 $\pm$ 0.0167 & 0.5339 $\pm$ 0.0171 \\
\bottomrule
\end{tabular}

\end{adjustbox}
\caption{Group F, Sensitivity, values after $\pm$ are standard deviations.}
\end{table}

\begin{table}[H]
\centering
\begin{adjustbox}{max width=\linewidth}
\begin{tabular}{lllllllll}
\toprule
 & BetaNB & BetaNB\_p & MCNB & MCNB\_p & NB & NB\_p & binom & binom\_beta \\
$ase\_coverage$ &  &  &  &  &  &  &  &  \\
\midrule
20 & 1.0000 $\pm$ 0.0000 & 1.0000 $\pm$ 0.0000 & 1.0000 $\pm$ 0.0000 & 1.0000 $\pm$ 0.0000 & 1.0000 $\pm$ 0.0000 & 1.0000 $\pm$ 0.0000 & 1.0000 $\pm$ 0.0000 & 1.0000 $\pm$ 0.0000 \\
30 & 0.9999 $\pm$ 0.0001 & 0.9999 $\pm$ 0.0001 & 1.0000 $\pm$ 0.0000 & 0.9999 $\pm$ 0.0001 & 0.9999 $\pm$ 0.0001 & 0.9999 $\pm$ 0.0001 & 1.0000 $\pm$ 0.0000 & 1.0000 $\pm$ 0.0000 \\
40 & 0.9996 $\pm$ 0.0002 & 0.9996 $\pm$ 0.0002 & 1.0000 $\pm$ 0.0001 & 0.9996 $\pm$ 0.0002 & 0.9996 $\pm$ 0.0002 & 0.9996 $\pm$ 0.0002 & 0.9999 $\pm$ 0.0001 & 1.0000 $\pm$ 0.0000 \\
50 & 0.9991 $\pm$ 0.0003 & 0.9990 $\pm$ 0.0003 & 0.9998 $\pm$ 0.0001 & 0.9991 $\pm$ 0.0003 & 0.9991 $\pm$ 0.0003 & 0.9990 $\pm$ 0.0003 & 0.9998 $\pm$ 0.0002 & 0.9999 $\pm$ 0.0001 \\
60 & 0.9986 $\pm$ 0.0004 & 0.9985 $\pm$ 0.0004 & 0.9996 $\pm$ 0.0002 & 0.9986 $\pm$ 0.0003 & 0.9986 $\pm$ 0.0004 & 0.9985 $\pm$ 0.0004 & 0.9996 $\pm$ 0.0002 & 0.9998 $\pm$ 0.0001 \\
70 & 0.9982 $\pm$ 0.0004 & 0.9980 $\pm$ 0.0004 & 0.9994 $\pm$ 0.0002 & 0.9981 $\pm$ 0.0004 & 0.9982 $\pm$ 0.0004 & 0.9980 $\pm$ 0.0004 & 0.9994 $\pm$ 0.0002 & 0.9998 $\pm$ 0.0002 \\
\bottomrule
\end{tabular}

\end{adjustbox}
\caption{Group F, Specificity, values after $\pm$ are standard deviations.}
\end{table}
\subsubsection*{Group G}
\begin{table}[H]
\centering
\begin{adjustbox}{max width=\linewidth}
\begin{tabular}{lllllllll}
\toprule
 & NB & NB\_p & BetaNB & BetaNB\_p & MCNB & MCNB\_p & binom & binom\_beta \\
$bias$ &  &  &  &  &  &  &  &  \\
\midrule
1.1000 & 0.6725 $\pm$ 0.0111 & 0.8982 $\pm$ 0.0068 & 0.6527 $\pm$ 0.0093 & 0.8855 $\pm$ 0.0070 & 0.6554 $\pm$ 0.0113 & 0.8787 $\pm$ 0.0085 & 0.2493 $\pm$ 0.0092 & 0.2020 $\pm$ 0.0190 \\
1.2000 & 0.4576 $\pm$ 0.0112 & 0.9044 $\pm$ 0.0062 & 0.4819 $\pm$ 0.0120 & 0.8939 $\pm$ 0.0065 & 0.4676 $\pm$ 0.0113 & 0.8913 $\pm$ 0.0063 & 0.2203 $\pm$ 0.0112 & 0.1905 $\pm$ 0.0080 \\
1.3000 & 0.3362 $\pm$ 0.0121 & 0.9007 $\pm$ 0.0065 & 0.4164 $\pm$ 0.0110 & 0.8914 $\pm$ 0.0072 & 0.3477 $\pm$ 0.0121 & 0.8884 $\pm$ 0.0097 & 0.1979 $\pm$ 0.0116 & 0.1886 $\pm$ 0.0085 \\
1.5000 & 0.1835 $\pm$ 0.0120 & 0.8942 $\pm$ 0.0078 & 0.3076 $\pm$ 0.0211 & 0.8857 $\pm$ 0.0075 & 0.1920 $\pm$ 0.0126 & 0.8837 $\pm$ 0.0083 & 0.1609 $\pm$ 0.0098 & 0.1863 $\pm$ 0.0058 \\
\bottomrule
\end{tabular}

\end{adjustbox}
\caption{Group G, PR AUC, values after $\pm$ are standard deviations.}
\end{table}

\begin{table}[H]
\centering
\begin{adjustbox}{max width=\linewidth}
\begin{tabular}{lllllllll}
\toprule
 & BetaNB & BetaNB\_p & MCNB & MCNB\_p & NB & NB\_p & binom & binom\_beta \\
$bias$ &  &  &  &  &  &  &  &  \\
\midrule
1.1000 & 0.3771 $\pm$ 0.0377 & 0.4398 $\pm$ 0.0235 & 0.4754 $\pm$ 0.0209 & 0.5026 $\pm$ 0.0195 & 0.5372 $\pm$ 0.0168 & 0.5787 $\pm$ 0.0180 & 0.9997 $\pm$ 0.0005 & 0.5279 $\pm$ 0.0142 \\
1.2000 & 0.3565 $\pm$ 0.0134 & 0.4564 $\pm$ 0.0212 & 0.5319 $\pm$ 0.0151 & 0.5215 $\pm$ 0.0188 & 0.5462 $\pm$ 0.0145 & 0.5885 $\pm$ 0.0140 & 1.0000 $\pm$ 0.0002 & 0.5520 $\pm$ 0.0151 \\
1.3000 & 0.3523 $\pm$ 0.0140 & 0.4519 $\pm$ 0.0220 & 0.5330 $\pm$ 0.0153 & 0.5149 $\pm$ 0.0173 & 0.5304 $\pm$ 0.0151 & 0.5783 $\pm$ 0.0203 & 1.0000 $\pm$ 0.0000 & 0.5821 $\pm$ 0.0149 \\
1.5000 & 0.3439 $\pm$ 0.0170 & 0.4375 $\pm$ 0.0217 & 0.5555 $\pm$ 0.0140 & 0.4992 $\pm$ 0.0203 & 0.5531 $\pm$ 0.0144 & 0.5563 $\pm$ 0.0189 & 1.0000 $\pm$ 0.0000 & 0.6352 $\pm$ 0.0112 \\
\bottomrule
\end{tabular}

\end{adjustbox}
\caption{Group G, Sensitivity, values after $\pm$ are standard deviations.}
\end{table}

\begin{table}[H]
\centering
\begin{adjustbox}{max width=\linewidth}
\begin{tabular}{lllllllll}
\toprule
 & BetaNB & BetaNB\_p & MCNB & MCNB\_p & NB & NB\_p & binom & binom\_beta \\
$bias$ &  &  &  &  &  &  &  &  \\
\midrule
1.1000 & 0.9968 $\pm$ 0.0023 & 0.9996 $\pm$ 0.0002 & 0.9905 $\pm$ 0.0018 & 0.9991 $\pm$ 0.0004 & 0.9816 $\pm$ 0.0013 & 0.9988 $\pm$ 0.0003 & 0.0000 $\pm$ 0.0000 & 0.1548 $\pm$ 0.0077 \\
1.2000 & 0.9911 $\pm$ 0.0013 & 0.9996 $\pm$ 0.0002 & 0.8505 $\pm$ 0.0073 & 0.9992 $\pm$ 0.0003 & 0.8097 $\pm$ 0.0046 & 0.9989 $\pm$ 0.0004 & 0.0000 $\pm$ 0.0000 & 0.1126 $\pm$ 0.0039 \\
1.3000 & 0.9805 $\pm$ 0.0020 & 0.9996 $\pm$ 0.0002 & 0.5521 $\pm$ 0.0046 & 0.9991 $\pm$ 0.0005 & 0.5414 $\pm$ 0.0044 & 0.9989 $\pm$ 0.0003 & 0.0000 $\pm$ 0.0000 & 0.0878 $\pm$ 0.0038 \\
1.5000 & 0.9365 $\pm$ 0.0205 & 0.9996 $\pm$ 0.0002 & 0.2506 $\pm$ 0.0034 & 0.9992 $\pm$ 0.0003 & 0.2655 $\pm$ 0.0036 & 0.9989 $\pm$ 0.0004 & 0.0000 $\pm$ 0.0000 & 0.0631 $\pm$ 0.0032 \\
\bottomrule
\end{tabular}

\end{adjustbox}
\caption{Group G, Specificity, values after $\pm$ are standard deviations.}
\end{table}
\subsubsection*{Group H}
\begin{table}[H]
\centering
\begin{adjustbox}{max width=\linewidth}
\begin{tabular}{lllllllll}
\toprule
 & NB & NB\_p & BetaNB & BetaNB\_p & MCNB & MCNB\_p & binom & binom\_beta \\
$r0\_noise$ &  &  &  &  &  &  &  &  \\
\midrule
2.0 & 0.9617 $\pm$ 0.0060 & 0.9709 $\pm$ 0.0051 & 0.9749 $\pm$ 0.0045 & 0.9751 $\pm$ 0.0045 & 0.9679 $\pm$ 0.0053 & 0.9762 $\pm$ 0.0044 & 0.9767 $\pm$ 0.0046 & 0.9782 $\pm$ 0.0040 \\
4.0 & 0.9266 $\pm$ 0.0058 & 0.9416 $\pm$ 0.0053 & 0.9505 $\pm$ 0.0047 & 0.9500 $\pm$ 0.0061 & 0.9406 $\pm$ 0.0056 & 0.9551 $\pm$ 0.0058 & 0.9452 $\pm$ 0.0054 & 0.9504 $\pm$ 0.0046 \\
6.0 & 0.8949 $\pm$ 0.0076 & 0.9151 $\pm$ 0.0063 & 0.9269 $\pm$ 0.0058 & 0.9240 $\pm$ 0.0070 & 0.9074 $\pm$ 0.0065 & 0.9346 $\pm$ 0.0061 & 0.9141 $\pm$ 0.0065 & 0.9232 $\pm$ 0.0061 \\
8.0 & 0.8709 $\pm$ 0.0092 & 0.8938 $\pm$ 0.0075 & 0.9081 $\pm$ 0.0083 & 0.9002 $\pm$ 0.0094 & 0.8844 $\pm$ 0.0116 & 0.9095 $\pm$ 0.0146 & 0.8891 $\pm$ 0.0071 & 0.9028 $\pm$ 0.0067 \\
10.0 & 0.8436 $\pm$ 0.0082 & 0.8726 $\pm$ 0.0078 & 0.8820 $\pm$ 0.0082 & 0.8778 $\pm$ 0.0094 & 0.8635 $\pm$ 0.0096 & 0.8895 $\pm$ 0.0107 & 0.8651 $\pm$ 0.0081 & 0.8818 $\pm$ 0.0070 \\
12.0 & 0.8317 $\pm$ 0.0086 & 0.8607 $\pm$ 0.0077 & 0.8679 $\pm$ 0.0098 & 0.8676 $\pm$ 0.0105 & 0.8512 $\pm$ 0.0087 & 0.8764 $\pm$ 0.0101 & 0.8490 $\pm$ 0.0083 & 0.8673 $\pm$ 0.0073 \\
\bottomrule
\end{tabular}

\end{adjustbox}
\caption{Group H, PR AUC, values after $\pm$ are standard deviations.}
\end{table}

\begin{table}[H]
\centering
\begin{adjustbox}{max width=\linewidth}
\begin{tabular}{lllllllll}
\toprule
 & BetaNB & BetaNB\_p & MCNB & MCNB\_p & NB & NB\_p & binom & binom\_beta \\
$r0\_noise$ &  &  &  &  &  &  &  &  \\
\midrule
2.0 & 0.8696 $\pm$ 0.0112 & 0.8722 $\pm$ 0.0110 & 0.8895 $\pm$ 0.0111 & 0.8913 $\pm$ 0.0113 & 0.9126 $\pm$ 0.0116 & 0.9073 $\pm$ 0.0111 & 0.9097 $\pm$ 0.0107 & 0.8693 $\pm$ 0.0110 \\
4.0 & 0.7886 $\pm$ 0.0150 & 0.7960 $\pm$ 0.0161 & 0.8203 $\pm$ 0.0094 & 0.8286 $\pm$ 0.0152 & 0.8396 $\pm$ 0.0126 & 0.8347 $\pm$ 0.0143 & 0.8230 $\pm$ 0.0150 & 0.7729 $\pm$ 0.0141 \\
6.0 & 0.7275 $\pm$ 0.0154 & 0.7459 $\pm$ 0.0230 & 0.7641 $\pm$ 0.0173 & 0.7797 $\pm$ 0.0118 & 0.7846 $\pm$ 0.0154 & 0.7770 $\pm$ 0.0138 & 0.7545 $\pm$ 0.0137 & 0.7037 $\pm$ 0.0137 \\
8.0 & 0.6874 $\pm$ 0.0178 & 0.7153 $\pm$ 0.0200 & 0.7258 $\pm$ 0.0152 & 0.7581 $\pm$ 0.0227 & 0.7456 $\pm$ 0.0148 & 0.7373 $\pm$ 0.0135 & 0.7039 $\pm$ 0.0170 & 0.6528 $\pm$ 0.0130 \\
10.0 & 0.6418 $\pm$ 0.0186 & 0.6694 $\pm$ 0.0202 & 0.6745 $\pm$ 0.0159 & 0.7016 $\pm$ 0.0226 & 0.7035 $\pm$ 0.0138 & 0.6949 $\pm$ 0.0147 & 0.6536 $\pm$ 0.0157 & 0.6006 $\pm$ 0.0155 \\
12.0 & 0.6170 $\pm$ 0.0147 & 0.6395 $\pm$ 0.0180 & 0.6575 $\pm$ 0.0161 & 0.6730 $\pm$ 0.0217 & 0.6805 $\pm$ 0.0142 & 0.6703 $\pm$ 0.0153 & 0.6231 $\pm$ 0.0153 & 0.5726 $\pm$ 0.0153 \\
\bottomrule
\end{tabular}

\end{adjustbox}
\caption{Group H, Sensitivity, values after $\pm$ are standard deviations.}
\end{table}

\begin{table}[H]
\centering
\begin{adjustbox}{max width=\linewidth}
\begin{tabular}{lllllllll}
\toprule
 & BetaNB & BetaNB\_p & MCNB & MCNB\_p & NB & NB\_p & binom & binom\_beta \\
$r0\_noise$ &  &  &  &  &  &  &  &  \\
\midrule
2.0 & 0.9996 $\pm$ 0.0002 & 0.9996 $\pm$ 0.0002 & 0.9970 $\pm$ 0.0005 & 0.9988 $\pm$ 0.0004 & 0.9920 $\pm$ 0.0010 & 0.9967 $\pm$ 0.0006 & 0.9980 $\pm$ 0.0005 & 0.9997 $\pm$ 0.0002 \\
4.0 & 0.9996 $\pm$ 0.0002 & 0.9994 $\pm$ 0.0005 & 0.9971 $\pm$ 0.0004 & 0.9986 $\pm$ 0.0006 & 0.9925 $\pm$ 0.0008 & 0.9970 $\pm$ 0.0004 & 0.9985 $\pm$ 0.0004 & 0.9998 $\pm$ 0.0002 \\
6.0 & 0.9997 $\pm$ 0.0001 & 0.9991 $\pm$ 0.0009 & 0.9964 $\pm$ 0.0008 & 0.9981 $\pm$ 0.0008 & 0.9926 $\pm$ 0.0012 & 0.9972 $\pm$ 0.0005 & 0.9988 $\pm$ 0.0002 & 0.9999 $\pm$ 0.0001 \\
8.0 & 0.9996 $\pm$ 0.0002 & 0.9987 $\pm$ 0.0008 & 0.9962 $\pm$ 0.0016 & 0.9946 $\pm$ 0.0058 & 0.9932 $\pm$ 0.0009 & 0.9975 $\pm$ 0.0006 & 0.9988 $\pm$ 0.0004 & 0.9999 $\pm$ 0.0002 \\
10.0 & 0.9996 $\pm$ 0.0003 & 0.9988 $\pm$ 0.0006 & 0.9971 $\pm$ 0.0009 & 0.9972 $\pm$ 0.0021 & 0.9930 $\pm$ 0.0014 & 0.9976 $\pm$ 0.0005 & 0.9990 $\pm$ 0.0003 & 0.9999 $\pm$ 0.0001 \\
12.0 & 0.9996 $\pm$ 0.0002 & 0.9989 $\pm$ 0.0006 & 0.9972 $\pm$ 0.0006 & 0.9972 $\pm$ 0.0036 & 0.9931 $\pm$ 0.0014 & 0.9977 $\pm$ 0.0004 & 0.9990 $\pm$ 0.0003 & 0.9999 $\pm$ 0.0001 \\
\bottomrule
\end{tabular}

\end{adjustbox}
\caption{Group H, Specificity, values after $\pm$ are standard deviations.}
\end{table}
\subsubsection*{Group I}
\begin{table}[H]
\centering
\begin{adjustbox}{max width=\linewidth}
\begin{tabular}{lllllllll}
\toprule
 & NB & NB\_p & BetaNB & BetaNB\_p & MCNB & MCNB\_p & binom & binom\_beta \\
$\tau$ &  &  &  &  &  &  &  &  \\
\midrule
0.0 & 0.9583 $\pm$ 0.0044 & 0.9590 $\pm$ 0.0044 & 0.9481 $\pm$ 0.0051 & 0.9499 $\pm$ 0.0052 & 0.9561 $\pm$ 0.0055 & 0.9361 $\pm$ 0.0111 & 0.9571 $\pm$ 0.0049 & 0.9382 $\pm$ 0.0054 \\
0.2 & 0.9559 $\pm$ 0.0052 & 0.9568 $\pm$ 0.0051 & 0.9468 $\pm$ 0.0054 & 0.9524 $\pm$ 0.0060 & 0.9500 $\pm$ 0.0050 & 0.9523 $\pm$ 0.0044 & 0.9545 $\pm$ 0.0050 & 0.9388 $\pm$ 0.0057 \\
0.4 & 0.9536 $\pm$ 0.0052 & 0.9544 $\pm$ 0.0052 & 0.9459 $\pm$ 0.0053 & 0.9481 $\pm$ 0.0072 & 0.9457 $\pm$ 0.0125 & 0.9388 $\pm$ 0.0121 & 0.9533 $\pm$ 0.0053 & 0.9394 $\pm$ 0.0053 \\
0.6 & 0.9497 $\pm$ 0.0041 & 0.9506 $\pm$ 0.0040 & 0.9427 $\pm$ 0.0046 & 0.9436 $\pm$ 0.0044 & 0.9401 $\pm$ 0.0109 & 0.9463 $\pm$ 0.0065 & 0.9503 $\pm$ 0.0040 & 0.9365 $\pm$ 0.0049 \\
0.8 & 0.9493 $\pm$ 0.0061 & 0.9501 $\pm$ 0.0060 & 0.9432 $\pm$ 0.0074 & 0.9439 $\pm$ 0.0066 & 0.9433 $\pm$ 0.0080 & 0.9436 $\pm$ 0.0090 & 0.9510 $\pm$ 0.0059 & 0.9396 $\pm$ 0.0066 \\
1.0 & 0.9482 $\pm$ 0.0048 & 0.9490 $\pm$ 0.0048 & 0.9422 $\pm$ 0.0051 & 0.9432 $\pm$ 0.0055 & 0.9466 $\pm$ 0.0046 & 0.9463 $\pm$ 0.0042 & 0.9502 $\pm$ 0.0048 & 0.9391 $\pm$ 0.0054 \\
\bottomrule
\end{tabular}

\end{adjustbox}
\caption{Group I, PR AUC, values after $\pm$ are standard deviations.}
\end{table}

\begin{table}[H]
\centering
\begin{adjustbox}{max width=\linewidth}
\begin{tabular}{lllllllll}
\toprule
 & BetaNB & BetaNB\_p & MCNB & MCNB\_p & NB & NB\_p & binom & binom\_beta \\
$\tau$ &  &  &  &  &  &  &  &  \\
\midrule
0.0 & 0.6779 $\pm$ 0.0175 & 0.7127 $\pm$ 0.0468 & 0.7644 $\pm$ 0.0122 & 0.8075 $\pm$ 0.0142 & 0.8228 $\pm$ 0.0116 & 0.8142 $\pm$ 0.0112 & 0.7143 $\pm$ 0.0140 & 0.4980 $\pm$ 0.0197 \\
0.2 & 0.6690 $\pm$ 0.0201 & 0.7489 $\pm$ 0.0540 & 0.7650 $\pm$ 0.0161 & 0.7665 $\pm$ 0.0173 & 0.8066 $\pm$ 0.0153 & 0.7984 $\pm$ 0.0154 & 0.7096 $\pm$ 0.0201 & 0.5097 $\pm$ 0.0174 \\
0.4 & 0.6602 $\pm$ 0.0194 & 0.7098 $\pm$ 0.0512 & 0.7201 $\pm$ 0.0189 & 0.7751 $\pm$ 0.0215 & 0.7931 $\pm$ 0.0150 & 0.7852 $\pm$ 0.0148 & 0.7018 $\pm$ 0.0176 & 0.5201 $\pm$ 0.0223 \\
0.6 & 0.6554 $\pm$ 0.0150 & 0.6790 $\pm$ 0.0352 & 0.7251 $\pm$ 0.0174 & 0.7324 $\pm$ 0.0148 & 0.7755 $\pm$ 0.0108 & 0.7680 $\pm$ 0.0108 & 0.6963 $\pm$ 0.0154 & 0.5296 $\pm$ 0.0162 \\
0.8 & 0.6558 $\pm$ 0.0199 & 0.6755 $\pm$ 0.0343 & 0.7063 $\pm$ 0.0194 & 0.7323 $\pm$ 0.0178 & 0.7741 $\pm$ 0.0155 & 0.7675 $\pm$ 0.0153 & 0.7039 $\pm$ 0.0183 & 0.5401 $\pm$ 0.0184 \\
1.0 & 0.6525 $\pm$ 0.0200 & 0.6691 $\pm$ 0.0401 & 0.6903 $\pm$ 0.0185 & 0.7173 $\pm$ 0.0176 & 0.7704 $\pm$ 0.0177 & 0.7627 $\pm$ 0.0171 & 0.6978 $\pm$ 0.0207 & 0.5408 $\pm$ 0.0184 \\
\bottomrule
\end{tabular}

\end{adjustbox}
\caption{Group I, Sensitivity, values after $\pm$ are standard deviations.}
\end{table}

\begin{table}[H]
\centering
\begin{adjustbox}{max width=\linewidth}
\begin{tabular}{lllllllll}
\toprule
 & BetaNB & BetaNB\_p & MCNB & MCNB\_p & NB & NB\_p & binom & binom\_beta \\
$\tau$ &  &  &  &  &  &  &  &  \\
\midrule
0.0 & 0.9991 $\pm$ 0.0002 & 0.9986 $\pm$ 0.0007 & 0.9983 $\pm$ 0.0004 & 0.9931 $\pm$ 0.0030 & 0.9968 $\pm$ 0.0006 & 0.9972 $\pm$ 0.0006 & 0.9993 $\pm$ 0.0003 & 0.9999 $\pm$ 0.0001 \\
0.2 & 0.9992 $\pm$ 0.0003 & 0.9984 $\pm$ 0.0007 & 0.9980 $\pm$ 0.0005 & 0.9981 $\pm$ 0.0006 & 0.9974 $\pm$ 0.0005 & 0.9978 $\pm$ 0.0005 & 0.9995 $\pm$ 0.0003 & 0.9999 $\pm$ 0.0001 \\
0.4 & 0.9993 $\pm$ 0.0002 & 0.9988 $\pm$ 0.0005 & 0.9981 $\pm$ 0.0018 & 0.9958 $\pm$ 0.0027 & 0.9977 $\pm$ 0.0005 & 0.9980 $\pm$ 0.0004 & 0.9993 $\pm$ 0.0002 & 0.9999 $\pm$ 0.0001 \\
0.6 & 0.9993 $\pm$ 0.0002 & 0.9990 $\pm$ 0.0004 & 0.9978 $\pm$ 0.0014 & 0.9982 $\pm$ 0.0006 & 0.9976 $\pm$ 0.0004 & 0.9979 $\pm$ 0.0004 & 0.9993 $\pm$ 0.0002 & 0.9998 $\pm$ 0.0001 \\
0.8 & 0.9994 $\pm$ 0.0002 & 0.9991 $\pm$ 0.0004 & 0.9985 $\pm$ 0.0006 & 0.9981 $\pm$ 0.0008 & 0.9979 $\pm$ 0.0004 & 0.9982 $\pm$ 0.0004 & 0.9993 $\pm$ 0.0003 & 0.9999 $\pm$ 0.0001 \\
1.0 & 0.9994 $\pm$ 0.0003 & 0.9992 $\pm$ 0.0005 & 0.9991 $\pm$ 0.0003 & 0.9987 $\pm$ 0.0004 & 0.9979 $\pm$ 0.0004 & 0.9982 $\pm$ 0.0004 & 0.9993 $\pm$ 0.0003 & 0.9998 $\pm$ 0.0001 \\
\bottomrule
\end{tabular}

\end{adjustbox}
\caption{Group I, Specificity, values after $\pm$ are standard deviations.}
\end{table}
\subsubsection*{Group J}
\begin{table}[H]
\centering
\begin{adjustbox}{max width=\linewidth}
\begin{tabular}{lllllllll}
\toprule
 & NB & NB\_p & BetaNB & BetaNB\_p & MCNB & MCNB\_p & binom & binom\_beta \\
$\tau$ &  &  &  &  &  &  &  &  \\
\midrule
0.0 & 0.1567 $\pm$ 0.0061 & 0.2062 $\pm$ 0.0083 & 0.2003 $\pm$ 0.0085 & 0.2362 $\pm$ 0.0100 & 0.2531 $\pm$ 0.0113 & 0.2574 $\pm$ 0.0109 & 0.0920 $\pm$ 0.0029 & 0.1937 $\pm$ 0.0084 \\
0.2 & 0.2373 $\pm$ 0.0121 & 0.2406 $\pm$ 0.0121 & 0.2955 $\pm$ 0.0143 & 0.2963 $\pm$ 0.0143 & 0.2406 $\pm$ 0.0119 & 0.2440 $\pm$ 0.0119 & 0.2440 $\pm$ 0.0123 & 0.2870 $\pm$ 0.0140 \\
0.4 & 0.2217 $\pm$ 0.0084 & 0.2251 $\pm$ 0.0086 & 0.2951 $\pm$ 0.0117 & 0.2958 $\pm$ 0.0118 & 0.2251 $\pm$ 0.0084 & 0.2290 $\pm$ 0.0085 & 0.2297 $\pm$ 0.0086 & 0.2942 $\pm$ 0.0118 \\
0.6 & 0.2103 $\pm$ 0.0090 & 0.2141 $\pm$ 0.0096 & 0.2973 $\pm$ 0.0153 & 0.2981 $\pm$ 0.0155 & 0.2141 $\pm$ 0.0091 & 0.2185 $\pm$ 0.0097 & 0.2192 $\pm$ 0.0097 & 0.3024 $\pm$ 0.0146 \\
0.8 & 0.2083 $\pm$ 0.0078 & 0.2124 $\pm$ 0.0081 & 0.3046 $\pm$ 0.0116 & 0.3048 $\pm$ 0.0117 & 0.2119 $\pm$ 0.0081 & 0.2168 $\pm$ 0.0083 & 0.2176 $\pm$ 0.0085 & 0.3106 $\pm$ 0.0120 \\
1.0 & 0.2056 $\pm$ 0.0099 & 0.2110 $\pm$ 0.0100 & 0.3122 $\pm$ 0.0174 & 0.3122 $\pm$ 0.0174 & 0.2096 $\pm$ 0.0103 & 0.2159 $\pm$ 0.0103 & 0.2151 $\pm$ 0.0102 & 0.3177 $\pm$ 0.0176 \\
\bottomrule
\end{tabular}

\end{adjustbox}
\caption{Group J, PR AUC, values after $\pm$ are standard deviations.}
\end{table}

\begin{table}[H]
\centering
\begin{adjustbox}{max width=\linewidth}
\begin{tabular}{lllllllll}
\toprule
 & BetaNB & BetaNB\_p & MCNB & MCNB\_p & NB & NB\_p & binom & binom\_beta \\
$\tau$ &  &  &  &  &  &  &  &  \\
\midrule
0.0 & 0.0542 $\pm$ 0.0083 & 0.0573 $\pm$ 0.0068 & 0.4793 $\pm$ 0.0166 & 0.4559 $\pm$ 0.0147 & 0.3998 $\pm$ 0.0146 & 0.3827 $\pm$ 0.0171 & 0.9976 $\pm$ 0.0014 & 0.0000 $\pm$ 0.0000 \\
0.2 & 0.1043 $\pm$ 0.0134 & 0.1042 $\pm$ 0.0134 & 0.4709 $\pm$ 0.0156 & 0.4353 $\pm$ 0.0146 & 0.5261 $\pm$ 0.0158 & 0.4933 $\pm$ 0.0167 & 0.4765 $\pm$ 0.0166 & 0.0664 $\pm$ 0.0107 \\
0.4 & 0.1078 $\pm$ 0.0124 & 0.1077 $\pm$ 0.0123 & 0.4677 $\pm$ 0.0165 & 0.4426 $\pm$ 0.0169 & 0.5212 $\pm$ 0.0159 & 0.4874 $\pm$ 0.0160 & 0.4775 $\pm$ 0.0177 & 0.0717 $\pm$ 0.0103 \\
0.6 & 0.1142 $\pm$ 0.0128 & 0.1149 $\pm$ 0.0123 & 0.4796 $\pm$ 0.0182 & 0.4387 $\pm$ 0.0171 & 0.5214 $\pm$ 0.0204 & 0.4937 $\pm$ 0.0193 & 0.4910 $\pm$ 0.0189 & 0.0748 $\pm$ 0.0106 \\
0.8 & 0.1204 $\pm$ 0.0127 & 0.1204 $\pm$ 0.0125 & 0.4745 $\pm$ 0.0138 & 0.4454 $\pm$ 0.0117 & 0.5282 $\pm$ 0.0112 & 0.5011 $\pm$ 0.0112 & 0.4999 $\pm$ 0.0117 & 0.0740 $\pm$ 0.0094 \\
1.0 & 0.1173 $\pm$ 0.0105 & 0.1176 $\pm$ 0.0112 & 0.4831 $\pm$ 0.0161 & 0.4481 $\pm$ 0.0163 & 0.5332 $\pm$ 0.0160 & 0.5042 $\pm$ 0.0138 & 0.5068 $\pm$ 0.0176 & 0.0763 $\pm$ 0.0087 \\
\bottomrule
\end{tabular}

\end{adjustbox}
\caption{Group J, Sensitivity, values after $\pm$ are standard deviations.}
\end{table}

\begin{table}[H]
\centering
\begin{adjustbox}{max width=\linewidth}
\begin{tabular}{lllllllll}
\toprule
 & BetaNB & BetaNB\_p & MCNB & MCNB\_p & NB & NB\_p & binom & binom\_beta \\
$\tau$ &  &  &  &  &  &  &  &  \\
\midrule
0.0 & 0.9941 $\pm$ 0.0008 & 0.9949 $\pm$ 0.0009 & 0.8521 $\pm$ 0.0151 & 0.8732 $\pm$ 0.0062 & 0.7992 $\pm$ 0.0040 & 0.8677 $\pm$ 0.0037 & 0.0008 $\pm$ 0.0002 & 1.0000 $\pm$ 0.0000 \\
0.2 & 0.9919 $\pm$ 0.0015 & 0.9920 $\pm$ 0.0015 & 0.8546 $\pm$ 0.0061 & 0.8793 $\pm$ 0.0049 & 0.8157 $\pm$ 0.0052 & 0.8416 $\pm$ 0.0046 & 0.8531 $\pm$ 0.0050 & 0.9962 $\pm$ 0.0004 \\
0.4 & 0.9919 $\pm$ 0.0010 & 0.9921 $\pm$ 0.0010 & 0.8424 $\pm$ 0.0058 & 0.8614 $\pm$ 0.0052 & 0.8018 $\pm$ 0.0036 & 0.8286 $\pm$ 0.0031 & 0.8371 $\pm$ 0.0034 & 0.9961 $\pm$ 0.0007 \\
0.6 & 0.9909 $\pm$ 0.0013 & 0.9909 $\pm$ 0.0013 & 0.8280 $\pm$ 0.0077 & 0.8584 $\pm$ 0.0045 & 0.7911 $\pm$ 0.0048 & 0.8173 $\pm$ 0.0048 & 0.8241 $\pm$ 0.0045 & 0.9958 $\pm$ 0.0007 \\
0.8 & 0.9912 $\pm$ 0.0013 & 0.9912 $\pm$ 0.0013 & 0.8275 $\pm$ 0.0042 & 0.8511 $\pm$ 0.0055 & 0.7819 $\pm$ 0.0043 & 0.8088 $\pm$ 0.0036 & 0.8138 $\pm$ 0.0039 & 0.9963 $\pm$ 0.0008 \\
1.0 & 0.9922 $\pm$ 0.0010 & 0.9922 $\pm$ 0.0010 & 0.8160 $\pm$ 0.0052 & 0.8452 $\pm$ 0.0041 & 0.7721 $\pm$ 0.0043 & 0.8010 $\pm$ 0.0045 & 0.8039 $\pm$ 0.0040 & 0.9968 $\pm$ 0.0007 \\
\bottomrule
\end{tabular}

\end{adjustbox}
\caption{Group J, Specificity, values after $\pm$ are standard deviations.}
\end{table}
\subsubsection*{Group K}
\begin{table}[H]
\centering
\begin{adjustbox}{max width=\linewidth}
\begin{tabular}{lllllllll}
\toprule
 & NB & NB\_p & BetaNB & BetaNB\_p & MCNB & MCNB\_p & binom & binom\_beta \\
$bias$ &  &  &  &  &  &  &  &  \\
\midrule
1.1000 & 0.9541 $\pm$ 0.0047 & 0.9561 $\pm$ 0.0047 & 0.9442 $\pm$ 0.0056 & 0.9451 $\pm$ 0.0056 & 0.9533 $\pm$ 0.0057 & 0.9283 $\pm$ 0.0217 & 0.9509 $\pm$ 0.0051 & 0.9312 $\pm$ 0.0058 \\
1.2000 & 0.9443 $\pm$ 0.0048 & 0.9491 $\pm$ 0.0046 & 0.9349 $\pm$ 0.0057 & 0.9347 $\pm$ 0.0056 & 0.9428 $\pm$ 0.0059 & 0.9250 $\pm$ 0.0114 & 0.9314 $\pm$ 0.0062 & 0.9040 $\pm$ 0.0084 \\
1.3000 & 0.9296 $\pm$ 0.0055 & 0.9374 $\pm$ 0.0052 & 0.9141 $\pm$ 0.0075 & 0.9185 $\pm$ 0.0067 & 0.9173 $\pm$ 0.0218 & 0.9050 $\pm$ 0.0208 & 0.8972 $\pm$ 0.0079 & 0.8392 $\pm$ 0.0107 \\
1.4000 & 0.9101 $\pm$ 0.0057 & 0.9191 $\pm$ 0.0058 & 0.8793 $\pm$ 0.0138 & 0.8909 $\pm$ 0.0176 & 0.8923 $\pm$ 0.0277 & 0.8977 $\pm$ 0.0142 & 0.8510 $\pm$ 0.0107 & 0.7011 $\pm$ 0.0184 \\
1.5000 & 0.8869 $\pm$ 0.0066 & 0.8925 $\pm$ 0.0074 & 0.8346 $\pm$ 0.0334 & 0.8426 $\pm$ 0.0459 & 0.8614 $\pm$ 0.0254 & 0.8685 $\pm$ 0.0154 & 0.7939 $\pm$ 0.0130 & 0.5467 $\pm$ 0.0206 \\
\bottomrule
\end{tabular}

\end{adjustbox}
\caption{Group K, PR AUC, values after $\pm$ are standard deviations.}
\end{table}

\begin{table}[H]
\centering
\begin{adjustbox}{max width=\linewidth}
\begin{tabular}{lllllllll}
\toprule
 & BetaNB & BetaNB\_p & MCNB & MCNB\_p & NB & NB\_p & binom & binom\_beta \\
$bias$ &  &  &  &  &  &  &  &  \\
\midrule
1.1000 & 0.6594 $\pm$ 0.0184 & 0.6728 $\pm$ 0.0337 & 0.7613 $\pm$ 0.0164 & 0.8069 $\pm$ 0.0183 & 0.8211 $\pm$ 0.0115 & 0.8126 $\pm$ 0.0120 & 0.7137 $\pm$ 0.0163 & 0.4747 $\pm$ 0.0191 \\
1.2000 & 0.6398 $\pm$ 0.0452 & 0.6203 $\pm$ 0.0381 & 0.7592 $\pm$ 0.0191 & 0.7977 $\pm$ 0.0168 & 0.8200 $\pm$ 0.0146 & 0.8106 $\pm$ 0.0139 & 0.7129 $\pm$ 0.0181 & 0.3994 $\pm$ 0.0234 \\
1.3000 & 0.5935 $\pm$ 0.0550 & 0.5408 $\pm$ 0.0413 & 0.7615 $\pm$ 0.0171 & 0.7919 $\pm$ 0.0198 & 0.8140 $\pm$ 0.0156 & 0.8071 $\pm$ 0.0161 & 0.7113 $\pm$ 0.0151 & 0.3014 $\pm$ 0.0180 \\
1.4000 & 0.5071 $\pm$ 0.0636 & 0.4591 $\pm$ 0.0621 & 0.7589 $\pm$ 0.0167 & 0.7760 $\pm$ 0.0178 & 0.8093 $\pm$ 0.0164 & 0.8008 $\pm$ 0.0166 & 0.7058 $\pm$ 0.0158 & 0.2220 $\pm$ 0.0175 \\
1.5000 & 0.3997 $\pm$ 0.0875 & 0.3796 $\pm$ 0.0813 & 0.7583 $\pm$ 0.0159 & 0.7692 $\pm$ 0.0176 & 0.8069 $\pm$ 0.0161 & 0.7970 $\pm$ 0.0175 & 0.7031 $\pm$ 0.0149 & 0.1977 $\pm$ 0.0161 \\
\bottomrule
\end{tabular}

\end{adjustbox}
\caption{Group K, Sensitivity, values after $\pm$ are standard deviations.}
\end{table}

\begin{table}[H]
\centering
\begin{adjustbox}{max width=\linewidth}
\begin{tabular}{lllllllll}
\toprule
 & BetaNB & BetaNB\_p & MCNB & MCNB\_p & NB & NB\_p & binom & binom\_beta \\
$bias$ &  &  &  &  &  &  &  &  \\
\midrule
1.1000 & 0.9991 $\pm$ 0.0003 & 0.9990 $\pm$ 0.0005 & 0.9983 $\pm$ 0.0005 & 0.9911 $\pm$ 0.0077 & 0.9965 $\pm$ 0.0007 & 0.9971 $\pm$ 0.0005 & 0.9991 $\pm$ 0.0004 & 0.9999 $\pm$ 0.0001 \\
1.2000 & 0.9989 $\pm$ 0.0006 & 0.9991 $\pm$ 0.0005 & 0.9976 $\pm$ 0.0005 & 0.9918 $\pm$ 0.0031 & 0.9952 $\pm$ 0.0008 & 0.9963 $\pm$ 0.0006 & 0.9983 $\pm$ 0.0004 & 0.9999 $\pm$ 0.0001 \\
1.3000 & 0.9983 $\pm$ 0.0009 & 0.9993 $\pm$ 0.0006 & 0.9938 $\pm$ 0.0047 & 0.9875 $\pm$ 0.0067 & 0.9927 $\pm$ 0.0009 & 0.9945 $\pm$ 0.0008 & 0.9960 $\pm$ 0.0007 & 0.9998 $\pm$ 0.0001 \\
1.4000 & 0.9975 $\pm$ 0.0016 & 0.9986 $\pm$ 0.0019 & 0.9895 $\pm$ 0.0071 & 0.9881 $\pm$ 0.0031 & 0.9883 $\pm$ 0.0012 & 0.9913 $\pm$ 0.0009 & 0.9910 $\pm$ 0.0011 & 0.9997 $\pm$ 0.0001 \\
1.5000 & 0.9966 $\pm$ 0.0030 & 0.9966 $\pm$ 0.0046 & 0.9823 $\pm$ 0.0075 & 0.9826 $\pm$ 0.0035 & 0.9815 $\pm$ 0.0013 & 0.9856 $\pm$ 0.0011 & 0.9805 $\pm$ 0.0016 & 0.9997 $\pm$ 0.0002 \\
\bottomrule
\end{tabular}

\end{adjustbox}
\caption{Group K, Specificity, values after $\pm$ are standard deviations.}
\end{table}
\subsubsection*{Group L}
\begin{table}[H]
\centering
\begin{adjustbox}{max width=\linewidth}
\begin{tabular}{lllllllll}
\toprule
 & NB & NB\_p & BetaNB & BetaNB\_p & MCNB & MCNB\_p & binom & binom\_beta \\
$\kappa$ &  &  &  &  &  &  &  &  \\
\midrule
16 & 0.1019 $\pm$ 0.0041 & 0.1029 $\pm$ 0.0041 & 0.1097 $\pm$ 0.0036 & 0.1096 $\pm$ 0.0036 & 0.1012 $\pm$ 0.0038 & 0.1029 $\pm$ 0.0040 & 0.1097 $\pm$ 0.0052 & 0.1093 $\pm$ 0.0036 \\
32 & 0.1133 $\pm$ 0.0055 & 0.1151 $\pm$ 0.0057 & 0.1328 $\pm$ 0.0103 & 0.1327 $\pm$ 0.0104 & 0.1133 $\pm$ 0.0053 & 0.1158 $\pm$ 0.0055 & 0.1150 $\pm$ 0.0055 & 0.1318 $\pm$ 0.0104 \\
64 & 0.1352 $\pm$ 0.0059 & 0.1384 $\pm$ 0.0060 & 0.1746 $\pm$ 0.0067 & 0.1745 $\pm$ 0.0067 & 0.1355 $\pm$ 0.0059 & 0.1395 $\pm$ 0.0061 & 0.1357 $\pm$ 0.0062 & 0.1726 $\pm$ 0.0070 \\
128 & 0.1859 $\pm$ 0.0074 & 0.1914 $\pm$ 0.0076 & 0.2485 $\pm$ 0.0123 & 0.2488 $\pm$ 0.0122 & 0.1869 $\pm$ 0.0073 & 0.1941 $\pm$ 0.0077 & 0.1894 $\pm$ 0.0074 & 0.2369 $\pm$ 0.0116 \\
256 & 0.3059 $\pm$ 0.0138 & 0.3133 $\pm$ 0.0136 & 0.3509 $\pm$ 0.0126 & 0.3518 $\pm$ 0.0126 & 0.3063 $\pm$ 0.0136 & 0.3134 $\pm$ 0.0151 & 0.3119 $\pm$ 0.0134 & 0.3161 $\pm$ 0.0148 \\
512 & 0.4649 $\pm$ 0.0179 & 0.4697 $\pm$ 0.0172 & 0.4546 $\pm$ 0.0247 & 0.4605 $\pm$ 0.0151 & 0.4604 $\pm$ 0.0177 & 0.4597 $\pm$ 0.0208 & 0.4617 $\pm$ 0.0168 & 0.4027 $\pm$ 0.0144 \\
1024 & 0.5696 $\pm$ 0.0150 & 0.5729 $\pm$ 0.0157 & 0.5213 $\pm$ 0.0369 & 0.5352 $\pm$ 0.0258 & 0.5539 $\pm$ 0.0240 & 0.5581 $\pm$ 0.0173 & 0.5532 $\pm$ 0.0162 & 0.4811 $\pm$ 0.0205 \\
\bottomrule
\end{tabular}

\end{adjustbox}
\caption{Group L, PR AUC, values after $\pm$ are standard deviations.}
\end{table}

\begin{table}[H]
\centering
\begin{adjustbox}{max width=\linewidth}
\begin{tabular}{lllllllll}
\toprule
 & BetaNB & BetaNB\_p & MCNB & MCNB\_p & NB & NB\_p & binom & binom\_beta \\
$\kappa$ &  &  &  &  &  &  &  &  \\
\midrule
16 & 0.1220 $\pm$ 0.0097 & 0.1206 $\pm$ 0.0096 & 0.7188 $\pm$ 0.0162 & 0.5923 $\pm$ 0.0174 & 0.7368 $\pm$ 0.0133 & 0.6372 $\pm$ 0.0161 & 0.7716 $\pm$ 0.0106 & 0.0924 $\pm$ 0.0091 \\
32 & 0.1210 $\pm$ 0.0166 & 0.1203 $\pm$ 0.0163 & 0.6764 $\pm$ 0.0151 & 0.6017 $\pm$ 0.0170 & 0.6952 $\pm$ 0.0143 & 0.6294 $\pm$ 0.0148 & 0.6961 $\pm$ 0.0134 & 0.0858 $\pm$ 0.0116 \\
64 & 0.1177 $\pm$ 0.0090 & 0.1173 $\pm$ 0.0089 & 0.6216 $\pm$ 0.0172 & 0.5729 $\pm$ 0.0180 & 0.6404 $\pm$ 0.0152 & 0.5932 $\pm$ 0.0136 & 0.6158 $\pm$ 0.0119 & 0.0773 $\pm$ 0.0089 \\
128 & 0.0979 $\pm$ 0.0094 & 0.0979 $\pm$ 0.0095 & 0.5475 $\pm$ 0.0181 & 0.5110 $\pm$ 0.0168 & 0.5751 $\pm$ 0.0145 & 0.5432 $\pm$ 0.0143 & 0.5336 $\pm$ 0.0150 & 0.0615 $\pm$ 0.0102 \\
256 & 0.0821 $\pm$ 0.0142 & 0.0818 $\pm$ 0.0144 & 0.4488 $\pm$ 0.0181 & 0.4310 $\pm$ 0.0217 & 0.5078 $\pm$ 0.0159 & 0.4779 $\pm$ 0.0179 & 0.4431 $\pm$ 0.0189 & 0.0451 $\pm$ 0.0109 \\
512 & 0.0747 $\pm$ 0.0337 & 0.0666 $\pm$ 0.0104 & 0.3490 $\pm$ 0.0210 & 0.3484 $\pm$ 0.0220 & 0.4373 $\pm$ 0.0166 & 0.4064 $\pm$ 0.0166 & 0.3494 $\pm$ 0.0155 & 0.0360 $\pm$ 0.0087 \\
1024 & 0.0852 $\pm$ 0.0518 & 0.0650 $\pm$ 0.0340 & 0.2863 $\pm$ 0.0265 & 0.2809 $\pm$ 0.0262 & 0.3788 $\pm$ 0.0143 & 0.3529 $\pm$ 0.0173 & 0.2885 $\pm$ 0.0194 & 0.0265 $\pm$ 0.0087 \\
\bottomrule
\end{tabular}

\end{adjustbox}
\caption{Group L, Sensitivity, values after $\pm$ are standard deviations.}
\end{table}

\begin{table}[H]
\centering
\begin{adjustbox}{max width=\linewidth}
\begin{tabular}{lllllllll}
\toprule
 & BetaNB & BetaNB\_p & MCNB & MCNB\_p & NB & NB\_p & binom & binom\_beta \\
$\kappa$ &  &  &  &  &  &  &  &  \\
\midrule
16 & 0.9202 $\pm$ 0.0036 & 0.9212 $\pm$ 0.0034 & 0.3171 $\pm$ 0.0102 & 0.4585 $\pm$ 0.0051 & 0.2950 $\pm$ 0.0075 & 0.4059 $\pm$ 0.0047 & 0.2542 $\pm$ 0.0045 & 0.9431 $\pm$ 0.0024 \\
32 & 0.9445 $\pm$ 0.0033 & 0.9448 $\pm$ 0.0034 & 0.3986 $\pm$ 0.0077 & 0.4881 $\pm$ 0.0113 & 0.3746 $\pm$ 0.0075 & 0.4550 $\pm$ 0.0050 & 0.3697 $\pm$ 0.0047 & 0.9640 $\pm$ 0.0029 \\
64 & 0.9691 $\pm$ 0.0020 & 0.9692 $\pm$ 0.0020 & 0.5337 $\pm$ 0.0075 & 0.6000 $\pm$ 0.0124 & 0.5148 $\pm$ 0.0035 & 0.5728 $\pm$ 0.0030 & 0.5433 $\pm$ 0.0035 & 0.9829 $\pm$ 0.0020 \\
128 & 0.9893 $\pm$ 0.0014 & 0.9894 $\pm$ 0.0014 & 0.7261 $\pm$ 0.0138 & 0.7672 $\pm$ 0.0086 & 0.6996 $\pm$ 0.0065 & 0.7382 $\pm$ 0.0052 & 0.7427 $\pm$ 0.0053 & 0.9951 $\pm$ 0.0010 \\
256 & 0.9970 $\pm$ 0.0008 & 0.9970 $\pm$ 0.0008 & 0.8961 $\pm$ 0.0096 & 0.9078 $\pm$ 0.0098 & 0.8647 $\pm$ 0.0031 & 0.8864 $\pm$ 0.0034 & 0.9015 $\pm$ 0.0030 & 0.9988 $\pm$ 0.0004 \\
512 & 0.9982 $\pm$ 0.0041 & 0.9992 $\pm$ 0.0005 & 0.9738 $\pm$ 0.0038 & 0.9733 $\pm$ 0.0055 & 0.9555 $\pm$ 0.0030 & 0.9643 $\pm$ 0.0028 & 0.9743 $\pm$ 0.0023 & 0.9997 $\pm$ 0.0003 \\
1024 & 0.9982 $\pm$ 0.0026 & 0.9993 $\pm$ 0.0015 & 0.9923 $\pm$ 0.0037 & 0.9927 $\pm$ 0.0021 & 0.9859 $\pm$ 0.0015 & 0.9889 $\pm$ 0.0015 & 0.9928 $\pm$ 0.0010 & 0.9999 $\pm$ 0.0001 \\
\bottomrule
\end{tabular}

\end{adjustbox}
\caption{Group L, Specificity, values after $\pm$ are standard deviations.}
\end{table}

		\section{Full benchmark figure}
		
%
\captionof{figure}{Uncensored 18+ version of the Figure~\ref{fig:benchmark}. }
\label{fig:benchmark_full}

	\end{appendix}
\end{document}